\documentclass[fleqn,usenatbib]{mnras}

\usepackage{newtxtext,newtxmath}

\usepackage[T1]{fontenc}

\DeclareRobustCommand{\VAN}[3]{#2}
\let\VANthebibliography\thebibliography
\def\thebibliography{\DeclareRobustCommand{\VAN}[3]{##3}\VANthebibliography}

\usepackage{graphicx}	
\usepackage{amsmath}	

\title{Interpretation of Radio Afterglows in the framework of the Standard Fireball and Energy Injection models}

\author[D. Levine et al.]{
D. Levine$^{1}$,
M. Dainotti,$^{2, 3, 4}$\thanks{E-mail: maria.dainotti@nao.ac.jp}
N. Fraija$^{5}$,
D. Warren$^{6}$,
P. Chandra $^{7, 8}$,
N. Lloyd-Ronning $^{9, 10}$
\\
$^1$ Department of Astronomy, University of Maryland, College Park, MD 20742, USA\\
$^{2}$National Astronomical Observatory of Japan, 2-21-1 Osawa, Mitaka, Tokyo 181-8588, Japan\\
$^{3}$Space Science Institute, Boulder, CO, USA\\
$^{4}$The Graduate University for Advanced Studies, SOKENDAI, Shonankokusaimura, Hayama, Miura District, Kanagawa 240-0193, Japan \\
$^5$Instituto de Astronomía, Universidad Nacional Autónoma de México Circuito Exterior, C.U., A. Postal 70-264, 04510 México D.F., México\\
$^6$ RIKEN Interdisciplinary Theoretical and Mathematical Sciences Program (iTHEMS), Wakō, Saitama, 351-0198 Japan \\
$^7$ National Centre for Radio Astrophysics, Tata Institute of Fundamental Research, Ghaneshkhind Pune 411007, India \\
$^8$ National Radio Astronomy Observatory, 520 Edgemont Rd, Charlottesville, VA, 22903 USA \\
$^9$ Computational Physics and Methods Group (CCS-2), Los Alamos National Lab, Los Alamos, NM, USA 87545 \\
$^10$ Department of Science and Engineering, University of New Mexico, Los Alamos, 87544 \\
}

\date{Accepted XXX. Received YYY; in original form ZZZ}

\pubyear{2022}

\begin{document}
\label{firstpage}
\pagerange{\pageref{firstpage}--\pageref{lastpage}}
\maketitle

\begin{abstract}
Gamma-Ray Bursts (GRBs) are panchromatic, highly energetic transients whose energy emission mechanism is still debated. One of the possible explanations is the standard fireball model, which can be tested with the closure relations (CRs), or relations between the temporal and spectral indices of a GRB. To test these, we compile an extensive sample of radio afterglow light curves (LCs) that span from 1997 to 2020, the most comprehensive analysis of GRBs with radio observations to date. We fit 202 LCs from 82 distinct GRBs with a broken power law, obtaining a sample of 26 that display a clear break and a subsample of 14 GRBs that present a radio plateau. We test these samples against CRs corresponding to a constant-density interstellar medium (ISM) or a stellar wind medium in both fast- and slow-cooling regimes, as well as three additional density profiles, $k = 1, 1.5, 2.5$, following $n \propto r^{-k}$, and consider sets of CRs both with and without energy injection. We find that 12 of the 26 GRBs (46\%), of which 7/12 present a radio plateau, fulfill at least one CR in the sets tested, suggesting our data is largely incompatible with the standard fireball model. Of the fulfilled CRs, the most preferred environment is the ISM, SC, $\nu_{\rm m} < \nu < \nu_{\rm c}$ without energy injection. Our results are consistent with previous studies that test the standard fireball model via the CRs in radio.
\end{abstract}




\section{Introduction}
\label{sec:intro} 

Gamma-Ray Bursts (GRBs) are highly energetic phenomena that release more energy within seconds than our Sun does in its lifetime. They can be observed in all wavelengths, ranging from TeV \citep{2019Natur.575..455M} to radio \citep{1997Natur.386..686V}. Though they were discovered more than 50 years ago \citep{1973ApJ...182L..85K}, they remain mysterious, with both their origin and emission mechanisms still under investigation by the scientific community. Observations of GRBs have found two main phases of emission, the initial ``prompt" emission and the longer-lasting ``afterglow" emission; and has led to their categorization into two main classes: long (LGRBs) with $T_{90} > 2$s \citep{1981Ap&SS..80....3M, 1993ApJ...413L.101K} and short (SGRBs) with $T_{90} < 2$s. The former group are thought to come from the core collapse of massive stars or a highly-magnetized newborn magnetar, while the latter group are thought to originate from the coalescence of a binary system, such as a neutron star and a black hole (NS-BH) or two neutron stars (NS-NS, \cite{1993ApJ...405..273W,1998ApJ...507L..59L, 2013ApJ...774...25K}).

Regardless of their origins, the prompt and afterglow phases of GRB emission, as well as the dynamics of GRB outflow and their radiation processes, can be explained by the standard fireball model \citep{1992MNRAS.257P..29M, 1993ApJ...405..278M, 1993MNRAS.263..861P, 1993ApJ...403L..67P, 1999ApJ...513..679G,1999ApJ...513..669K, 1999PhR...314..575P, SPH99, 1999ApJ...517L.109S,  2000ApJ...541L...5M, 2002ApJ...568..820G, 2006ApJ...651.1005S,  2006ARA&A..44..507W, 2007PhR...442..166N, 2011ApJ...738L..32G, 2014MNRAS.437L...6K, 2015MNRAS.450.1430H}. The dynamics of the relativistic jet and the non-relativistic ejected materials are described by the Blandford-Mckee \citep{1976PhFl...19.1130B} and Sedov-Taylor \citep{1950RSPSA.201..159T} self-similar solution, respectively. The relativistic outflow produces multi-wavelength afterglow observations (including radio, though it may be absorbed by synchrotron self-absorption early on) in a timescale from seconds to hours, weeks, or even months.

The validity of this fireball model can be tested through closure relations (CRs), a set of theoretical relationships between the afterglow's temporal index ($\alpha$) and the synchrotron emission's spectral index ($\beta$) \citep{1998ApJ...497L..17S, 2002ApJ...571..779P}.\footnote{The $F_{\nu} \propto t^{-\alpha}\nu^{-\beta}$ convention is used throughout this paper \citep{1998ApJ...497L..17S,2002ApJ...568..820G}.} CRs are typically tested in two environments, the interstellar medium (ISM) and stellar wind environment. CRs have previously been extensively tested in high-energy $\gamma$-rays by \citet{2019ApJ...883..134T, 2020ApJ...905..112F, 2021ApJS..255...13D, 2021ApJ...907...78F, 2022arXiv220611490F}, in X-ray samples (observed by \textit{Swift}\footnote{\label{foot:swfit}Detected by the \textit{Neil Gehrels Swift Observatory}) \citep{2004ApJ...611.1005G}.} \citet{2007ApJ...662.1093W, 2009ApJ...698...43R, 2020ApJ...903...18S, 2021PASJ..tmp...63D} and in optical by \citet{2015ApJS..219....9W, 2017ApJ...844...92F, 2022arXiv221003870D}, \citep[for further review see:][]{2004IJMPA..19.2385Z,2013NewAR..57..141G, 2020ApJ...896..166R}. 

More recently, CRs have also been tested in radio. Though the relative scarcity of radio data has made such investigations more challenging, one study has been able to test CRs against a group of radio GRBs, while other authors have examined specific GRBs. \citet{2021ApJ...911...14K} (KF21) tested a sample of 21 \textit{Swift} GRBs, using standard CRs to compare the break and decay in radio LCs with those in X-ray and optical. They found that most of their sample was incompatible with expectations of the standard fireball model: the late-time decay in the radio band did not agree with the decay seen in X-ray and optical wavelengths, and did not match the expected post-break power law of $t^{-2}$ in most cases. The authors found a few GRBs where the radio decay matched the pre-break expectations, but at times much later than the observed jet break.

Regarding specific GRBs in radio, \citet{2021MNRAS.504.5685M} studied GRB 190114C, testing the X-ray, optical, and radio afterglows against the set of CRs for the standard model. They found that the data are incompatible with the standard model, and instead proposed that the time evolution of the microphysical parameters is needed to explain the behavior of the afterglow. \citet{2019ApJ...871..123F} studied the standard CRs in the X-ray, optical, and radio LCs of GRB 170817 and found them compatible with a slow-cooling regime. A study by \citet{2015ApJ...810...31V} of GRB 130907A comparing CRs in X-ray and radio found contradicting results; the X-ray relations support a Wind environment, while the radio relations support an ISM environment.

In the current investigation, we examine whether the previously-seen deviation of radio afterglows from the expectations of the standard fireball model can be reconciled by invoking energy injection. In the literature, CRs are more commonly tested in a scenario without energy injection, which is expected to describe GRB light curves (LCs) with a power law decay phase immediately following the prompt emission. However, CRs can also be tested in a scenario that considers energy injection, which is more suitable to explain the GRB plateau feature, or an observed flattening in the LC that follows the prompt episode \citep{2006ApJ...642..389N, 2006ApJ...647.1213O, 2007ApJ...669.1115S, 2019ApJ...883...97Z} and precedes the power law decay. Theoretically, this plateau emission has been explained as continuous energy injection from a central engine \citep{1998A&A...333L..87D,1998ApJ...496L...1R,2000ApJ...535L..33S,2001ApJ...552L..35Z,2006ApJ...642..389N, 2006MNRAS.373..729Z,2007ApJ...670..565L}, either the accretion of the progenitor's stellar envelope onto a newly-formed black hole \citep{2008Sci...321..376K,2009ApJ...700.1047C,2011ApJ...734...35C,2017MNRAS.472.3058B,2018ApJ...857...95M}, or the spin-down luminosity of a millisecond neutron star, called a magnetar \citep{2001ApJ...552L..35Z,  2007ApJ...659.1420T, 2007ApJ...665..599T, 2011A&A...526A.121D, 2013ApJ...776..106H, 2013MNRAS.430.1061R, 2014MNRAS.443.1779R, 2015ApJ...813...92R, 2017A&A...605A..60B, 2018ApJ...857...95M, 2018ApJ...869..155S, 2020arXiv200311252F}. 

Thus, in order for us to investigate the role of the energy injection in relation to the deviation of the radio afterglows from the standard fireball model we test CRs both with and without energy injection.

The paper is structured as follows: in Section \ref{sec:sample} we present our sample selection, with a brief comparison between the LCs displaying a plateau in radio to the plateaus in X-ray wavelengths, and in Section \ref{sec:method} we explain the process used in the CR analysis. In Section \ref{sec:results}, we present the results of testing the CRs against the plateau subsample (\ref{sec:plateau}) and the full sample (\ref{sec:full}). In Section \ref{sec:discussion}, we discuss our results and present our conclusions.

\section{Data Sample} \label{sec:sample}

Following the same methodology as in \citet{2022ApJ...925...15L}, we compile a sample of 404 GRBs from the literature with observed radio afterglow, adding 100 GRBs gathered from the literature between 2011 and 2020 to the \citet{2012ApJ...746..156C} sample of 304 GRBs observed from 1997 to 2011. We then discard 193 GRBs reporting only upper limits on observed energy flux, 127 GRBs with $< 5$ observations within the same frequency, and 2 without known redshift, leaving us with 82 GRBs. Associated with these 82 distinct GRBs are 202 light curves (LCs), as various GRBs carried observations in multiple frequencies.

The 202 LCs were fit to the broken power law (BPL) model using the formulation of \citet{1999A&A...352L..26B}: 
\begin{equation}
F(t)  = \begin{cases}
F_a (\frac{t}{T_a})^{-\alpha_1} & t < T_a \\
F_a (\frac{t}{T_a})^{-\alpha_2} & t \geq T_a
\end{cases}
\label{eq:BPL}
\end{equation}
where $F_{a}$ is the observed energy flux at the end of the plateau emission (if $\alpha<0.5$) or a break if this condition is not fulfilled in erg cm$^{-2}$s$^{-1}$, $T_{\rm a}$ is the time of the break in seconds at the end of the plateau emission, and $\alpha_{1}$ and $\alpha_{2}$ correspond to the temporal and decay indices of the power law before and after the break, respectively. The fitting parameters for all 202 GRBs are given in Table \ref{tab:fitting} (full table is available online).

To ``approve'' a fit, we require a break to be present and for the fitting results to convey an acceptable $\Delta \chi^2$ analysis (varying $\Delta \chi^2$ until the $1\sigma$ bounds can be determined, assuming the $\Delta \chi^2$ is parabolic) according to the \citet{1978ApJ...222L.113A} prescription\footnote{Please refer to \citet{1978ApJ...222L.113A} for a complete discussion.} leading us to reject 137 LCs for not fulfilling this criteria. We further reject 3 GRBs for which $\sigma_{\alpha_2}/\alpha_2 > 90\%$ (for the $\alpha_1$ parameter, when $|\alpha_1| < 0.5$, we do not remove GRBs for which $\sigma_{\alpha_1}/\alpha_1 > 90\%$ as this creates a bias against LCs that display a plateau), and 20 LCs for displaying a shape that disagrees with the expectations of the simple BPL (i.e. simple PL). We also require only one LC for each GRB in our analysis, so for GRBs that have well-fitted LCs in multiple frequencies, we choose the LC with the best coverage. After removing 11 LCs from GRBs with different frequencies\footnote{A note about bursts with LCs in different frequencies - for the majority of these, $\alpha_1$, $\alpha_2$, and their respective errors can only be obtained for one (if any) of the LCs. For the few GRBs with successful fits in multiple frequencies, there is no clear trend regarding the agreement of the temporal indices for each GRB.} we are left with 31 GRBs that pass the fitting criteria.

As we seek to investigate the relationship between the temporal and spectral indices of the GRB, we collect the radio spectral index, $\beta$ value and its uncertainties, from the literature. We assume a flat spectrum (constant spectral index for the duration of the emission) and record $\beta$ values from the literature observed at a time within 1$\sigma$ of our fitted $T_{\rm a}$, in seconds at the end of the plateau emission. If the radio spectral index could not be found in the literature, we estimate $\beta$ using observational data. For GRBs with observations in multiple frequencies, we use one set of coincident observations in all available frequencies to obtain the flux density vs. frequency relation and perform a linear regression analysis, taking the slope as the $\beta$ value. Five GRBs in our sample only have available data in one frequency, therefore, the radio spectral index could not be estimated and we remove them from our sample, leaving us with a final sample of 26 GRBs.

\begin{figure*}
\begin{tabular}{ccc}
\includegraphics[width = 0.28\textwidth]{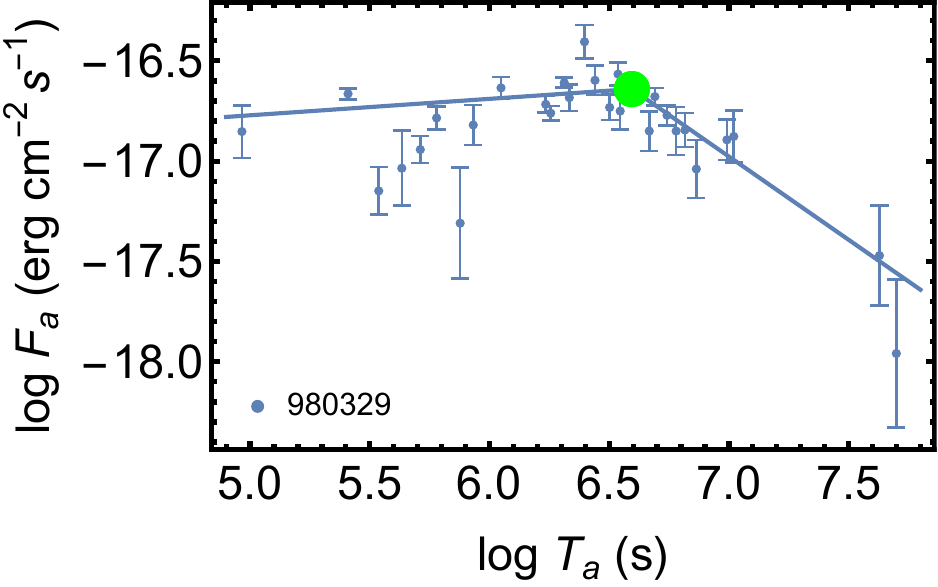} &
\includegraphics[width = 0.28\textwidth]{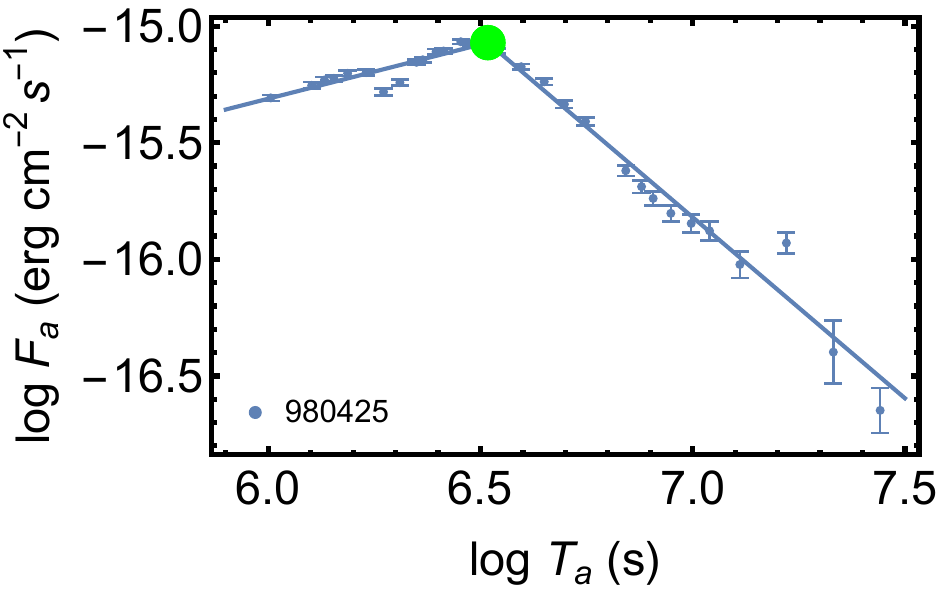} & 
\includegraphics[width = 0.28\textwidth]{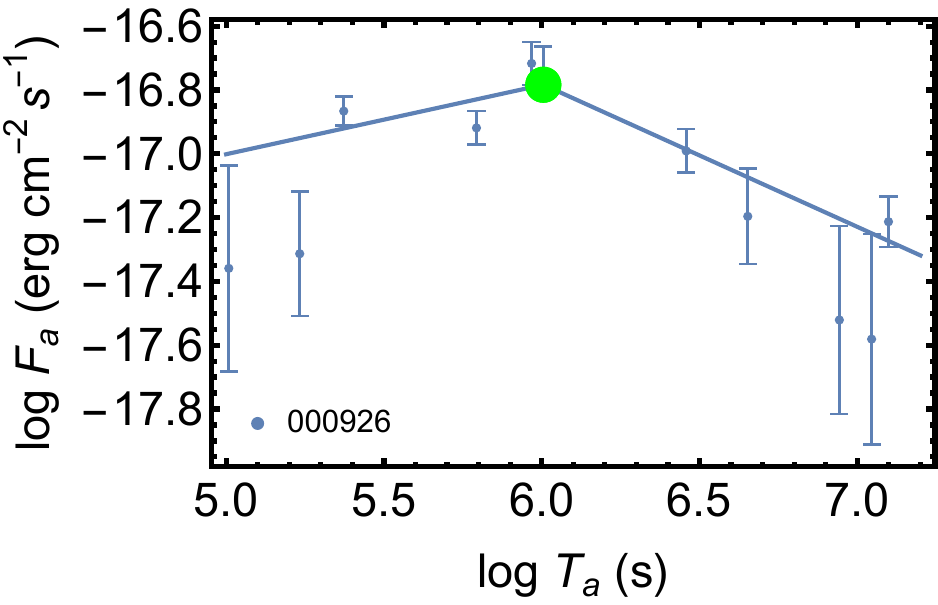} \\
\includegraphics[width = 0.28\textwidth]{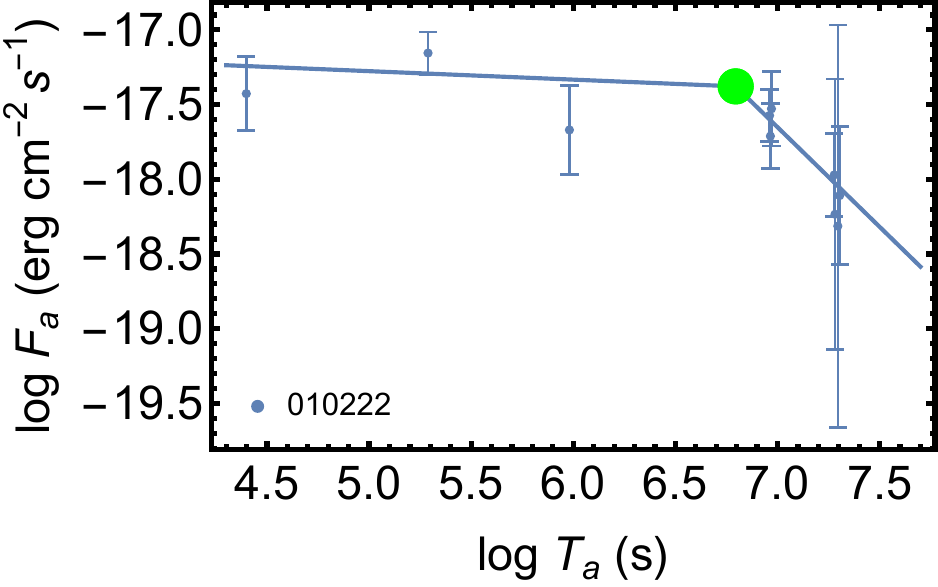} &
\includegraphics[width = 0.28\textwidth]{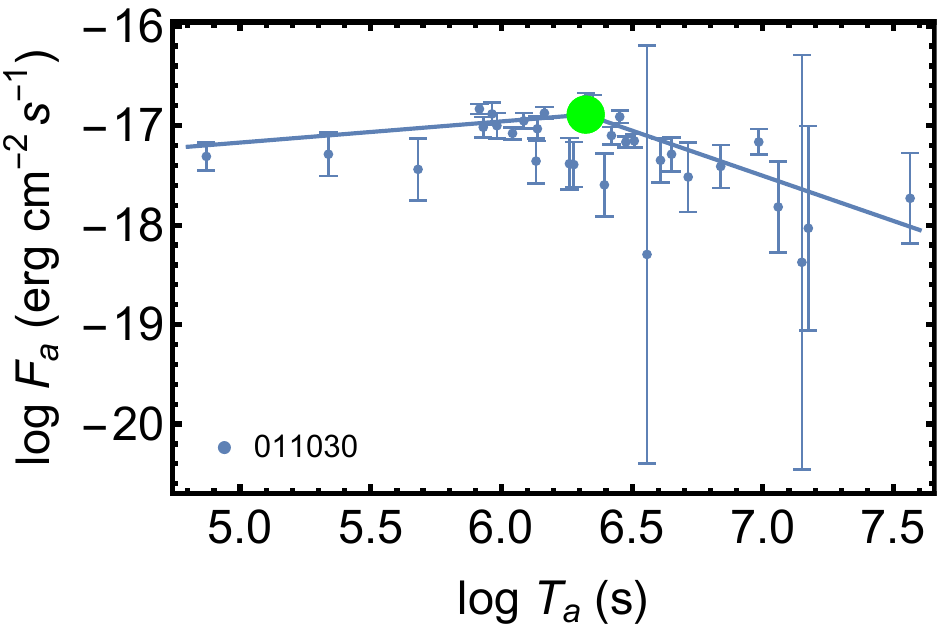} &
\includegraphics[width = 0.28\textwidth]{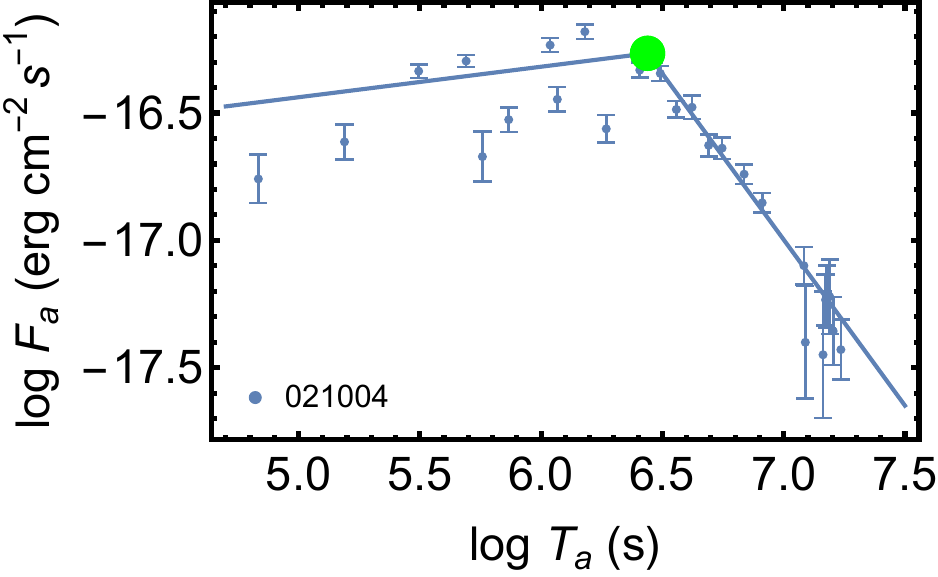} \\
\includegraphics[width = 0.28\textwidth]{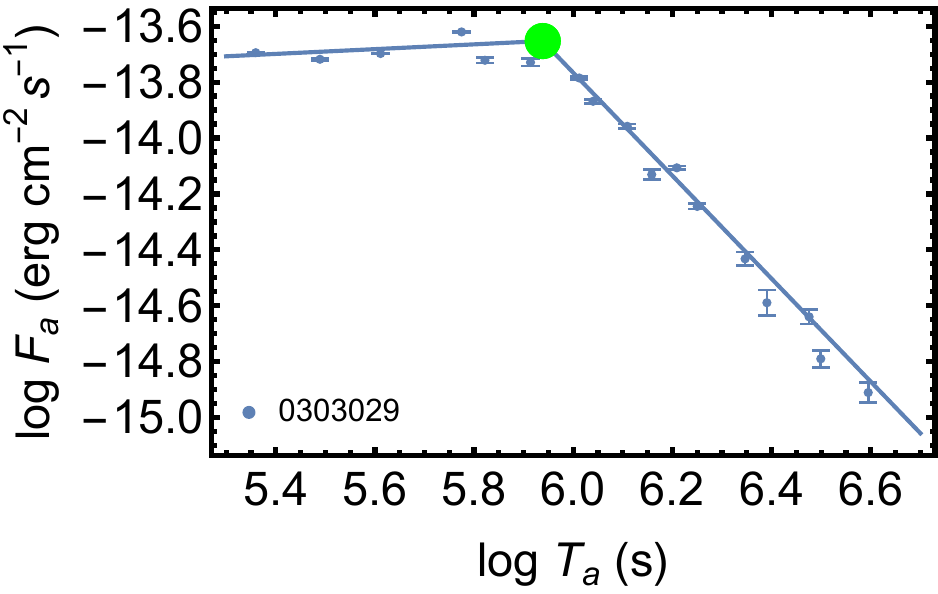} &
\includegraphics[width = 0.28\textwidth]{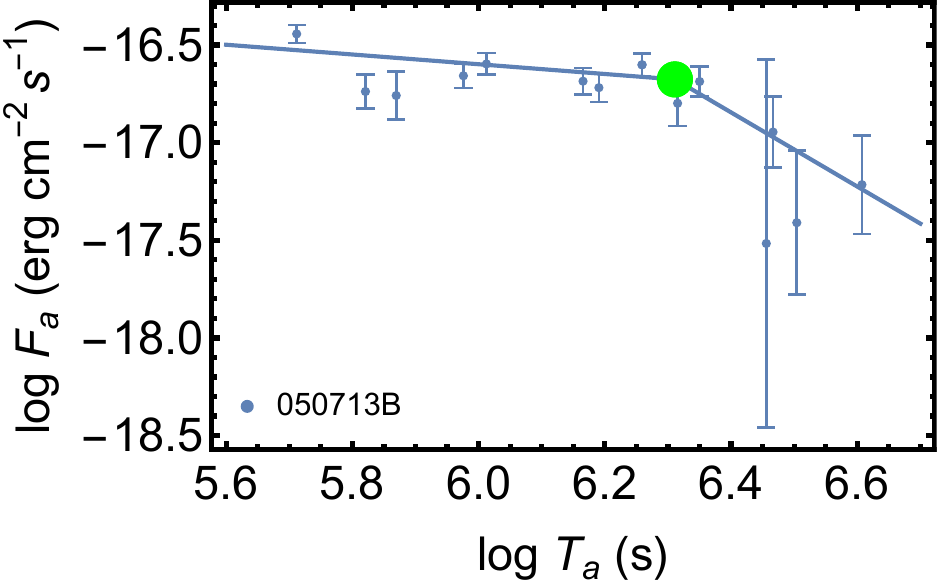} &
\includegraphics[width = 0.28\textwidth]{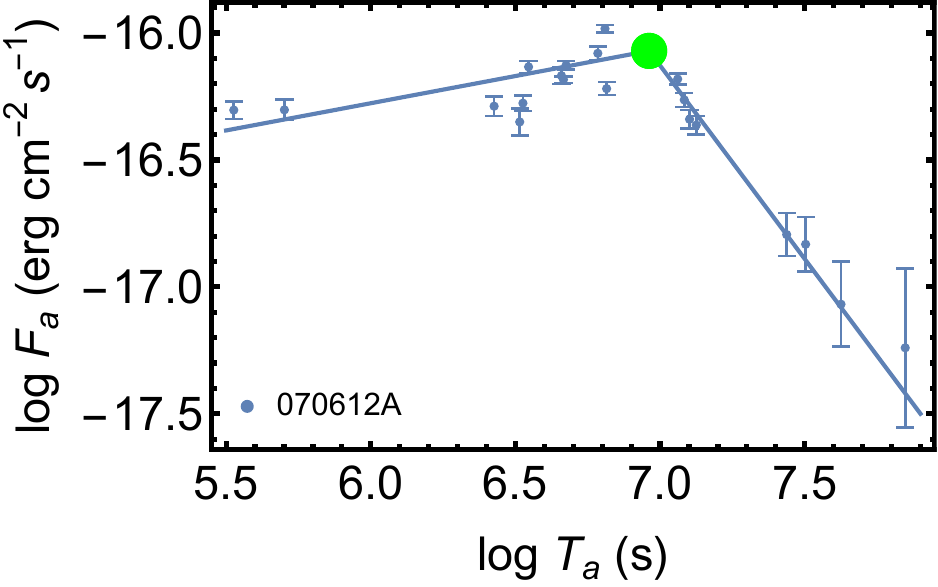} \\
\includegraphics[width = 0.28\textwidth]{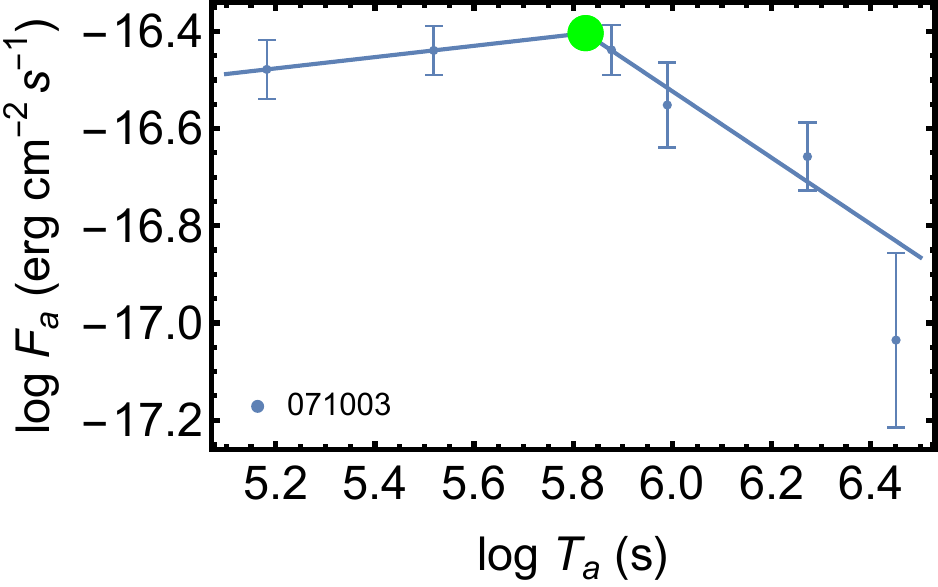} &
\includegraphics[width = 0.28\textwidth]{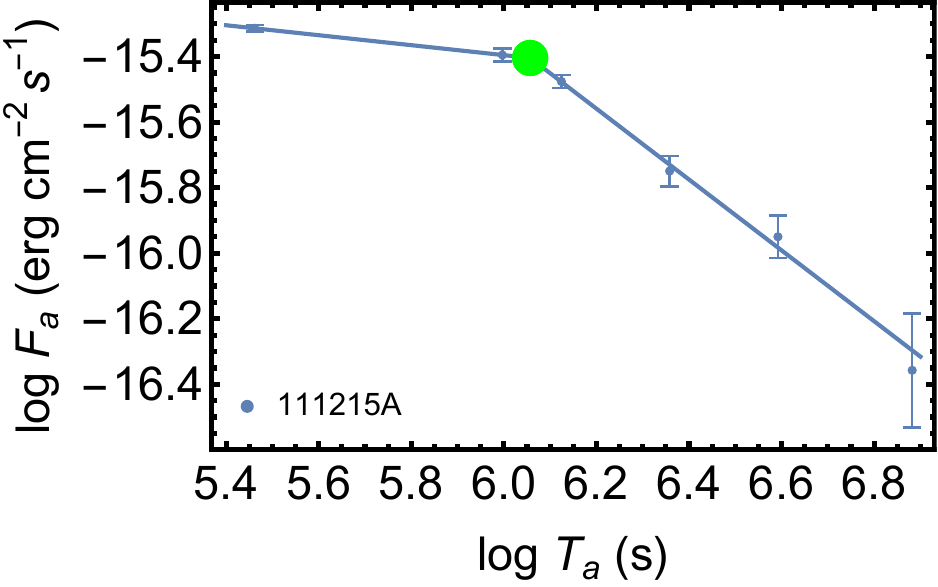} & 
\includegraphics[width = 0.28\textwidth]{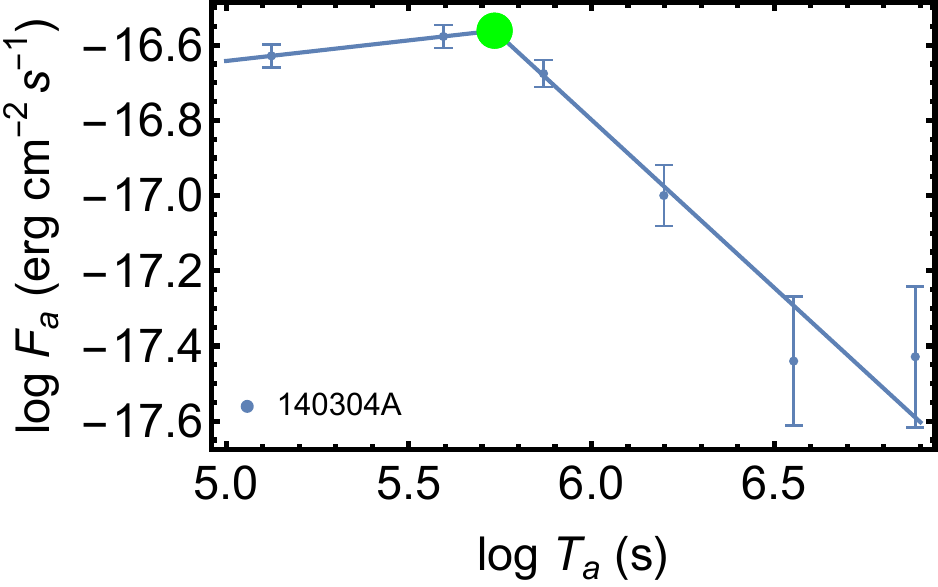} \\
\includegraphics[width = 0.28\textwidth]{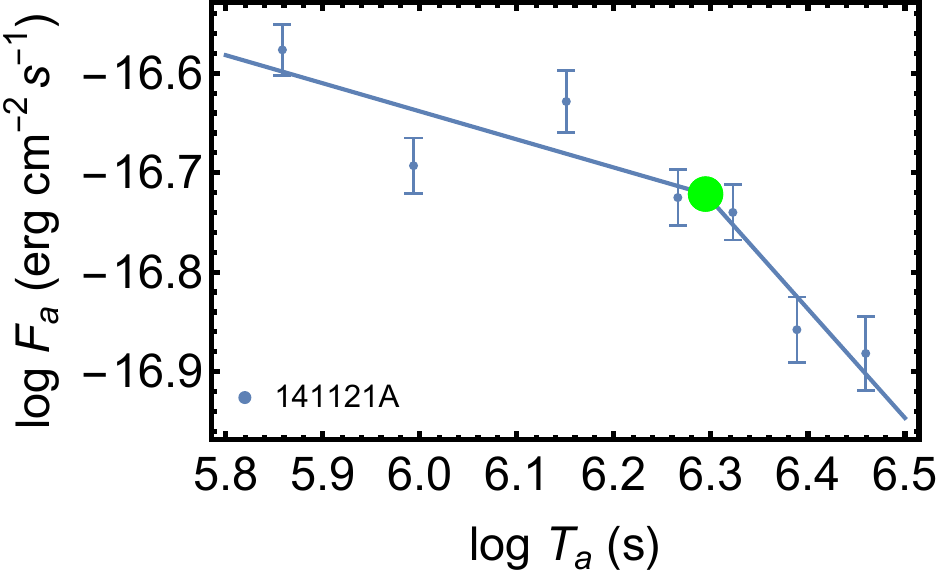} &
\includegraphics[width = 0.28\textwidth]{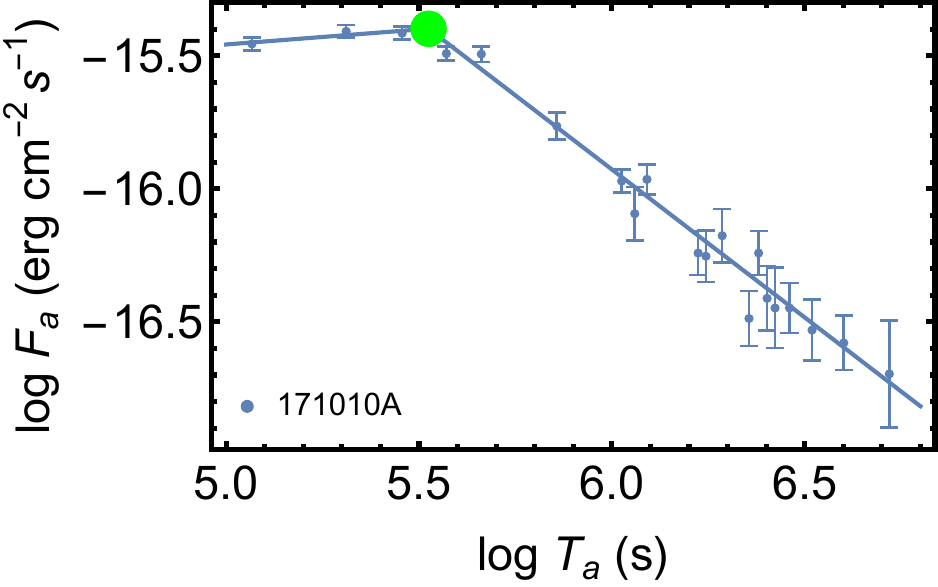} \\
\end{tabular}
\caption{Sample of 14 LCs that display a radio plateau.}\label{fig:plateaus}
\end{figure*}

\begin{figure*}
\begin{tabular}{ccc}
\includegraphics[width = 0.28\textwidth]{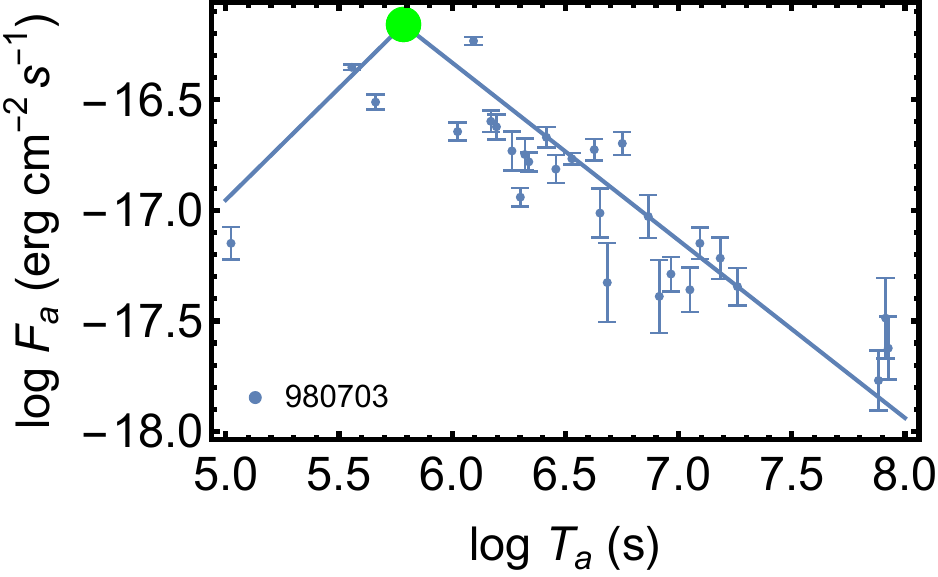} &
\includegraphics[width = 0.28\textwidth]{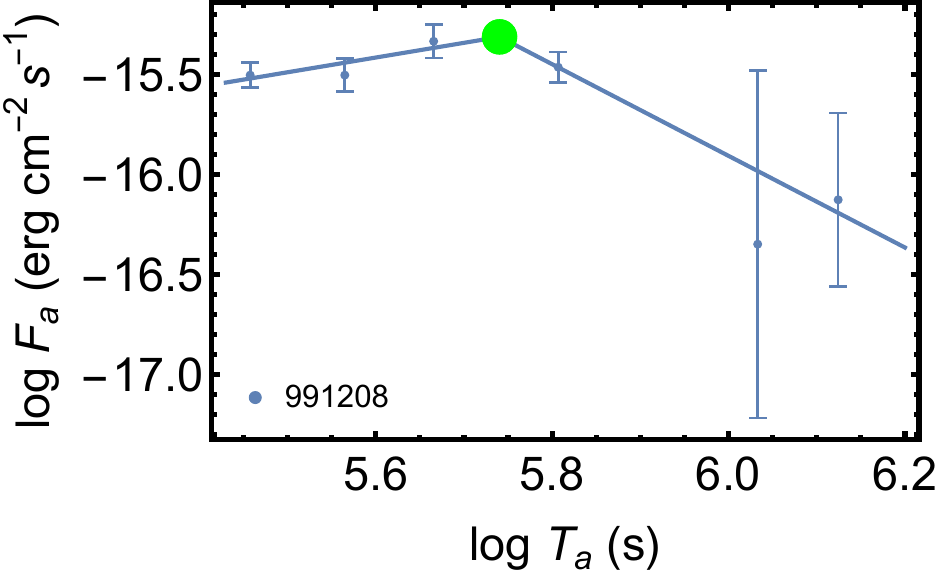} &
\includegraphics[width = 0.28\textwidth]{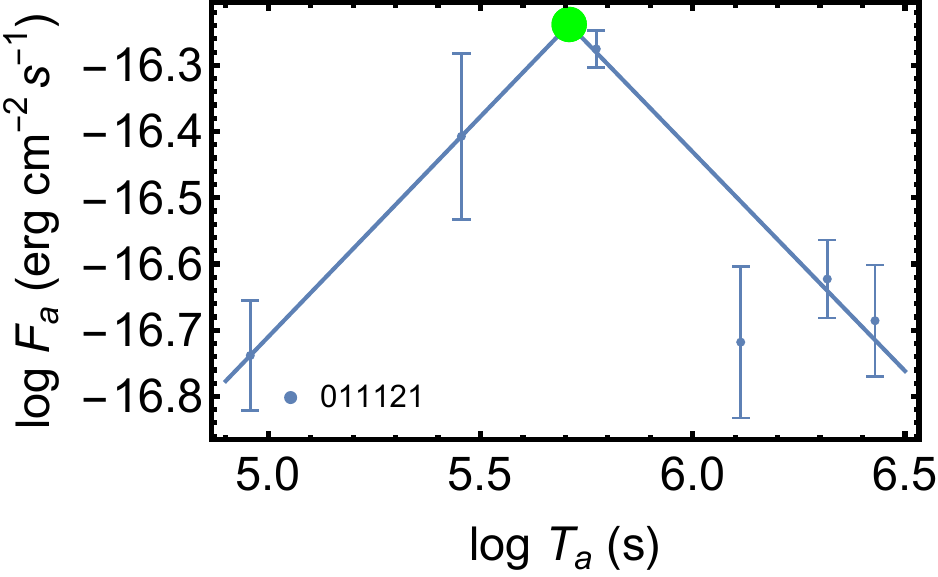} \\
\includegraphics[width = 0.28\textwidth]{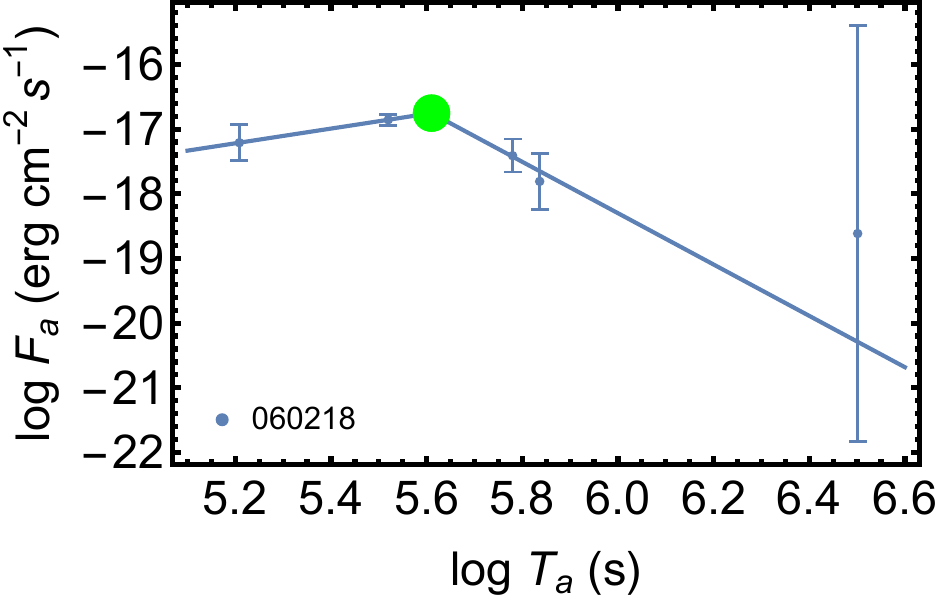} &
\includegraphics[width = 0.28\textwidth]{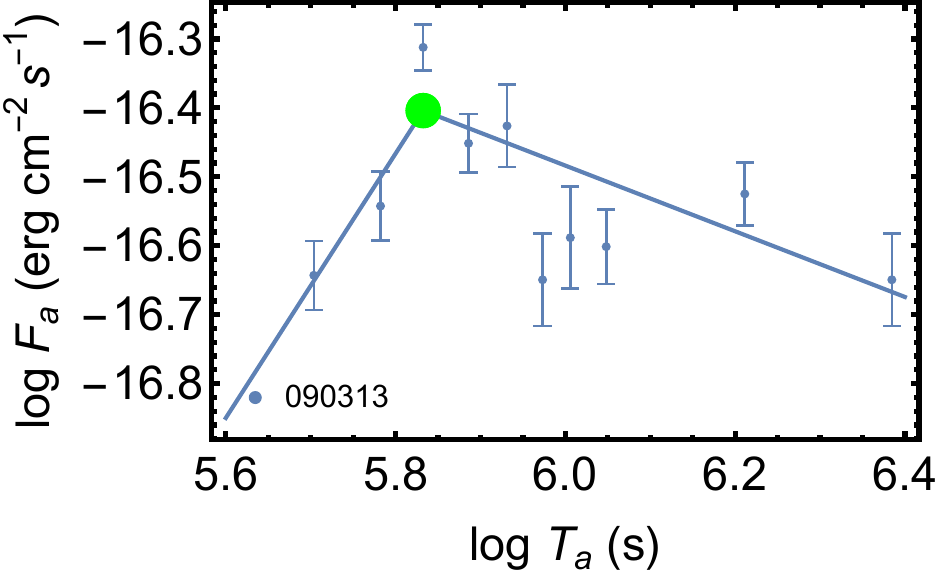} &
\includegraphics[width = 0.28\textwidth]{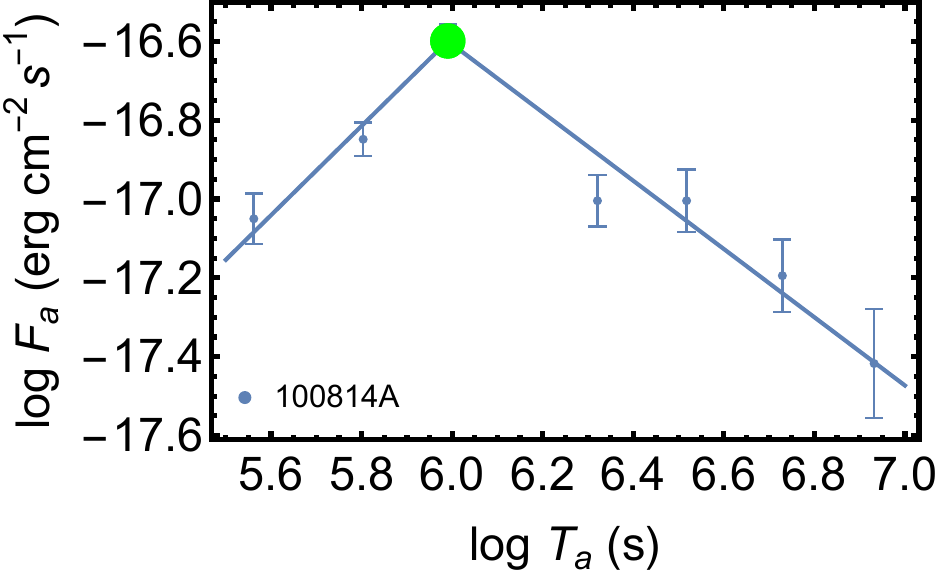} \\
\includegraphics[width = 0.28\textwidth]{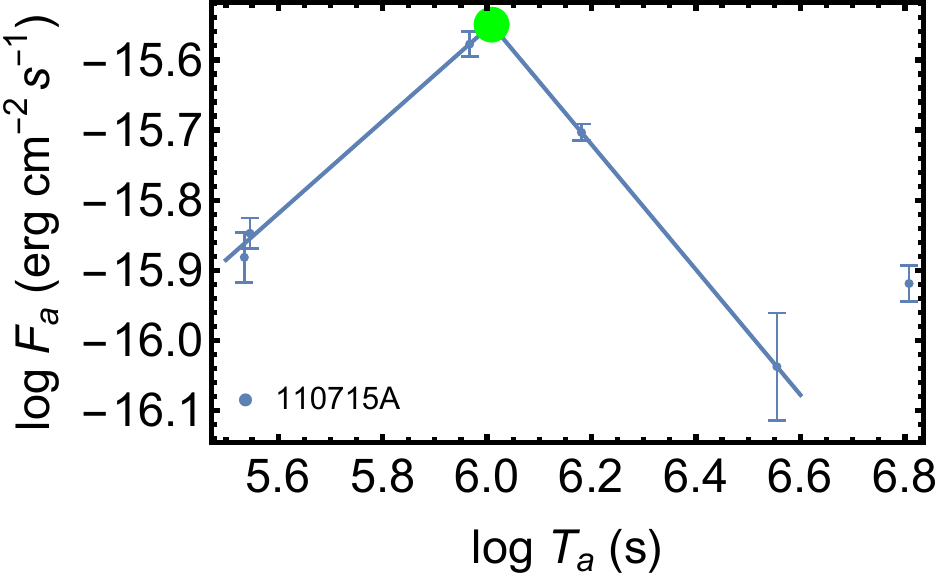} &
\includegraphics[width = 0.28\textwidth]{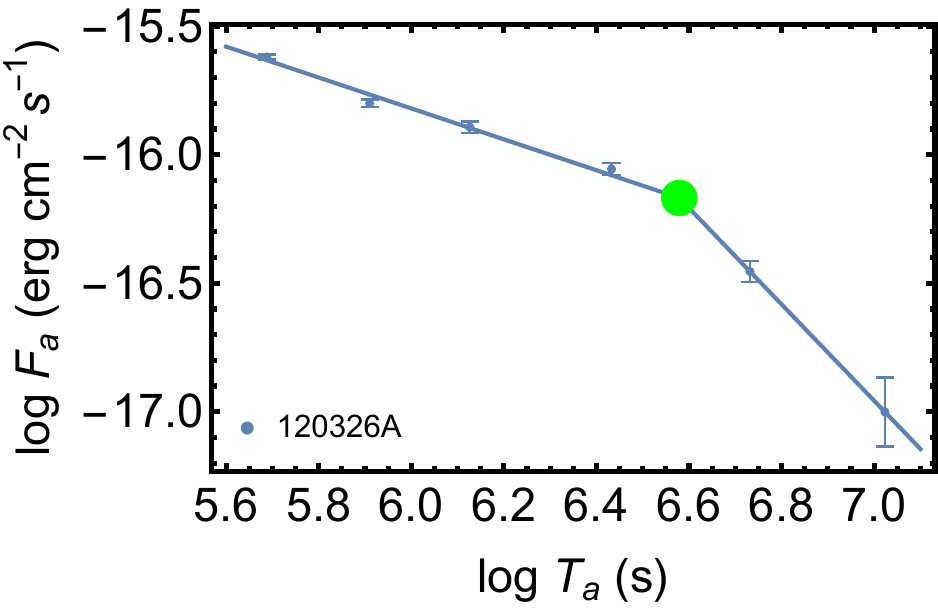} &
\includegraphics[width = 0.28\textwidth]{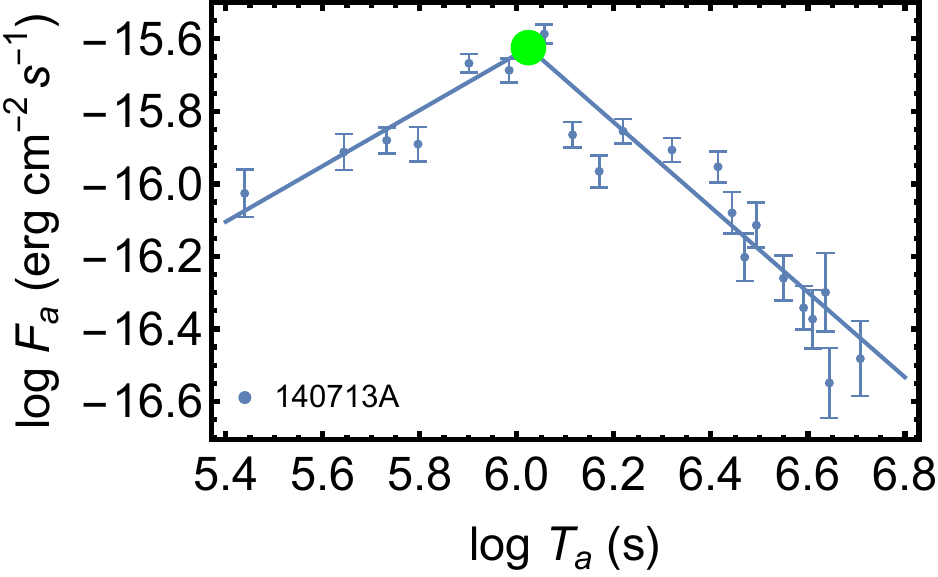} \\
\includegraphics[width = 0.28\textwidth]{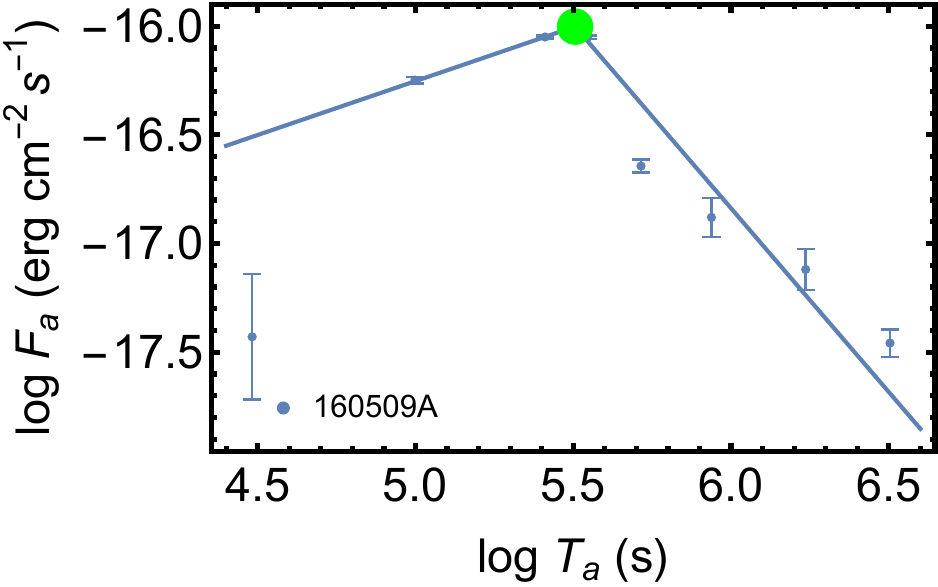} &
\includegraphics[width = 0.28\textwidth]{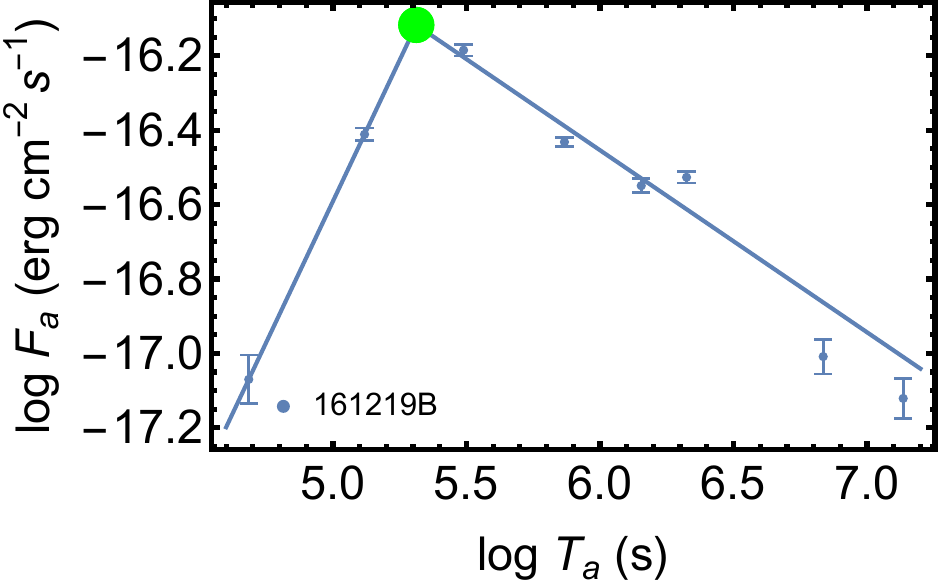} &
\includegraphics[width = 0.28\textwidth]{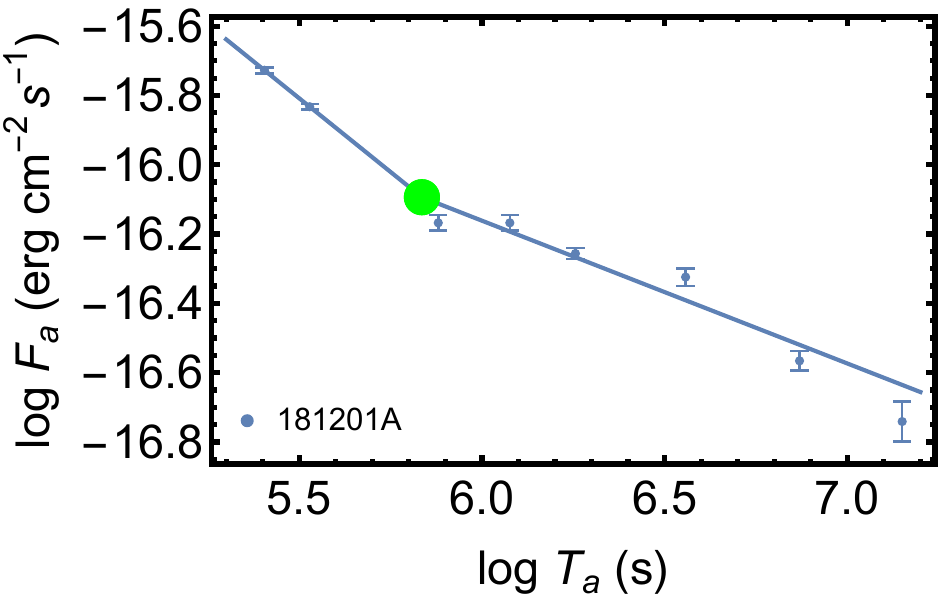} \\
\end{tabular}
\caption{Sample of LCs that passed the fitting criteria but do not display a radio plateau}.\label{fig:breaks}
\end{figure*}

\begin{figure*}
\begin{tabular}{ccc}
\includegraphics[width = 0.28\textwidth]{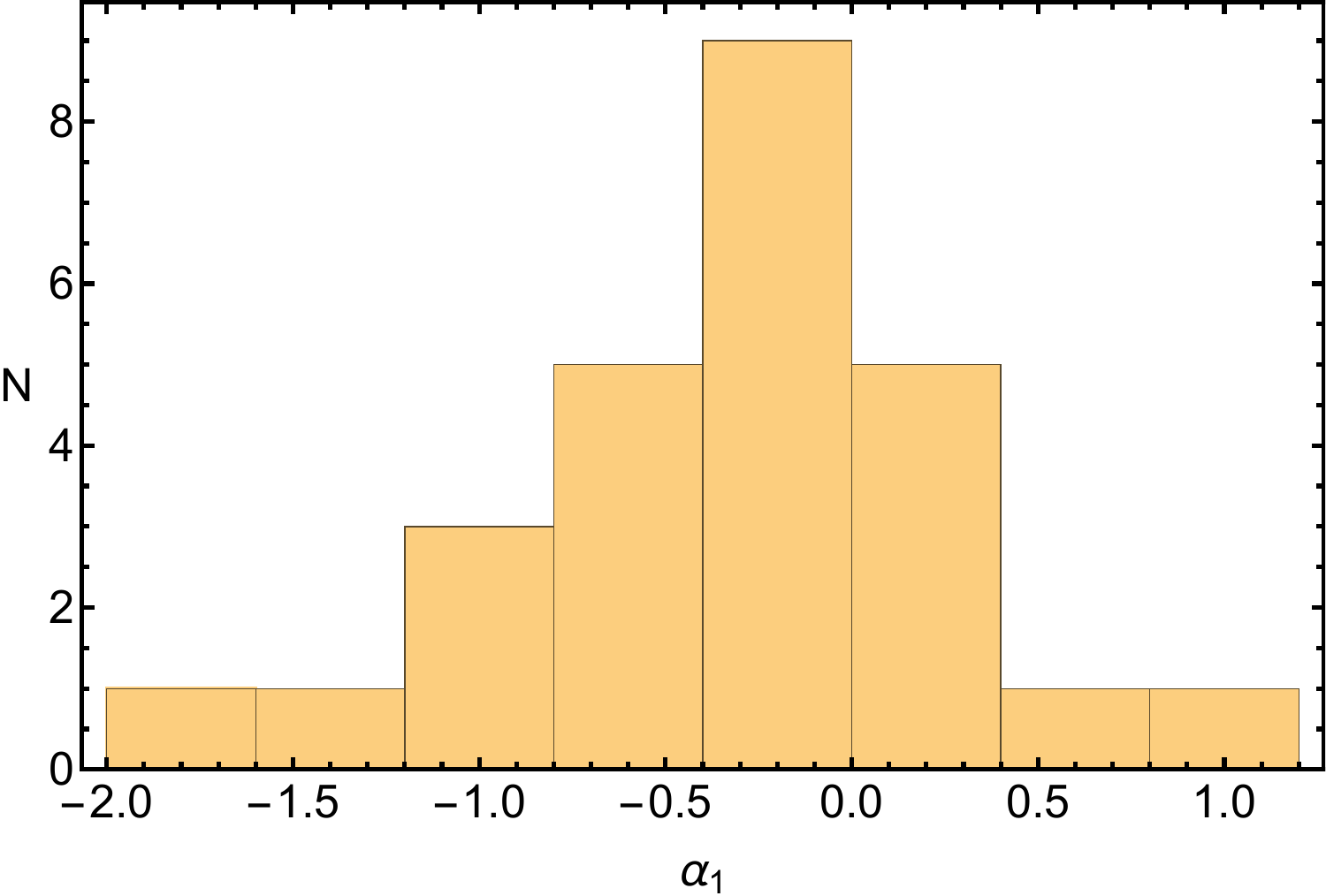} &
\includegraphics[width = 0.28\textwidth]{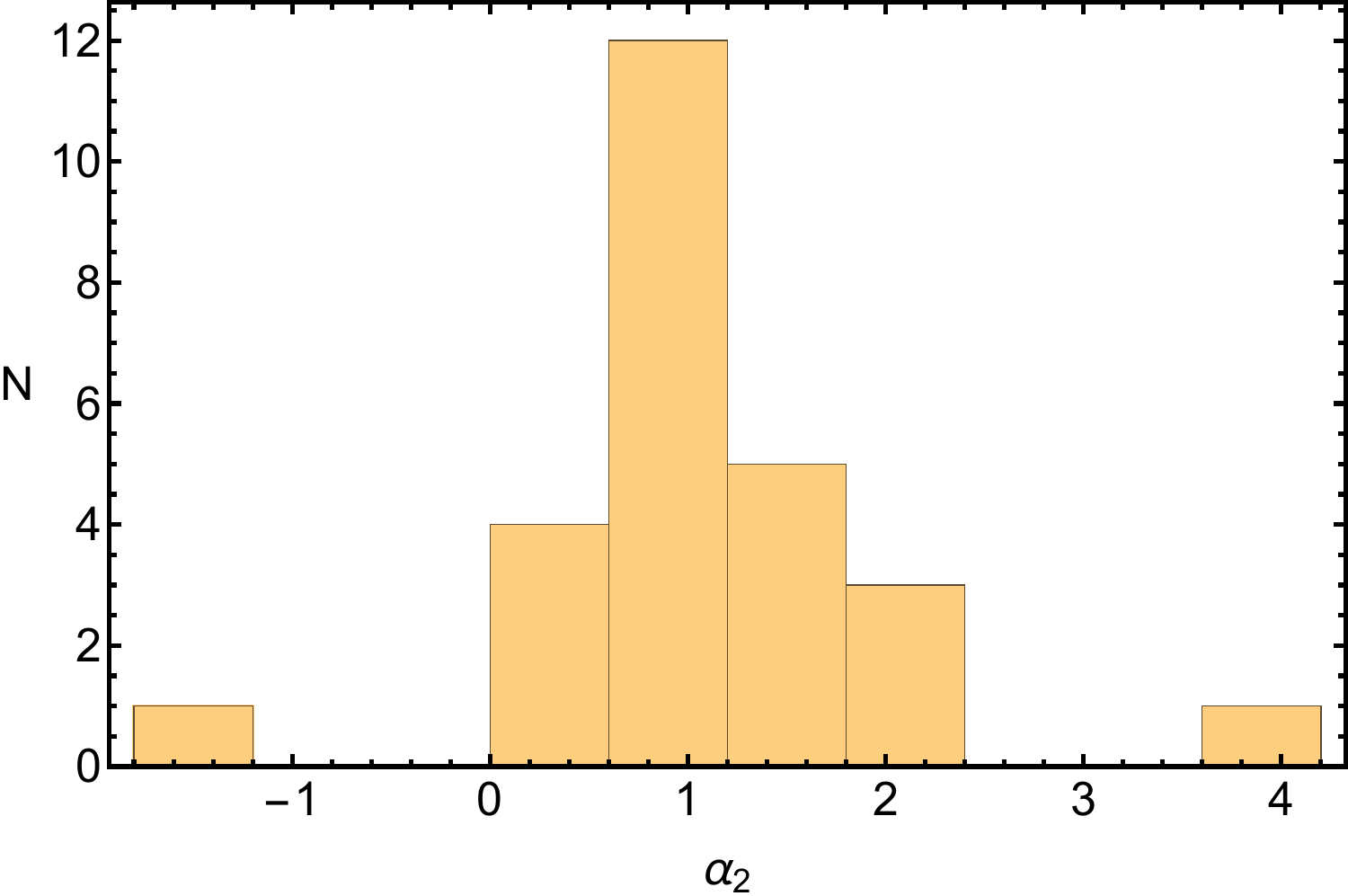} &
\includegraphics[width = 0.28\textwidth]{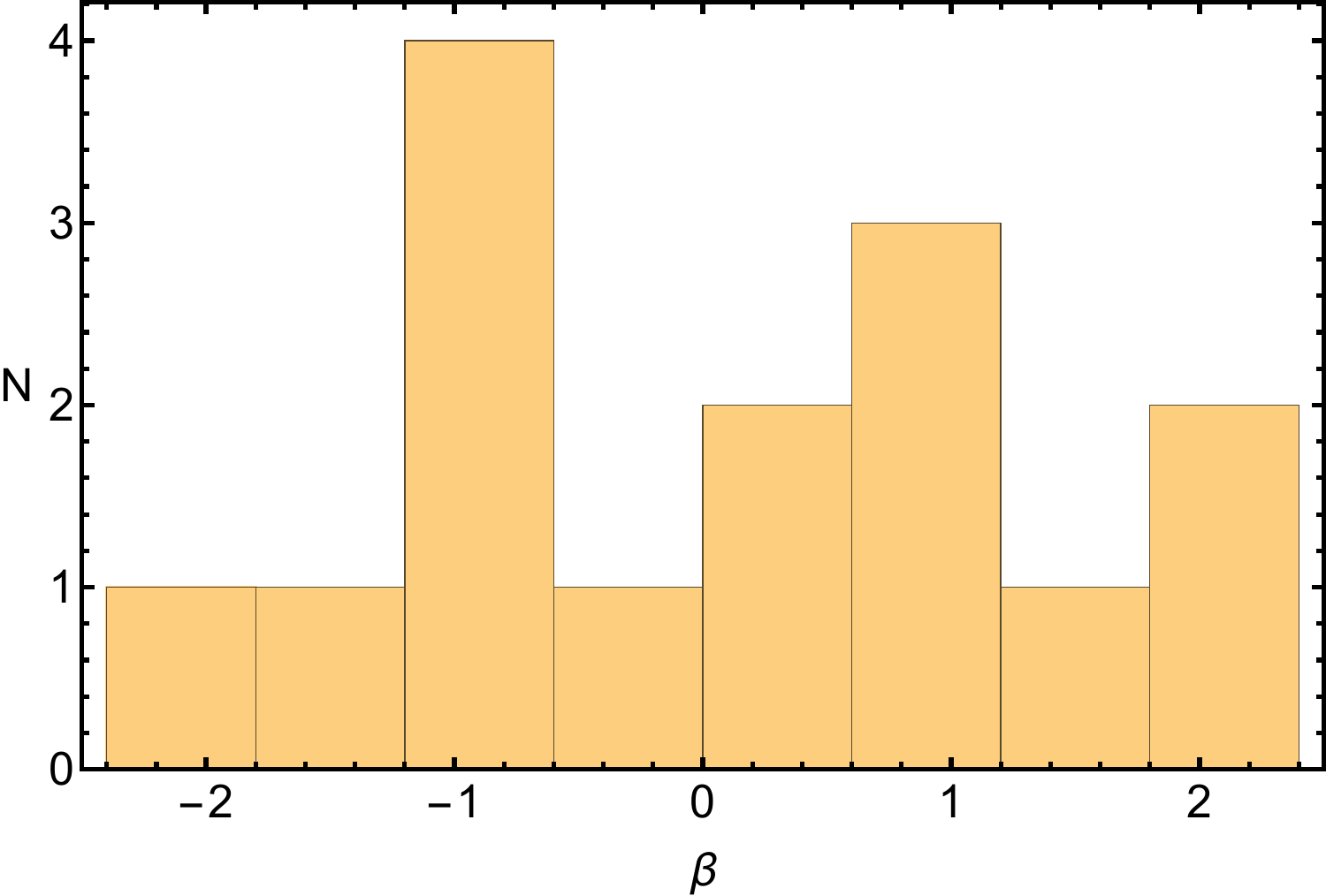} \\
\includegraphics[width = 0.28\textwidth]{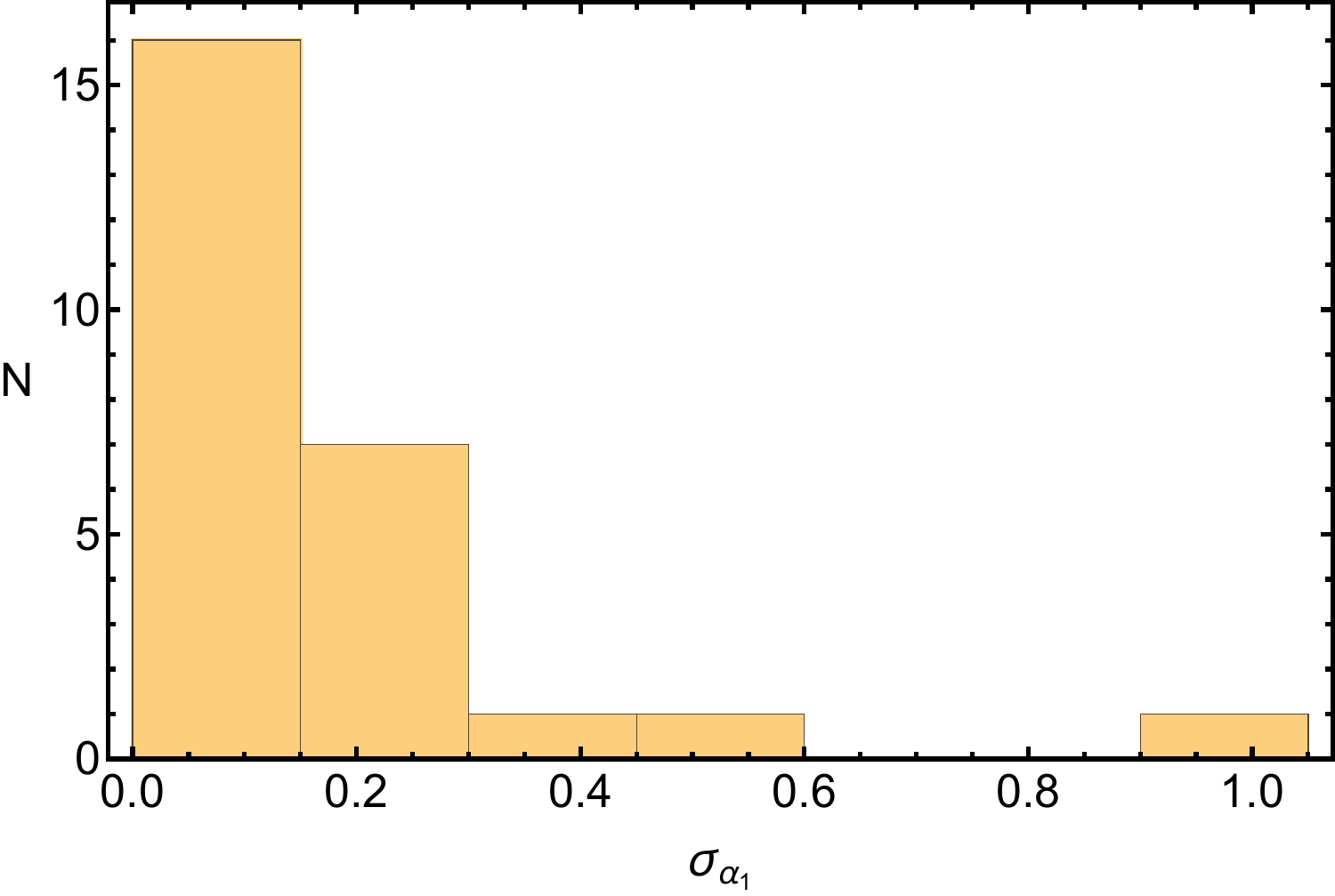} &
\includegraphics[width = 0.28\textwidth]{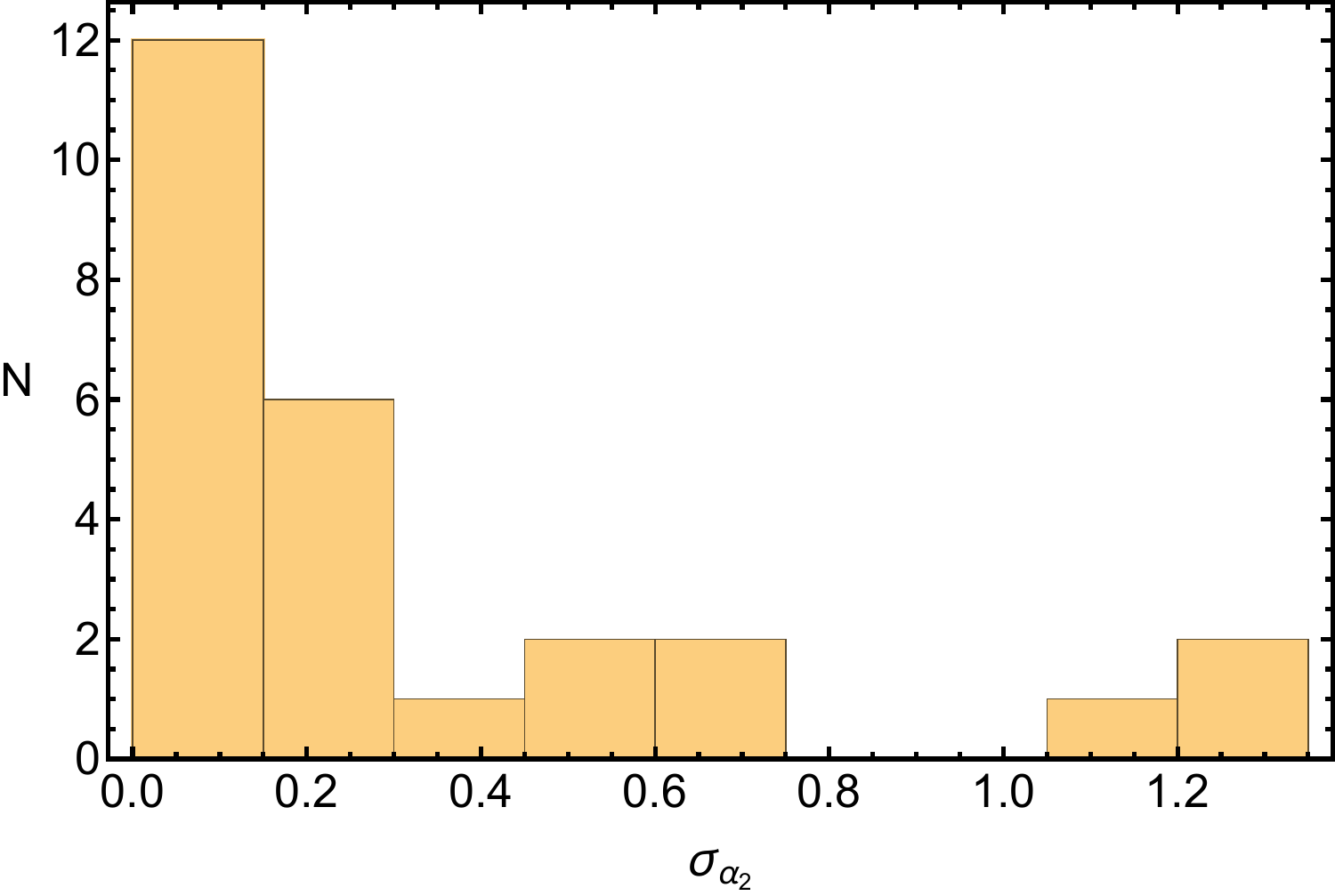} &
\includegraphics[width = 0.28\textwidth]{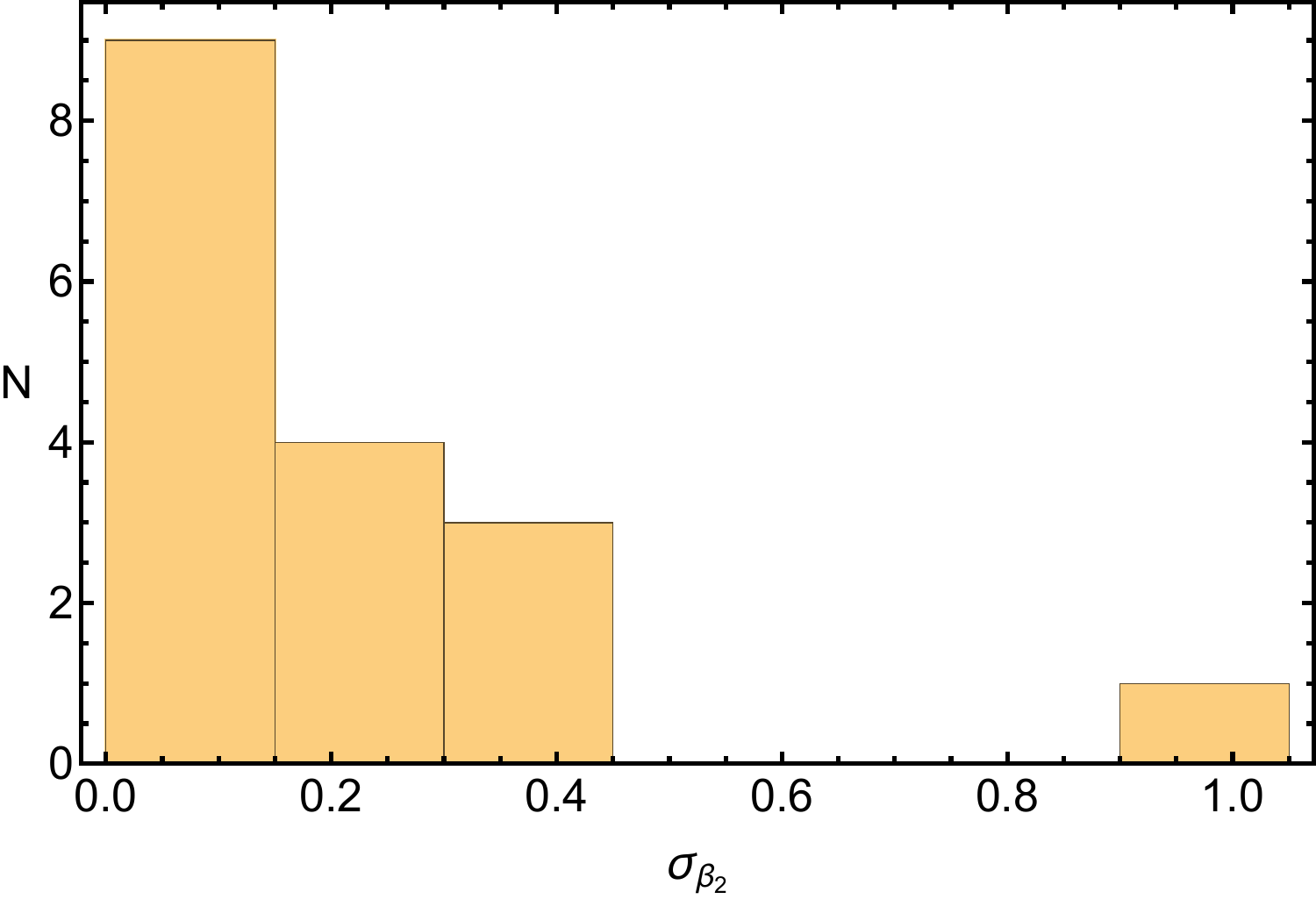} \\
\end{tabular}
\caption{Distribution of temporal and spectral indices (top) and their respective errors (bottom) for the full sample of 26 GRBs used in this analysis. Left: $\alpha_1$, or pre-break slope; middle: $\alpha_2$, or post-break slope; right: $\beta$, or spectral index.}\label{fig:dist}
\end{figure*}

In addition to the full sample of 26 GRBs that can be fit with a simple BPL, we are also interested in testing the subsample of GRBs that display a radio plateau, which we define as a flat region in the radio afterglow similar to the plateaus previously seen in X-ray LCs. Since the definition in the literature differs between several authors, we define these radio plateaus with the condition that the slope of the plateau, $|\alpha_1|$,is $< 0.5$. We choose this criterion to be consistent with the previous analysis of closure relations in optical wavelengths \citep{2022arXiv221003870D}, and in prior studies of plateaus seen in X-ray \citep{2013ApJ...774..157D}, optical \citep{2020ApJ...905L..26D, 2022ApJS..261...25D}, and radio \citep{2022ApJ...925...15L} wavelengths. In our sample, 14 of the 26 GRBs (54\%) display a radio plateau. The temporal and spectral indices for the full sample of GRBs used in this study are given in Table \ref{tab:Sample}. The fitted LCs for the plateau subsample are given in Figure \ref{fig:plateaus}, and the fitted LCs for the other 12 GRBs that display a break in their LC but do not display a plateau (hereafter referred to as the ``break" sample) are given in Figure \ref{fig:breaks}. We show the distribution of $\alpha_1$, $\alpha_2$, $\beta$, and their respective errors for the full sample of 26 GRBs in Figure \ref{fig:dist}.

\begin{table*}
\begin{tabular}{lccccccccc}
\hline
GRB	&	z	&	$T_{90}$	&	Freq  	&	$\log{F_a}  	\pm	\sigma_{F_a}$	&	$\log{T_a^*}	\pm	\sigma_{T_a^*}$	&	$\alpha_1	\pm	\sigma_{\alpha_1} $	&	$\alpha_2	\pm	\sigma_{alpha_2}$ 	&	Reason for Rejection	&	LC ref	\\
 & & (s) & (GHz) & ($\text{erg/s*cm}^2$) & (s) & & & & \\
\hline
GRB980329	&	3.9	&	58	&	8.46	&	$-16.64	\pm	0.04$	&	$6.59	\pm	0.09$	&	$-0.08	\pm	0.06$	&	$0.83	\pm	0.32$	&	none	&	[1]	\\									
&		&		&	350	&	$-14.18	\pm	\text{error}$	&	$5.89	\pm \text{error}$	&	$1.29	\pm	\text{error}$	&	$0.38	\pm	\text{error}$	&	scattered	&		\\									
&		&		&	1.43	&	$-17.74	\pm	0.33$	&	$6.39	\pm	0.7$	&	$-0.7	\pm	1.08$	&	$0.3	\pm	\text{error}$	&	bad chi sq 	&		\\									
&		&		&	4.86	&	$-16.92	\pm	0.05$	&	$6.98	\pm	0.044$	&	$-0.41	\pm	0.12$	&	$4.72	\pm	3.1$	&	scattered	&		\\		

GRB980425	&	0.0085	&	31	&	2.5	&	$-15.07	\pm	0.01$	&	$6.52	\pm	0.01$	&	$-0.46	\pm	0.05$	&	$1.55	\pm	0.08$	&	none	&	[1]	\\									
&		&		&	1.38	&	$-15.38	\pm	0.01$	&	$6.58	\pm	0.005$	&	$-1.19	\pm	0.04$	&	$1.3	\pm	0.04$	&	too steep for plateau	&		\\									
&		&		&	4.8	&	$-14.61	\pm	0.01$	&	$6.04	\pm	0.01$	&	$-1.28	\pm	0.06$	&	$0.82	\pm	0.02$	&	too steep for plateau	&		\\									
&		&		&	8.64	&	$-14.34	\pm	0.01$	&	$5.97	\pm	0.01$	&	$-1.21	\pm	0.05$	&	$1.04	\pm	0.01$	&	too steep for plateau	&		\\	
\hline
\end{tabular}
\caption{Fitting parameters for a sample of 202 LCs fitted with a BPL. Table additionally includes redshift, $T_{90}$, observational frequency, and reason for rejection. The LCs for the first two GRBs are given here, the full table is available online. \\
$[1]$ \citet{2012ApJ...746..156C}} \label{tab:fitting}
\end{table*}

\begin{table*}
\begin{tabular}{lccccccccc}
\hline
GRB & z & $\alpha_1$ & $\alpha_2$ & $\beta$  & $\log{T_a}$ & Frequency & Plateau? & LC ref & $\beta$ ref\\
&  &  &  & & (s) & (GHz) & & & \\
\hline
GRB980329	&	3.9	&	$	-0.08	\pm	0.06	$	&	$	0.83	\pm	0.32	$	&	$	-1.7	\pm	0.17	$	&	$	6.59	\pm	0.09	$	&	8.46	&	Y	&	[1]	&	[2]	\\
GRB980425	&	0.0085	&	$	-0.46	\pm	0.05	$	&	$	1.55	\pm	0.08	$	&	$	0.85	\pm	0.18	$	&	$	6.52	\pm	0.01	$	&	2.5	&	Y	&	[1]	&	estimated	\\
GRB000926	&	2.039	&	$	-0.22	\pm	0.2	$	&	$	0.45	\pm	0.15	$	&	$	2.03	\pm	0.06	$	&	$	5.79	\pm	0.02	$	&	4.86	&	Y	&	[1]	&	estimated	\\
GRB010222	&	1.477	&	$	0.06	\pm	0.19	$	&	$	1.33	\pm	0.64	$	&	$	1.65	\pm	0.14	$	&	$	6.01	\pm	0.05	$	&	4.86	&	Y	&	[1]	&	estimated	\\
GRB011030	&	3	&	$	-0.21	\pm	0.2	$	&	$	0.91	\pm	0.25	$	&	$	0.6	\pm	0.15	$	&	$	6.8	\pm	0.28	$	&	8.46	&	Y	&	[1]	&	[3]	\\
GRB021004	&	2.33	&	$	-0.12	\pm	0.08	$	&	$	1.3	\pm	0.14	$	&	$	-0.9	\pm	0.17	$	&	$	6.32	\pm	0.07	$	&	8.46	&	Y	&	[1]	&	[4] 	\\
GRB030329	&	0.168	&	$	-0.09	\pm	0.05	$	&	$	1.84	\pm	0.16	$	&	$	0.54	\pm	0.02	$	&	$	6.44	\pm	0.04	$	&	43.3	&	Y	&	[1]	&	[5] 	\\
GRB050713B	&	0.55	&	$	0.25	\pm	0.14	$	&	$	1.9	\pm	1.05	$	&	$	-0.9	\pm	0.09	$	&	$	5.94	\pm	0.02	$	&	8.46	&	Y	&	[1]	&	estimated	\\
GRB070612A	&	0.617	&	$	-0.21	\pm	0.07	$	&	$	1.52	\pm	0.45	$	&	$	-0.44	\pm	0.14	$	&	$	6.31	\pm	0.07	$	&	8.46	&	Y	&	[1]	&	estimated	\\
GRB071003	&	1.1	&	$	-0.12	\pm	0.31	$	&	$	0.68	\pm	0.25	$	&	$	-1.15	\pm	0.42	$	&	$	6.96	\pm	0.05	$	&	8.46	&	Y	&	[1]	&	[6] \\
GRB111215A	&	2.06	&	$	0.15	\pm	0.03	$	&	$	1.08	\pm	0.08	$	&	$	1.03	\pm	0.13	$	&	$	5.82	\pm	0.2	$	&	24.4	&	Y	&	[7]	& estimated	\\
GRB140304A	&	5.283	&	$	-0.11	\pm	0.11	$	&	$	0.89	\pm	0.16	$	&	$	-1.1	\pm	0.4	$	&	$	6.58	\pm	0.08	$	&	11	&	Y	&	[8]	&	[9]	\\
GRB141121A	&	1.47	&	$	0.28	\pm	0.18	$	&	$	1.1	\pm	0.71	$	&	$	0.18	\pm	0.07	$	&	$	5.73	\pm	0.07	$	&	13	&	Y	&	[10]	&	[10]	\\
GRB171010A	&	0.3285	&	$	-0.11	\pm	0.09	$	&	$	1.11	\pm	0.05	$	&	$	-1.9	\pm	0.05	$	&	$	6.29	\pm	0.11	$	&	15.5	&	Y	&	[11]	&	[11]	\\
\hline
GRB980703	&	0.966	&	$	-1.01	\pm	0.13	$	&	$	0.8	\pm	0.03	$	&	$	0.32	\pm	0.12	$	&	$	5.52	\pm	0.03	$	&	4.86	&	N	&	[1]	&	[12]	\\
GRB991208	&	0.706	&	$	0.26	\pm	0.14	$	&	$	-1.46	\pm	1.3	$	&	$	-0.68	\pm	0.06	$	&	$	5.76	\pm	0.08	$	&	15	&	N	&	[1] & [13]	\\
GRB011121	&	0.362	&	$	-0.67	\pm	0.28	$	&	$	0.66	\pm	0.1	$	&	$	1.3	\pm	0.13	$	&	$	5.71	\pm	0.13	$	&	8.7	&	N	&	[1]	& estimated	\\
GRB060218	&	0.033	&	$	-1.14	\pm	0.92	$	&	$	3.97	\pm	1.31	$	&	$	0.71	\pm	0.27	$	&	$	5.61	\pm	error	$	&	4.86	&	N	&	[1]	&	estimated	\\
GRB090313	&	3.375	&	$	-1.92	\pm	0.52	$	&	$	0.48	\pm	0.08	$	&	$	1.84	\pm	0.13	$	&	$	5.83	\pm	error	$	&	8.46	&	N	&	[1]	&	estimated	\\
GRB100814A	&	1.44	&	$	-1.13	\pm	0.15	$	&	$	0.87	\pm	0.08	$	&	$	0.85	\pm	0.08	$	&	$	5.99	\pm	0.03	$	&	4.5	&	N	&	[1]	&	estimated	\\
GRB110715A	&	0.82	&	$	-0.66	\pm	0.04	$	&	$	0.89	\pm	0.21	$	&	$	1.32	\pm	0.11	$	&	$	6.01	\pm	0.03	$	&	18	&	N	&	[14]	& estimated	\\
GRB120326A	&	1.798	&	$	0.6	\pm	0.04	$	&	$	1.87	\pm	0.57	$	&	$	0.92	\pm	0.23	$	&	$	6.06	\pm	0.02	$	&	19.2	&	N	&	[8]	&	[8]	\\
GRB140713A	&	0.935	&	$	-0.77	\pm	0.09	$	&	$	1.17	\pm	0.05	$	&	$	0.99	\pm	0.13	$	&	$	6.02	\pm	0.02	$	&	15.7	&	N	&	[15]	&	estimated	\\
GRB160509A	&	1.17	&	$	-0.5	\pm	0.03	$	&	$	1.69	\pm	0.04	$	&	$	-0.8	\pm	0.15	$	&	$	5.51	\pm	0.01	$	&	8.5	&	N	&	[16, 17]	& [18]	\\
GRB161219B	&	0.1475	&	$	-1.52	\pm	0.15	$	&	$	0.49	\pm	0.02	$	&	$	0.19	\pm	0.03	$	&	$	5.31	\pm	0.02	$	&	5	&	N	&	[9]	&	[9]	\\
GRB181201A	&	0.45	&	$	0.85	\pm	0.09	$	&	$	0.41	\pm	0.03	$	&	$	0.35	\pm	0.03	$	&	$	5.84	\pm	0.06	$	&	11	&	N	&	[19]	&	[20] 	\\

\hline
\end{tabular}
\caption{\label{tab:Sample}The table includes GRB ID, redshift taken from Greiner (2021) [\url{https://www.mpe.mpg.de/~jcg/grbgen.html}] and GCNs, best-fit parameters $\alpha_1$ and $\alpha_2$ computed from BPL, and spectral index $\beta$ taken from literature. References: [1] \citet{2012ApJ...746..156C}, [2] \citet{1998ApJ...502L.115T}, [3] \citet{2001GCN..1124....1R} [4] \citet{2002GCN..1612....1B, 2002GCN..1613....1B}, [5]  \citet{2008A&A...480...35V} [6] \citet{2008ApJ...688..470P}, [7] \citet{2013ApJ...767..161Z} [8] \citet{2015ApJ...814....1L} [9] \citet{2018ApJ...859..134L} [10] \citet{2015ApJ...812..122C} [11] \citet{2019MNRAS.486.2721B} [12] \citet{2001ApJ...560..652B} [13] \citet{2001A&A...370..398C} [14] \citet{2017MNRAS.464.4624S} [15] \citet{2019MNRAS.484.5245H} [16] \citet{2020ApJ...894...43K} [17] \citet{2016ApJ...833...88L} [18] \citet{2016GCN.19428....1C} [19] \citet{2020MNRAS.492.1919M} [20] \citet{2019ApJ...884..121L}}
\end{table*}

\subsection{X-ray vs. radio plateaus}\label{sec:Xray}

The plateau feature seen in radio LCs were first observed in X-ray wavelengths, using data taken from the \textit{Swift} observatory. For the subsample of 14 GRBs that display a plateau in radio, we compare the end-time of the plateau $T_a$ to the end-time of the plateau observed in X-ray. We take the X-ray LC data from the \textit{Swift} XRT repository \footnote{\url{https://www.swift.ac.uk/xrt_curves/}}; thus, we can only draw the comparison between GRBs observed post-\textit{Swift}. We also note that GRB 140304A could not be successfully fitted with a simple BPL, so the end-time of the X-ray plateau could not be reliably determined.

\begin{figure}
\includegraphics[width = 0.5\textwidth]{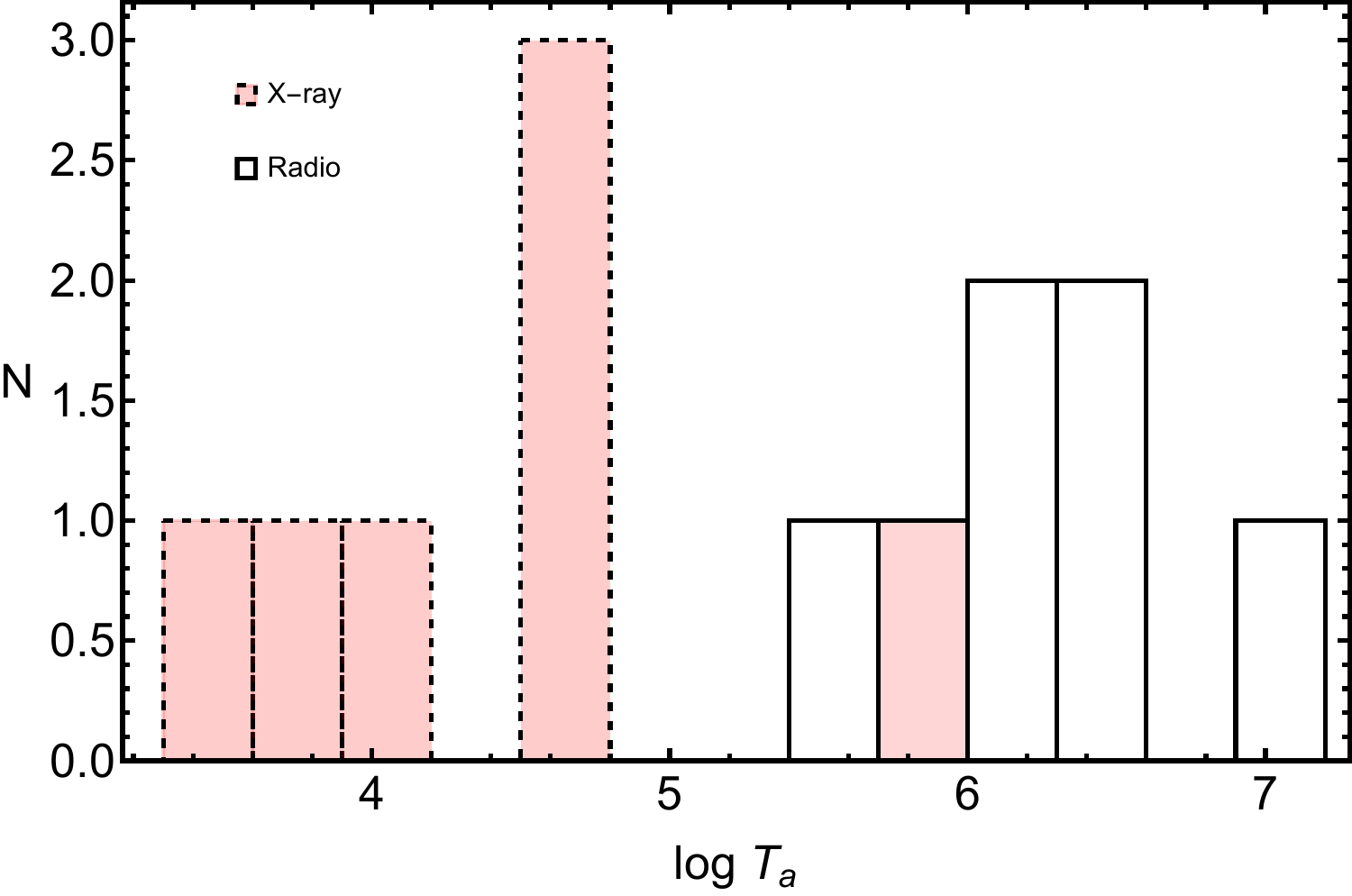}
\caption{Distributions of end-time of plateau $T_a$ for 7 post-\textit{Swift} GRBs in X-ray (red) and radio (white).}\label{fig:coinc}
\end{figure}

We see that for all 7 of the remaining GRBs in the plateau sub-sample, the radio $T_a$ occurs later than the X-ray $T_a$ by approximately two orders of magnitude. On average, the end-time of the X-ray plateaus occurs at $\approx 10^{4.43}$ seconds, and the end-time of the radio plateaus occurs at $\approx 10^{6.22}$ seconds. To determine the statistical significance of this difference, we conduct a Kolmogorov-Smirnov (KS) test between the $T_a$'s of the radio LCs and the X-ray LCs, finding a KS value of $KS = 0.86$ and a p-value of $p = 0.01$. This indicates that the two distributions were not drawn from the same parent sample, as expected from prior studies of radio plateaus as compared to plateaus in other wavelengths \citep{2022ApJ...925...15L}. We show the distributions of $T_a$ in X-ray and radio in Figure \ref{fig:coinc}.

\section{Methodology}
\label{sec:method}

The standard fireball model predicts distinct relationships between the temporal, $\alpha$, and spectral, $\beta$, indices of the GRB's afterglow measured from observations, known as CRs \citep{1978MNRAS.183..359C,1986ApJ...308L..47G,1986ApJ...308L..43P, 2004IJMPA..19.2385Z,2006ApJ...642..354Z}. These relationships can take the form of either a point or a line, depending on the characteristics of the spectral regime. For regimes with fixed 
$\alpha$ and $\beta$, the CR takes the form of a point, and for regimes with varying $\alpha$ and $\beta$, the regime takes the form of a line. There are various scenarios that can be considered when testing CRs of GRB afterglow observations, such as the inclusion or omission of energy injection, ISM or stellar wind density-profiles, fast or slow cooling regimes, spectral regimes defined by several characteristic frequencies, and physical regimes such as the reverse shock crossing phase, self-similar phase, post-jet-break phase, and Newtonian phase \citep[for a complete review see][and the references therein.]{2013NewAR..57..141G}

We adopt the segmented LC model using the \citet{2006ApJ...642..354Z} and \citet{2006ApJ...642..389N} approach, where phase I refers to the steep decay of the prompt emission following the fast rising phase, phase II to the plateau phase immediately following the prompt emission, phase III to the normal decay phase characteristic of afterglow emission, and phase IV to a possible steeper decay phase following the normal decay. We consider two classes of CRs, one without energy injection, and one with energy injection. The rate of injection is controlled with a parameter $q$ - we assume $q=0$, which corresponds to continuous energy injection. For the relations without injection, assuming that the dominant mechanism of the decay phase is synchrotron radiation, we study phase III, when the deceleration of the adiabatic fireball produces the afterglow, as done in \citet{2009ApJ...698...43R}. We investigate how many GRBs satisfy a given CR between $\alpha_2$ and $\beta$, as this corresponds to the post-plateau phase. For the relations with energy injection, we consider phase II, which is during the radio plateau. We therefore test the relation between $\alpha_1$ and $\beta$, as $\alpha_1$ corresponds to the slope of the plateau.

We consider both the ISM environment, which assumes the relativistic ejecta expands into a constant-density circumburst medium and the Wind environment, where the ejecta expands into $\propto r^{-2}$ as a result of a core-collapse supernova \citep{1998ApJ...497L..17S}. Assuming $\gamma_c$ is the critical Lorentz factor where synchrotron cooling becomes significant, within the aforementioned environments we study both the fast-cooling (FC), when all electrons in the shocked ejecta are able to cool down to $\gamma_c$ and the slow-cooling (SC), where only some electrons are able to cool \citep{1998ApJ...497L..17S}. Breaks in the synchrotron spectrum occur at the emission frequencies defined by the aforementioned electron Lorentz factors, at $\nu_{\rm a}$ (self-absorption frequency), $\nu_c$ (cooling frequency), and $\nu_{\rm m}$ (frequency at the characteristic spectral break). Unlike in the investigations of X-ray, optical, and high-energy afterglow, studying the break in the synchrotron spectrum corresponding to self-absorption, $\nu_a$, is a necessary area of consideration for sub-millimeter and radio observations \citep{2013NewAR..57..141G}. 

We expand the CRs from the following tables published in \citet{2013NewAR..57..141G} to include scenarios accounting for energy injection by setting $q=0$; a variable that characterizes the flatness of the plateau emission \citep{2021PASJ..tmp...63D}. We here summarize the tables presented in this paper:
\begin{itemize}
    \item Table \ref{tab:2} taken from Table 7 of \citet{2013NewAR..57..141G} discusses the thick shell forward shock model for $\nu_a <$ min($\nu_m,\nu_c$).
    \item Table \ref{tab:3} taken from Table 8 of \citet{2013NewAR..57..141G} discusses the thick shell forward shock model for $\nu_m < \nu_a < \nu_c$.
    \item Table \ref{tab:4} taken from Table 13 and 15 of \citet{2013NewAR..57..141G} discusses the relativistic, isotropic, self-similar deceleration phase within $\nu_a <$ min($\nu_m,\nu_c$) for $p>2$ and $1<p<2$, respectively.
    \item Table \ref{tab:5} taken from Table 14 and 16 of \citet{2013NewAR..57..141G} discusses the relativistic, isotropic, self-similar deceleration phase within $\nu_m<\nu_a<\nu_c$ for $p>2$ and $1<p<2$, respectively.
    \item Table \ref{tab:6} taken from Table 18 of \citet{2013NewAR..57..141G} consider edge effect after the jet break for $\nu_a <$ min($\nu_m,\nu_c$).\footnote{For a complete discussion concerning the jet-break phase refer to \citep{2013NewAR..57..141G} Section 3.3.}
    \item Table \ref{tab:7} taken from Table 19 of \citet{2013NewAR..57..141G} consider edge effect after the jet break for $\nu_m < \nu_a < \nu_c$.
    \item Table \ref{tab:8} assumes a thick shell model for varying density profiles $n(r) \propto r^k$ for $k = 0,-1,-1.5,-2,-2.5$; including stellar wind and ISM environments.\footnote{See \citet{2019ApJ...884...71F, 2021ApJ...907...78F} for reference.}
\end{itemize}

We then test our sample from Table \ref{tab:Sample} against the full set of CRs presented in Tables \ref{tab:2},
\ref{tab:3},
\ref{tab:4},
\ref{tab:5},
\ref{tab:6},  \ref{tab:7}, and  \ref{tab:8} considering scenarios both with and without energy injection.\footnote{When considering energy injection we use $q=0$. We denote single-point CRs, whose $\beta$ and $\alpha$ are constant valued, with $\alpha^*$.} To determine how many GRBs are fulfilled by a given CR, we represent the uncertainties on both $\alpha$ and $\beta$ by using an elliptical region and require that this region intersects the line representing the CR on a $\alpha$--$\beta$ diagram. Those with a non-zero measure of intersection are classified as fulfilling the CR within 1 $\sigma$ and further discussed in Section \ref{sec:results}.

\begin{table*}
\begin{tabular}{lccccc} 
\hline
& & $p>2$ & & $1<p<2$ & \\
& $\beta$ & $\alpha$ & $\alpha(\beta)$& $\alpha$ & $\alpha(\beta)$ \\
\hline ISM & slow cooling & & && \\
$\nu<\nu_{a}$ & $-2$ & $-1$ & $\alpha^*=\frac{\beta}{2}$ & $\frac{11p-14}{8(p-1)}$ &--- \\
$\nu_{a}<\nu<\nu_{m}$ & $-\frac{1}{3}$ & $-\frac{4}{3}$ & $\alpha^*=4\beta$& $\frac{13p-10}{12(p-1)}$&--- \\
$\nu_{m}<\nu<\nu_{c}$ & $\frac{p-1}{2}$ & $\frac{p-3}{2}$ & $\alpha=\beta-1$ & $\frac{p-6}{8}$&$\alpha= \frac{2\beta-5}{8}$\\
\hline
ISM & fast cooling & &&& \\
$\nu<\nu_{a}$ & $-2$ & $-1$ & $\alpha^*=\frac{\beta}{2}$ & $-1$& $\alpha^*=\frac{\beta}{2}$\\
$\nu_{a}<\nu<\nu_{c}$ & $-\frac{1}{3}$ & $-\frac{4}{3}$ & $\alpha^*=4\beta$ & $-\frac{4}{3}$&$\alpha^*=4\beta$\\
$\nu_{c}<\nu<\nu_{m}$ & $\frac{1}{2}$ & $-\frac{1}{2}$ & $\alpha^*=-\beta$ & $-\frac{1}{2}$&$\alpha^* = -\beta$\\
$\nu>\nu_{m}$ & $\frac{p}{2}$ & $\frac{p-2}{2}$ & $\alpha= \beta-1$ & $0$&---\\
\hline
Wind & slow cooling & &&& \\
$\nu<\nu_{a}$ & $-2$ & $-2$ & $\alpha^*=\beta$ &$\frac{5p-6}{2(1-p)}$&--- \\
$\nu_{a}<\nu<\nu_{m}$ & $-\frac{1}{3}$ &  $-\frac{1}{3}$  & $\alpha^*=\beta$ &$-\frac{1}{3(p-1)}$ &---\\
$\nu_{m}<\nu<\nu_{c}$ & $\frac{p-1}{2}$ &  $\frac{p-1}{2}$  & $\alpha=\beta$ & $\frac{1}{2}$&---\\
\hline
Wind & fast cooling & &&& \\
$\nu<\nu_{a}$ & $-2$ & $-3$ & $\alpha^*=\frac{3\beta}{2}$&$-3$&$\alpha^*=\frac{3\beta}{2}$\\
$\nu_{a}<\nu<\nu_{c}$ & $-\frac{1}{3}$  & $\frac{1}{3}$ & $\alpha^*=-\beta$&$\frac{1}{3}$&$\alpha^*=-\beta$\\
$\nu_{c}<\nu<\nu_{m}$ & $\frac{1}{2}$ & $-\frac{1}{2}$ & $\alpha^*=-\beta$ & $-\frac{1}{2}$ & $\alpha^*=-\beta$\\
$\nu > \nu_{m}$& $\frac{p}{2}$ & $\frac{p-2}{2}$ & $\alpha = \beta - 1$&$0$ & ---\\
\hline
\end{tabular}
\caption{\label{tab:2} Taken from \citet{2013MNRAS.435.2520G} Table 7: the thick shell forward shock model for $\nu_a <$ min($\nu_m,\nu_c$). We denote single-point CRs, whose \(\alpha\) and \(\beta\) are constant valued, with \(\alpha^*\).}
\end{table*}

\begin{table*}
\centering
\begin{tabular}{lccccc} 
\hline
& & &$p>2$ &$1<p<2$\\
& $\beta$ & $\alpha$ & $\alpha(\beta)$& $\alpha$ &$\alpha(\beta)$ \\
\hline 
ISM & slow cooling & & & &  \\
$\nu<\nu_{a}$ & $-2$ & $-1$ &$\alpha^*=\frac{\beta}{2}$& $\frac{11p-14}{8(1-p)}$ & ---\\
$\nu_{a}<\nu<\nu_{m}$ & $-\frac{1}{3}$ & $-\frac{3}{2}$ & $\alpha^*=\frac{9\beta}{2}$& $-\frac{3}{2}$& $\alpha^*=\frac{9\beta}{2}$  \\
$\nu_{m}<\nu<\nu_{c}$ & $\frac{p-1}{2}$ & $\frac{p-3}{2}$ & $\alpha = \beta-1$&$\frac{p-6}{8}$  &$\alpha=\frac{2\beta-5}{8}$\\
\hline
Wind & slow cooling & & & & \\
$\nu<\nu_{a}$ & $-2$ & $-2$ &  $\alpha^*=\beta$  & $\frac{5p-6}{2(1-p)}$ &---\\
$\nu_{a}<\nu<\nu_{m}$ & $-\frac{1}{3}$ & $-\frac{5}{2}$ & $\alpha^*= \frac{15\beta}{2}$&$-\frac{5}{2}$ &$\alpha^*=\frac{15\beta}{2}$ \\
$\nu_{m}<\nu<\nu_{c}$ & $\frac{p-1}{2}$ & $\frac{p-1}{2}$ & $\alpha=\beta $& $\frac{1}{2}$&---\\
\hline
\end{tabular}
\caption{\label{tab:3} Taken from \citet{2013MNRAS.435.2520G} Table 8: the thick shell forward shock model for $\nu_m < \nu_a < \nu_c$. We denote single-point CRs, whose \(\alpha\) and \(\beta\) are constant valued, with \(\alpha^*\).}
\end{table*}

\begin{table*}
\begin{tabular}{lccccccc} 
\hline
& & No Injection &$1<p<2$ &$p>2$ &  { Injection } &$1<p<2$&$p>2$\\
& $\beta$ & $\alpha$ & $\alpha(\beta)$ &$\alpha(\beta)$ & $\alpha$ & $\alpha(\beta)$& $\alpha(\beta)$ \\
\hline ISM & slow cooling & & & & & \\
$\nu<\nu_{a}$ & $-2$ & $-\frac{1}{2}$ & --- & $\alpha^*=\frac{\beta}{4}$ & $\frac{q}{2}-1$  & ---& $\alpha^*=\frac{\beta}{2}$\\
$\nu_{a}<\nu<\nu_{m}$ & $-\frac{1}{3}$ & $-\frac{1}{2}$ & ---& $\alpha^*=\frac{3 \beta}{2}$ & $\frac{5 q-8}{6}$& --- & $\alpha^*=4\beta$ \\
$\nu_{m}<\nu<\nu_{c}$ & $\frac{p-1}{2}$ & $\frac{3(p-1)}{4}$ & $\alpha=\frac{6 \beta + 9}{16}$& $\alpha=\frac{3 \beta}{2}$ & $\frac{(2 p-6)+(p+3) q}{4}$ &  $\alpha=\frac{2\beta -5}{8}$&$\alpha=\beta - 1$ \\
\hline
ISM & fast cooling & & & & & \\
$\nu<\nu_{a}$ & $-2$ & $-1$ & $\alpha^* = \frac{\beta}{2}$&$\alpha^*=\frac{\beta}{2}$ & $-1$ & $\alpha^* = \frac{\beta}{2}$& $\alpha^*=\frac{\beta}{2}$ \\
$\nu_{a}<\nu<\nu_{c}$ & $-\frac{1}{3}$ & $-\frac{1}{6}$ & $\alpha^* = \frac{\beta}{2}$&$\alpha^*=\frac{\beta}{2}$ & $\frac{7 q-8}{6}$ & ---&$\alpha^*=4\beta$ \\
$\nu_{c}<\nu<\nu_{m}$ & $\frac{1}{2}$ & $\frac{1}{4}$ & $\alpha^* = \frac{\beta}{2}$&$\alpha^*=\frac{\beta}{2}$ & $\frac{3 q-2}{4}$ & ---&$\alpha^*=-\beta $ \\
$\nu>\nu_{m}$ & $\frac{p}{2}$ & $\frac{3 p-2}{4}$ &$\alpha=\frac{3 \beta+5}{8}$ &$\alpha=\frac{3 \beta-1}{2}$ & $\frac{(2 p-4)+(p+2) q}{4}$ & $\alpha = \frac{2\beta - 2}{8}$&$\alpha=\beta - 1$ \\
\hline
Wind & slow cooling & &  & & & \\
$\nu<\nu_{a}$ & $-2$ & $-1$ & ---& $\alpha^*=\frac{\beta}{2}$ & $q-2$ &---& $\alpha^*=\beta$ \\
$\nu_{a}<\nu<\nu_{m}$ & $-\frac{1}{3}$ & 0 &---& $\alpha^*=0$ & $-$ & ---&$\alpha=0$ \\
$\nu_{m}<\nu<\nu_{c}$ & $\frac{p-1}{2}$ & $\frac{3 p-1}{4}$ & $\alpha = \frac{2\beta +9}{8}$&$\alpha=\frac{3 \beta+1}{2}$ & $\frac{(2 p-2)+(p+1) q}{4}$ & $\alpha = \frac{1}{2}$& $\alpha= \beta$ \\
\hline
Wind & fast cooling & & & &  & \\
$\nu<\nu_{a}$ & $-2$ & $-2$& $\alpha^*=\beta$ & $\alpha^*=\beta$ & $q-3$ &---& $\alpha^*=\frac{3\beta}{2}$ \\
$\nu_{a}<\nu<\nu_{c}$ & $-\frac{1}{3}$ & $\frac{2}{3}$& $\alpha^*=-2 \beta$ & $\alpha^*=-2 \beta$ & $\frac{(1+q)}{3}$ &---& $\alpha^*=-\beta$ \\
$\nu_{c}<\nu<\nu_{m}$ & $\frac{1}{2}$ & $\frac{1}{4}$& $\alpha^*=\frac{\beta}{2}$ & $\alpha^*=\frac{\beta}{2}$ & $\frac{3 q-2}{4}$&--- & $\alpha^*=-\beta$ \\
$\nu>\nu_{m}$ & $\frac{p}{2}$ & $\frac{3 p-2}{4}$& $\alpha^*=\frac{2\beta+7}{8}$ & $\alpha^*=\frac{3 \beta-1}{2}$ & $\frac{(2 p-4)+(p+2) q}{4}$ & $\alpha=0$& $\alpha^*=\beta - 1$ \\
\hline
\end{tabular}
\caption{\label{tab:4} Taken from \citet{2013MNRAS.435.2520G} Gao Table 13: in relativistic, isotropic, self-similar deceleration phase for $\nu_a <$ min($\nu_m,\nu_c$) ($p>2$) and Gao Table 15: in relativistic, isotropic, self-similar deceleration phase for $\nu_a <$ min($\nu_m,\nu_c$) ($1<p<2$).We denote single-point CRs, whose \(\alpha\) and \(\beta\) are constant valued, with \(\alpha^*\).}
\end{table*}

\begin{table*}
\begin{tabular}{lccccccc} 
\hline
& & No Injection &$1<p<2$ &$p>2$ &  { Injection } &$1<p<2$&$p>2$\\
& $\beta$ & $\alpha$ & $\alpha(\beta)$ &$\alpha(\beta)$ & $\alpha$ & $\alpha(\beta)$& $\alpha(\beta)$ \\
\hline ISM & slow cooling & & & & & \\
$\nu<\nu_{m}$ & $-2$ & $-\frac{1}{2}$ &---& $\alpha^*=\frac{\beta}{4}$ & $\frac{q}{2}-1$ & ---&$\alpha^*=\frac{\beta}{2}$ \\
$\nu_{m}<\nu<\nu_{a}$ & $-\frac{5}{2}$ & $-\frac{5}{4}$ & $\alpha^*=\frac{\beta}{2}$& $\alpha^*=\frac{\beta}{2}$ & $\frac{q-6}{4}$ & ---&$\alpha^*=\frac{3 \beta}{5}$ \\
$\nu_{a}<\nu<\nu_{c}$ & $\frac{p-1}{2}$ & $\frac{3(p-1)}{4}$ & $\alpha = \frac{6\beta +9}{16}$&$\alpha=\frac{3 \beta}{2}$ & $\frac{(2 p-6)+(p+3) q}{4}$ & $\alpha = \frac{2\beta - 5}{8} $& $\alpha=\beta - 1$ \\
\hline
Wind & slow cooling & & &  & & \\
$\nu<\nu_{m}$ & $-2$ & $-1$ & ---& $\alpha^*=\frac{\beta}{2}$ & $q-2$ &---& $\alpha^*=\beta $ \\
$\nu_{m}<\nu<\nu_{a}$ & $-\frac{5}{2}$ & $-\frac{7}{4}$ & $\alpha^*= \frac{7\beta}{10}$&$\alpha^*=\frac{7 \beta}{10}$ & $\frac{3 q-10}{4}$ &--- & $\alpha^*=\beta $ \\
$\nu_{a}<\nu<\nu_{c}$ & $\frac{p-1}{2}$ & $\frac{3 p-1}{4}$ & $\alpha=\frac{2\beta + 9}{8}$&$\alpha=\frac{3 \beta+1}{2}$ & $\frac{(2 p-2)+(p+1) q}{4}$ &$\alpha=\frac{1}{2}$& $\alpha=\beta $ \\
\hline
\end{tabular}
\caption{\label{tab:5} Taken from \citet{2013MNRAS.435.2520G} Gao Table 14: in relativistic, isotropic, self-similar deceleration phase for $\nu_m<\nu_a<\nu_c$ ($p>2$) and Gao Table 16: in relativistic, isotropic, self-similar deceleration phase for $\nu_m<\nu_a<\nu_c$ ($1<p<2$). We denote single-point CRs, whose \(\alpha\) and \(\beta\) are constant valued, with \(\alpha^*\).}
\end{table*}

\begin{table*}
\begin{tabular}{lccccc} 
\hline
& & $p>2$ & & $1<p<2$ & \\
& $\beta$ & $\alpha$ & $\alpha(\beta)$& $\alpha$ & $\alpha(\beta)$ \\
\hline ISM & no injection & & && \\
$\nu<\nu_{a}$ & $-2$ & $\frac{1}{4}$ & $\alpha^*=\frac{\beta}{8}$ & $\frac{14-5p}{16(p-1)}$ &--- \\
$\nu_{a}<\nu<\nu_{m}$ & $-\frac{1}{3}$ & $\frac{1}{4}$ & $\alpha^*=\frac{3 \beta}{4}$ & $\frac{5p-8}{8(p-1)}$&--- \\
$\nu_{m}<\nu<\nu_{c}$ & $\frac{p-1}{2}$ & $\frac{3 p}{4}$ & $\alpha=\frac{6 \beta+3}{4}$ & $\frac{3(p+6)}{16}$&$\alpha= \frac{3(2\beta+7)}{16}$\\
\hline
Wind & no injection & &&& \\
$\nu<\nu_{a}$ & $-2$ & $-\frac{1}{2}$ & $\alpha^*=\frac{\beta}{4}$ & $\frac{14-9p}{8(p-1)}$&---\\
$\nu_{a}<\nu<\nu_{m}$ & $-\frac{5}{2}$ & $\frac{1}{2}$ & $\alpha^*=\frac{\beta}{5}$ & $\frac{11p-16}{12(p-1)}$&---\\
$\nu_{m}<\nu<\nu_{c}$ & $\frac{p-1}{2}$ & $\frac{3 p+1}{4}$ & $\alpha=\frac{3 \beta+2}{2}$ & $\frac{p+12}{8}$&$\alpha = \frac{2\beta + 13}{8}$\\
\hline
ISM & injection & &&& \\
$\nu<\nu_{a}$ & $-2$ & $\frac{3 q-2}{4}$ & $\alpha^*=\frac{\beta}{4}$ &$\frac{20-14p - 6q +9pq}{16(p-1)}$&--- \\
$\nu_{a}<\nu<\nu_{m}$ & $-\frac{1}{3}$ & $\frac{13 q-10}{12}$ & $\alpha^*=\frac{5\beta}{2}$ &$\frac{8-14p-32q+29pq}{24(p-1)}$ &---\\
$\nu_{m}<\nu<\nu_{c}$ & $\frac{p-1}{2}$ & $\frac{p(q+2)-4(1-q)}{4}$ & $\alpha=\beta-\frac{1}{2}$ & $\frac{22q - 4 + p(q+2)}{16}$&$\alpha= \frac{\beta - 1}{4} $\\
\hline
Wind & injection & &&& \\
$\nu<\nu_{a}$ & $-2$ & $\frac{3 q-4}{2}$ & $\alpha^*=\beta$&$\frac{24-20p-10q+11pq}{8(p-1)}$&---\\
$\nu_{a}<\nu<\nu_{m}$ & $-\frac{5}{2}$  & $\frac{5 q-2}{6}$ & $\alpha^*=\frac{2\beta}{15}$&$\frac{11pq-12q-4}{12(p-1)}$&---\\
$\nu_{m}<\nu<\nu_{c}$ & $\frac{p-1}{2}$ & $\frac{3 q-2+p(q+2)}{4}$ & $\alpha=\beta$ &$\frac{pq+8q+4}{8}$& $\alpha=\frac{1}{2} + \frac{2\beta +9}{8}$\\
\hline
\end{tabular}
\caption{\label{tab:6}Taken from \citet{2013MNRAS.435.2520G} Gao Table 18: after jet break for $\nu_a <$ min($\nu_m,\nu_c$), considering edge effect only. We denote single-point CRs, whose \(\alpha\) and \(\beta\) are constant valued, with \(\alpha^*\).}
\end{table*}

\begin{table*}
\begin{tabular}{lccccc} 
\hline
& & $p>2$ & & $1<p<2$ & \\
& $\beta$ & $\alpha$ & $\alpha(\beta)$& $\alpha$ & $\alpha(\beta)$ \\
\hline ISM & no injection & & && \\
$\nu<\nu_{m}$ & $-2$ & $\frac{1}{4}$ & $\alpha^*=\frac{\beta}{8}$ & $\frac{14-5p}{16(p-1)}$ &--- \\
$\nu_{m}<\nu<\nu_{a}$ & $-\frac{1}{3}$ & $-\frac{1}{2}$ & $\alpha^*=\frac{3 \beta}{2}$ & $-\frac{1}{2}$&$\alpha=\frac{3 \beta}{2}$\\
$\nu_{a}<\nu<\nu_{c}$ & $\frac{p-1}{2}$ & $\frac{3 p}{4}$ & $\alpha=\frac{6 \beta+3}{4}$ & $\frac{3(p+6)}{16}$&$\alpha= \frac{3(2\beta+7)}{16}$\\
\hline
Wind & no injection & &&& \\
$\nu<\nu_{m}$ & $-2$ & $-\frac{1}{2}$ & $\alpha^*=\frac{\beta}{4}$ & $\frac{14-9p}{8(p-1)}$&---\\
$\nu_{m}<\nu<\nu_{a}$ & $-\frac{5}{2}$ & $-\frac{5}{4}$ & $\alpha^*=\frac{\beta}{2}$ & $-\frac{5}{4}$&$\alpha^*=\frac{\beta}{2}$\\
$\nu_{a}<\nu<\nu_{c}$ & $\frac{p-1}{2}$ & $\frac{3 p+1}{4}$ & $\alpha=\frac{3 \beta+2}{2}$ & $\frac{p+12}{8}$&$\alpha = \frac{2\beta + 13}{8}$\\
\hline
ISM & injection & &&& \\
$\nu<\nu_{m}$ & $-2$ & $\frac{3 q-2}{4}$ & --- &$\frac{20-14p - 6q +9pq}{16(p-1)}$&--- \\
$\nu_{m}<\nu<\nu_{a}$ & $-\frac{1}{3}$ & $\frac{q-2}{2}$ & --- &$\frac{q-2}{2}$ &---\\
$\nu_{a}<\nu<\nu_{c}$ & $\frac{p-1}{2}$ & $\frac{p(q+2)-4(1-q)}{4}$ & $\alpha=\beta-\frac{1}{2}$ & $\frac{22q - 4 + p(q+2)}{16}$&$\alpha= \frac{\beta - 1}{4} $\\
\hline
Wind & injection & &&& \\
$\nu<\nu_{m}$ & $-2$ & $\frac{3 q-4}{2}$ & ---&$\frac{24-20p-10q+11pq}{8(p-1)}$&---\\
$\nu_{m}<\nu<\nu_{a}$ & $-\frac{5}{2}$  & $\frac{5 (q-2)}{4}$ & ---&$\frac{5(q-2)}{4}$&---\\
$\nu_{a}<\nu<\nu_{c}$ & $\frac{p-1}{2}$ & $\frac{3 q-2+p(q+2)}{4}$ & $\alpha=\beta$ &$\frac{pq+8q+4}{8}$& $\alpha=\frac{1}{2} + \frac{2\beta +9}{8}$\\
\hline
\end{tabular}
\caption{\label{tab:7}Taken from \citet{2013MNRAS.435.2520G} Gao Table 19: after jet break for $\nu_m < \nu_a < \nu_c$, considering edge effect only. We denote single-point CRs, whose \(\alpha\) and \(\beta\) are constant valued, with \(\alpha^*\).}
\end{table*}

\begin{table*}
\begin{center}
\begin{tabular}{ l c c c c c c}
\hline
 $n(r)$ & $\beta$ & $\alpha(\beta)$  & $\beta$& $\alpha(\beta)$ & $\beta$ & $\alpha(\beta)$ \\
 \hline
 Fast Cooling & &$\nu < \nu_c$ && $\nu_c < \nu < \nu_m $&&  $\nu > \nu_m$\\
 \hline
 $r^0$ & $-\frac{1}{3}$ & $\alpha^* = 4\beta$ & $\frac{1}{2}$& $\alpha^* = -\beta$ & $\frac{p}{2}$& $\alpha = \beta-1$  \\
 $r^{-1}$ & $-\frac{1}{3}$ & $\alpha^* = \frac{7\beta}{3}$ & $\frac{1}{2}$& $\alpha^* = -\beta$& $\frac{p}{2}$ & $\alpha = \beta-1$ \\
 $r^{-1.5}$ & $-\frac{1}{3}$ & $\alpha^* = \beta$ & $\frac{1}{2}$& $\alpha^* = -\beta$ & $\frac{p}{2}$&  $\alpha = \beta - 1$ \\
 $r^{-2}$ & $-\frac{1}{3}$ & $\alpha^* = -\beta$ & $\frac{1}{2}$& $\alpha^* = -\beta$ & $\frac{p}{2}$& $\alpha = \beta - 1$\\
  $r^{-2.5}$ & $-\frac{1}{3}$ & $\alpha^* = -\frac{13\beta}{3}$ & $\frac{1}{2}$& $\alpha^* = -\beta$ & $\frac{p}{2}$& $\alpha = \beta - 1$\\
 \hline
 Slow Cooling && $\nu < \nu_m$&&  $\nu_m < \nu <
 \nu_c$&& \\ 
 \hline
 $r^0$ & $-\frac{1}{3}$ & $\alpha^* = 4\beta$& $\frac{p-1}{2}$ & $\alpha = \beta-1$& &  \\
 $r^{-1}$ & $-\frac{1}{3}$ & $\alpha^* = 3\beta$ & $\frac{p-1}{2}$& $\alpha = \beta-\frac{2}{3}$ & & \\
 $r^{-1.5}$ & $-\frac{1}{3}$ & $\alpha^* = \frac{11\beta}{5}$ & $\frac{p-1}{2}$& $\alpha = \beta-\frac{2}{5}$ & &  \\
  $r^{-2}$ & $-\frac{1}{3}$ & $\alpha^* = \beta$ & $\frac{p-1}{2}$& $\alpha = \beta$ & &  \\
 $r^{-2.5}$ & $-\frac{1}{3}$ & $\alpha^* = \beta$ & $\frac{p-1}{2}$& $\alpha = \beta+\frac{2}{3}$ & & \\
 \hline
\end{tabular}
\end{center}
\caption{\label{tab:8}Assumes a thick shell model for varying density profiles $n(r) \propto r^k$ for $k = 0,-1,-1.5,-2,-2.5$; including stellar wind and ISM mediums. We denote single-point CRs, whose \(\alpha\) and \(\beta\) are constant valued, with \(\alpha^*\).}
\end{table*}

\section{Results}
\label{sec:results}

\subsection{The plateau sample}\label{sec:plateau}
For the subsample of 14 GRBs that display a radio plateau, we find that 7/14 GRBs (50\%) fulfill at least one CR in our set. The most preferred regime among our set of CRs is the SC, $\nu_{\rm m} < \nu < \nu_{\rm c}$ regime, for which 6 GRBs among the plateau sample fulfill the relation in at least one case. The preference for this regime is likely due to the variation of $\alpha$ and $\beta$, which presents visually as a linear CR, and is thus more likely to be fulfilled by multiple GRBs. The FC, $\nu < \nu_{\rm c}$ regime in the $k=2.5$ profile only satisfied by 1 GRB in the plateau sample, and is the only other regime satisfied by any GRB in our sample. The majority of CRs with a constant value of $\beta$ are not satisfied by any GRB in our sample as the point-CRs almost always have negative alpha values (i.e., rising LCs). These CRs fall at the bottom left end of the $\alpha$-$\beta$ plane. Almost no bursts in our sample have a negative $\alpha$ value.

We now consider the details of GRBs which fulfill at least one CR. GRB 980425 fulfills the SC, $\nu_{\rm m} < \nu < \nu_{\rm c}$ regime without injection from Table \ref{tab:4} (which matches the SC, $\nu_{\rm a} < \nu < \nu_{\rm c}$ regime from Table \ref{tab:5}) for both the ISM and Wind environment. It also fulfills the same regime from Table \ref{tab:8} for $k=2.5$.
GRB 010222 fulfills the SC, $\nu_{\rm m} < \nu < \nu_{\rm c}$ regime without injection from Tables \ref{tab:2} and \ref{tab:3} in an ISM environment, as well as the same regime from Table \ref{tab:8} for $k=1, 1.5, 2$. 
GRB 011030 fulfills the SC, $\nu_{\rm m} < \nu < \nu_{\rm c}$ regime without injection from Tables \ref{tab:2}, \ref{tab:3}, \ref{tab:4}, and \ref{tab:5} in an ISM environment. It also satisfies the SC, $\nu_{\rm m} < \nu < \nu_{\rm c}$ regime with injection from Tables \ref{tab:4} and \ref{tab:5} in an ISM environment. 
GRB 030329 fulfills the SC, $\nu_{\rm m} < \nu < \nu_{\rm c}$ regime without injection from Table \ref{tab:6} (which matches the SC, $\nu_{\rm a} < \nu < \nu_{\rm c}$ regime from Table \ref{tab:7}) in a Wind environment. 
GRB 070612A satisfies the FC, $\nu < \nu_{\rm c}$ without injection from Table \ref{tab:8} for $k=2.5$. 
GRB 111215A satisfies the SC, $\nu_{\rm m} < \nu < \nu_{\rm c}$ without injection in Tables \ref{tab:2} and \ref{tab:3} in a Wind environment; in Table \ref{tab:8} in a Wind environment; and in Tables \ref{tab:4} and \ref{tab:5} with energy injection for an ISM environment. 
GRB 141121A satisfies the SC, $\nu_{\rm m} < \nu < \nu_{\rm c}$ regime without injection in Tables \ref{tab:2} and \ref{tab:3} for an ISM environment. It satisfies the same regime without injection in Tables \ref{tab:4}, \ref{tab:5}, \ref{tab:6}, and \ref{tab:7} for both an ISM and Wind environment.

\begin{figure*} 
\centering
\begin{tabular}{ccc}
    \includegraphics[width =0.28\textwidth]{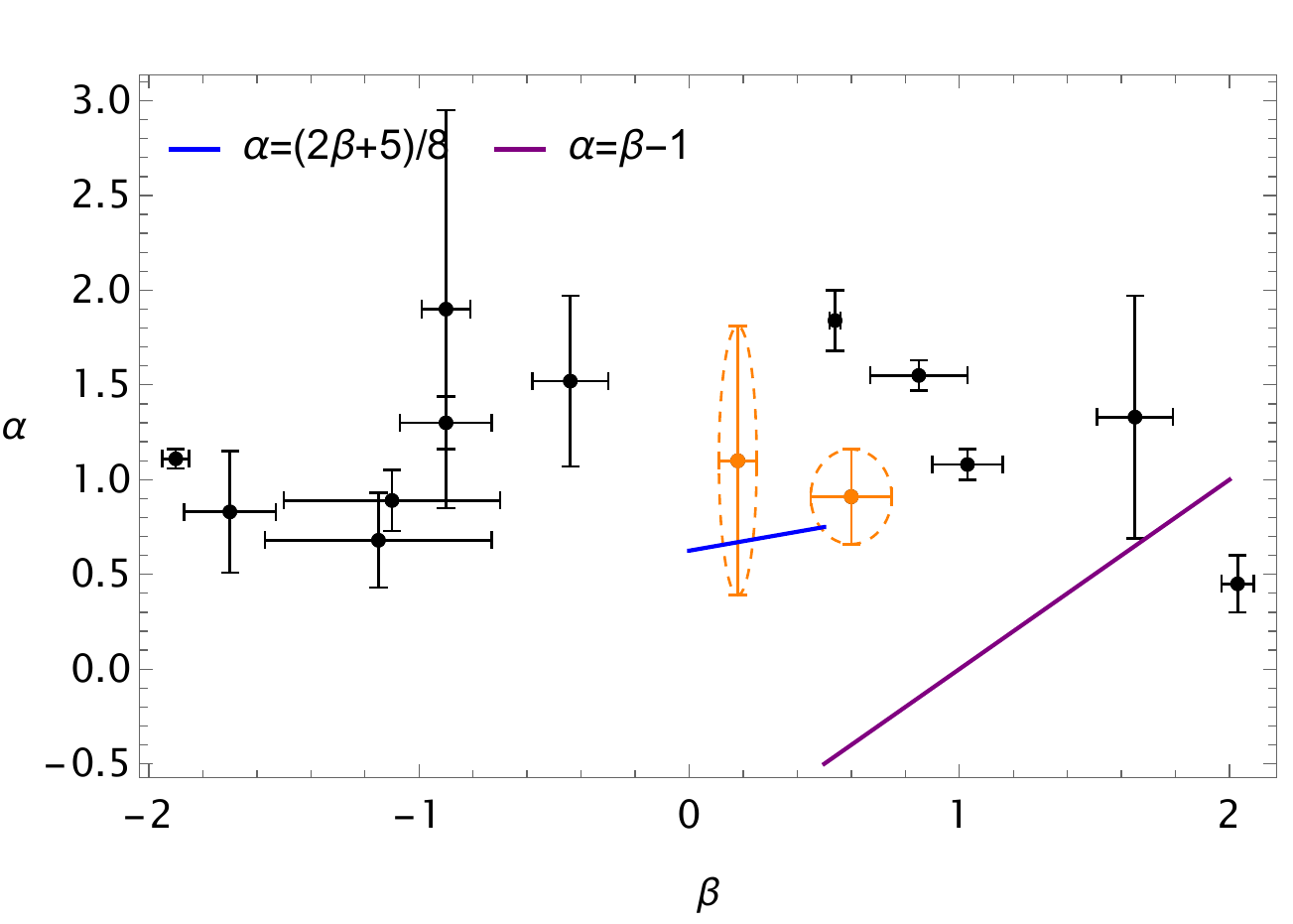} &
    \includegraphics[width=0.28\textwidth]{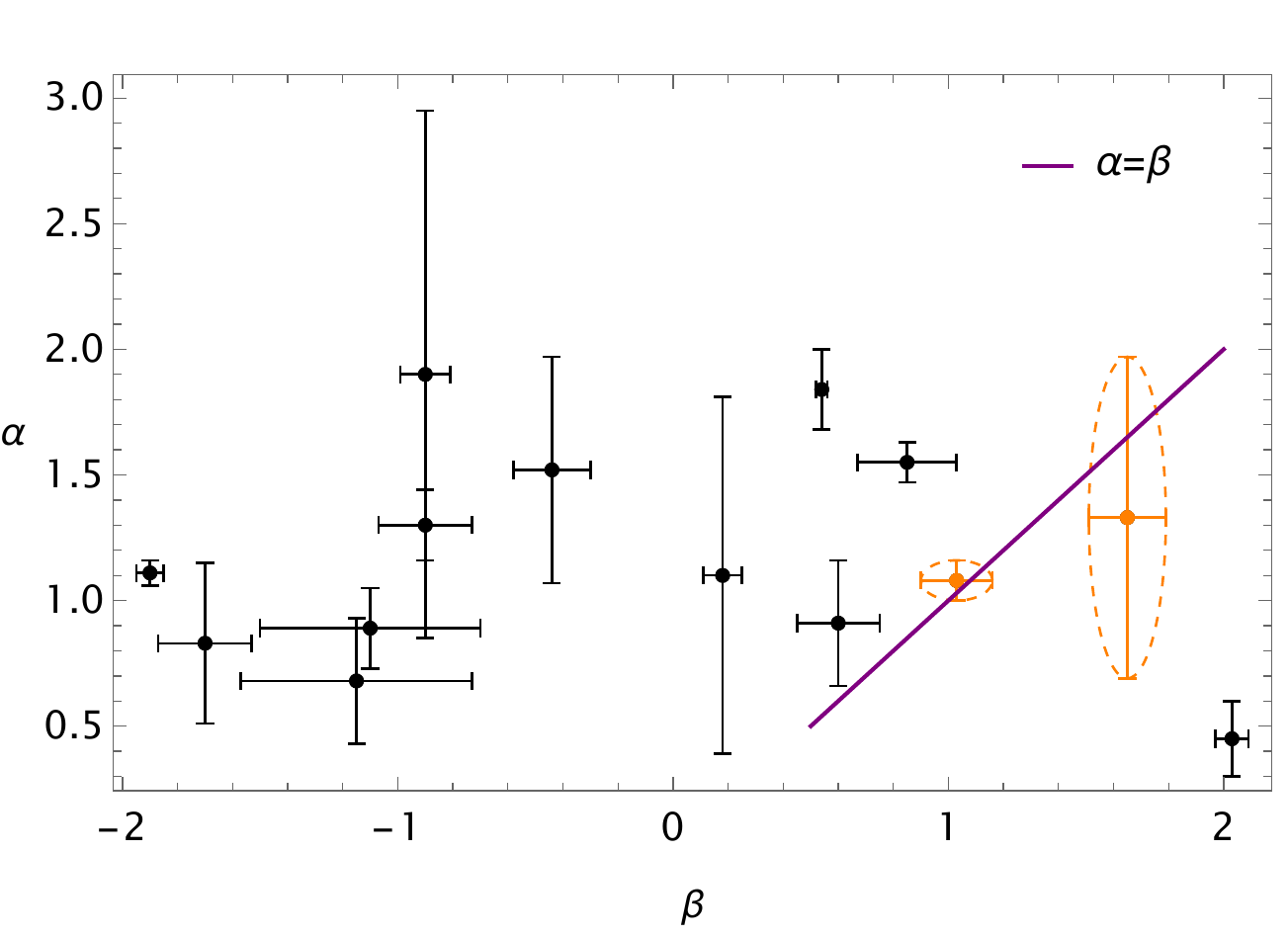} &
    \includegraphics[width = 0.28\textwidth]{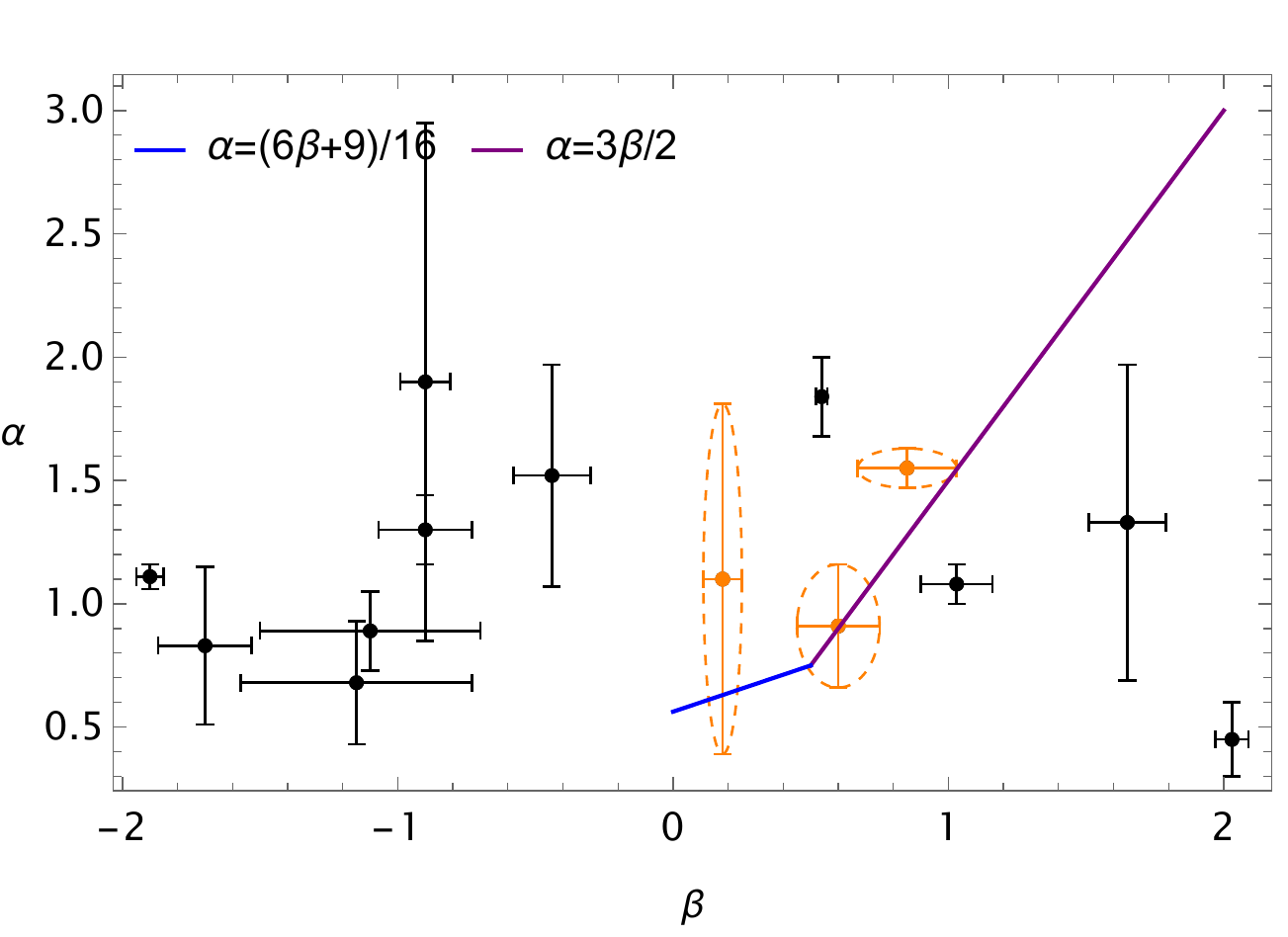} \\
    
    \includegraphics[width = 0.28\textwidth]{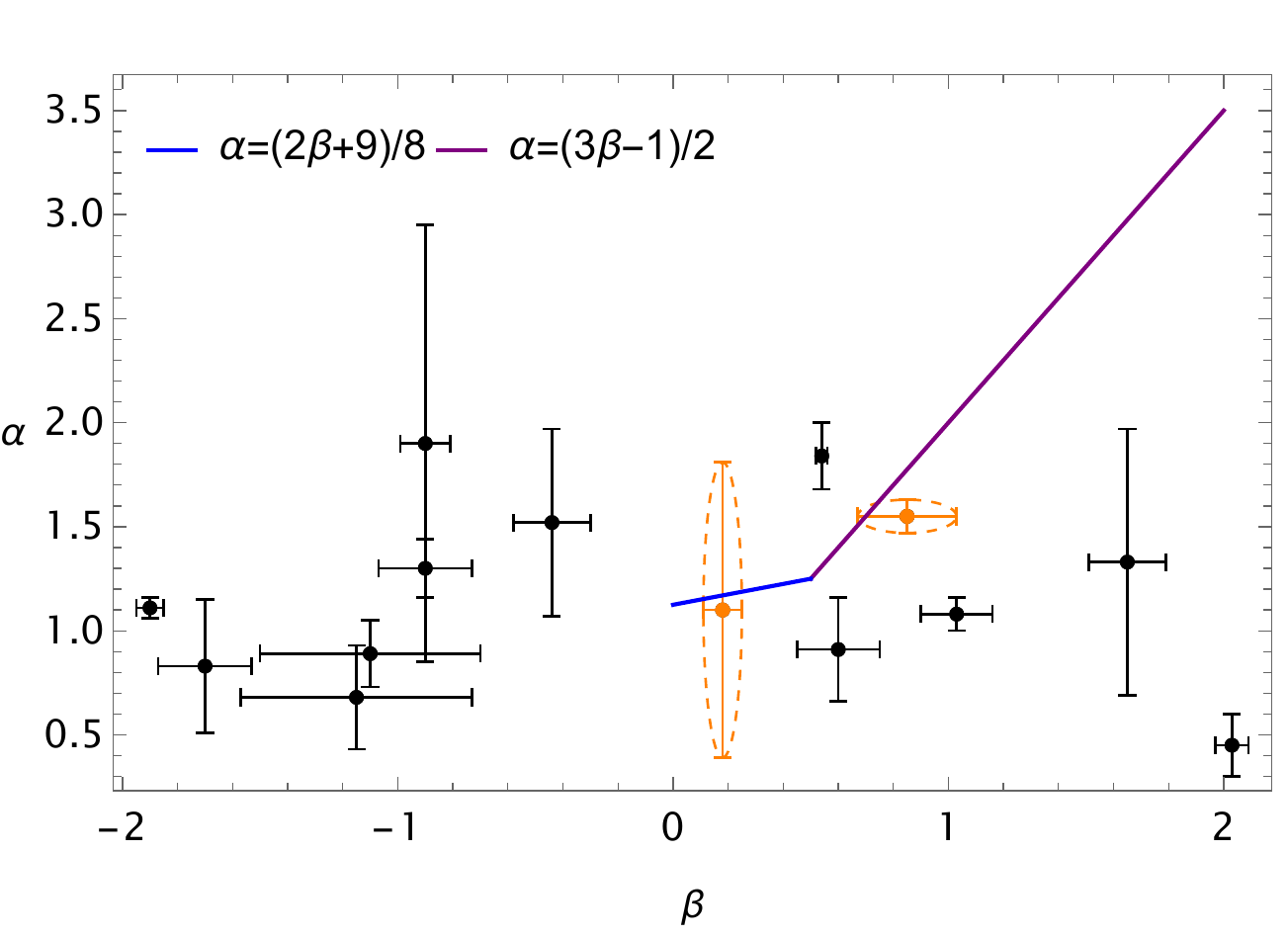} &
    \includegraphics[width = 0.28\textwidth]{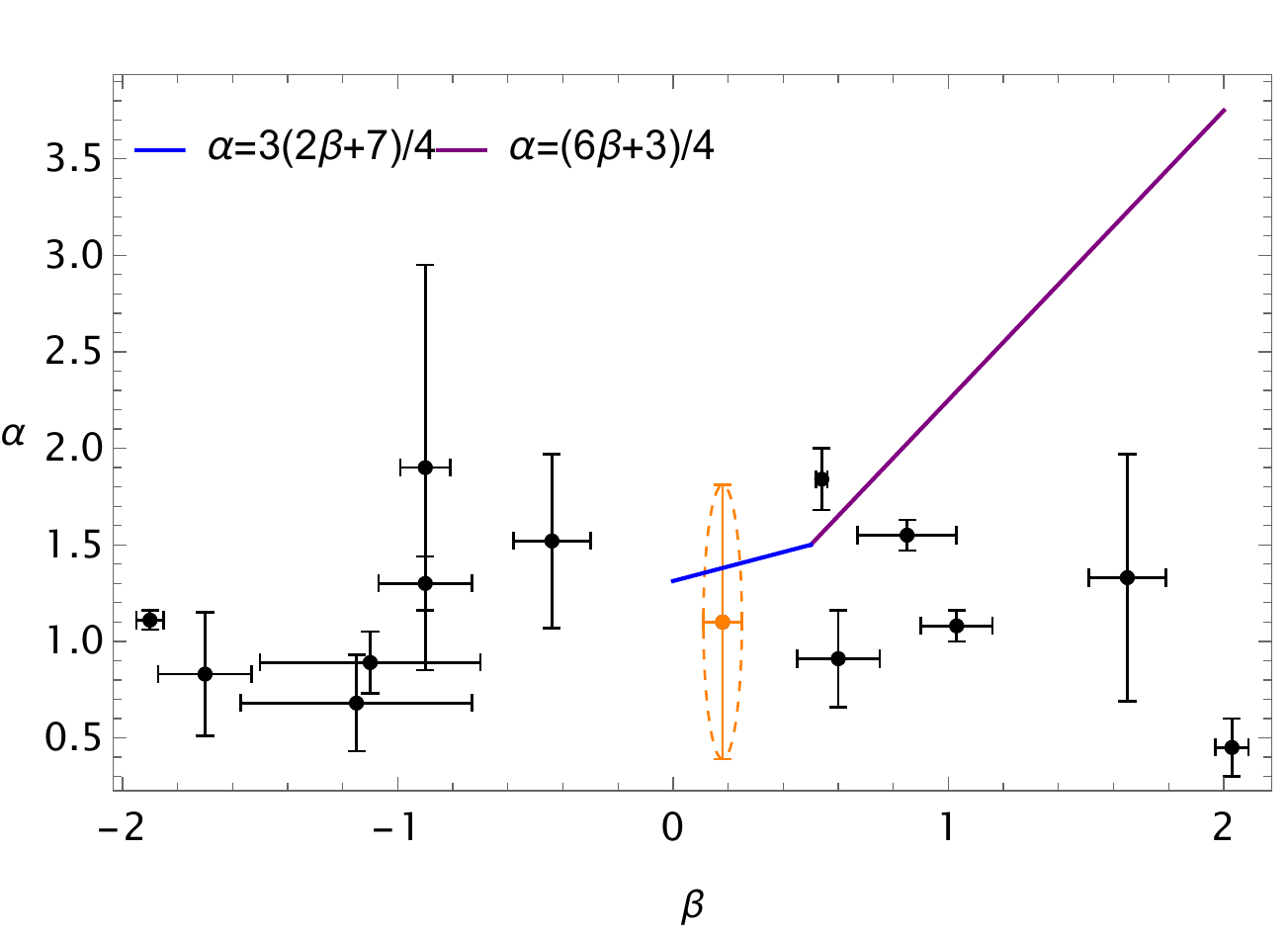} &
    \includegraphics[width = 0.28\textwidth]{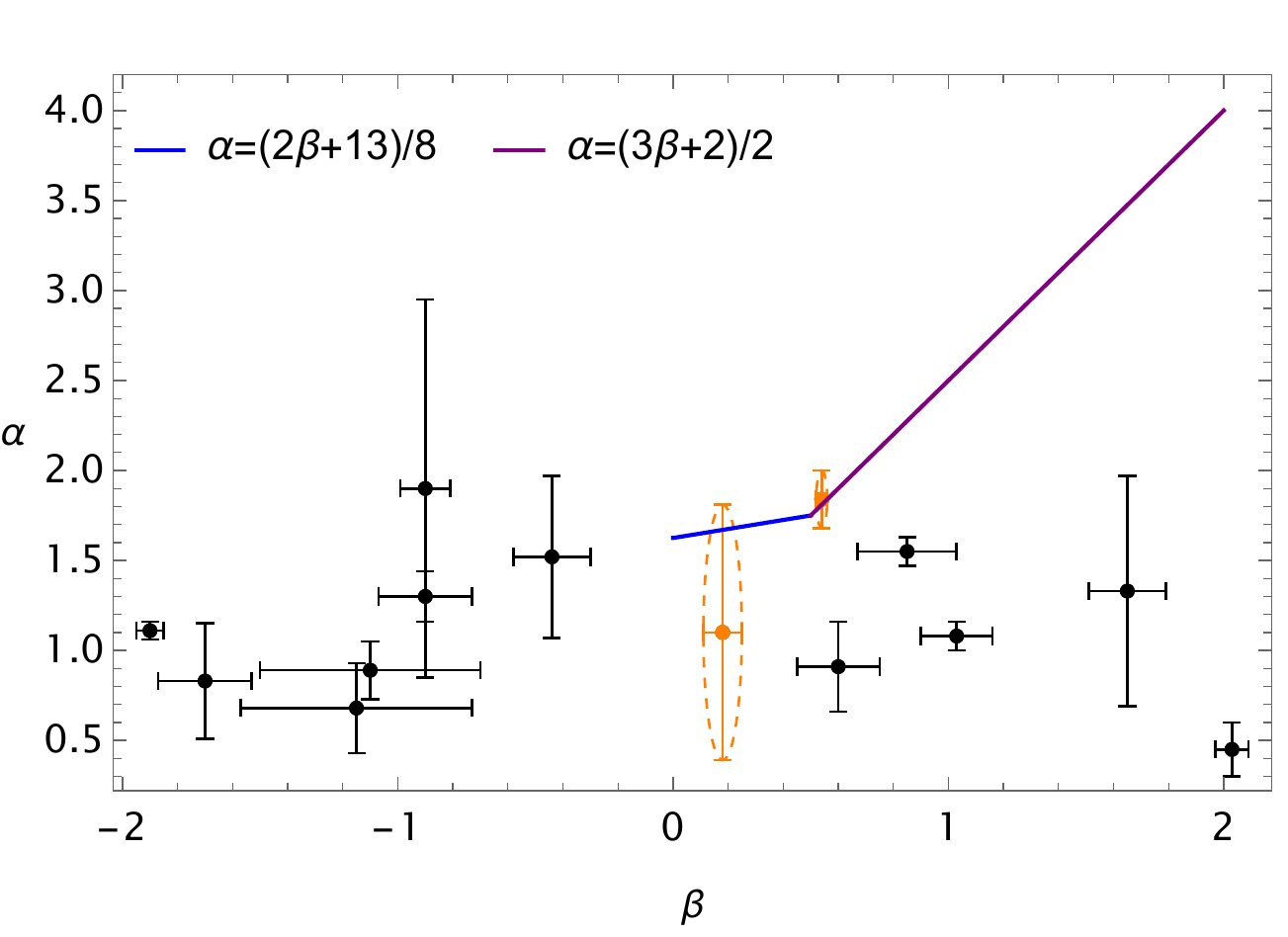} \\
    
    \includegraphics[width = 0.28\textwidth]{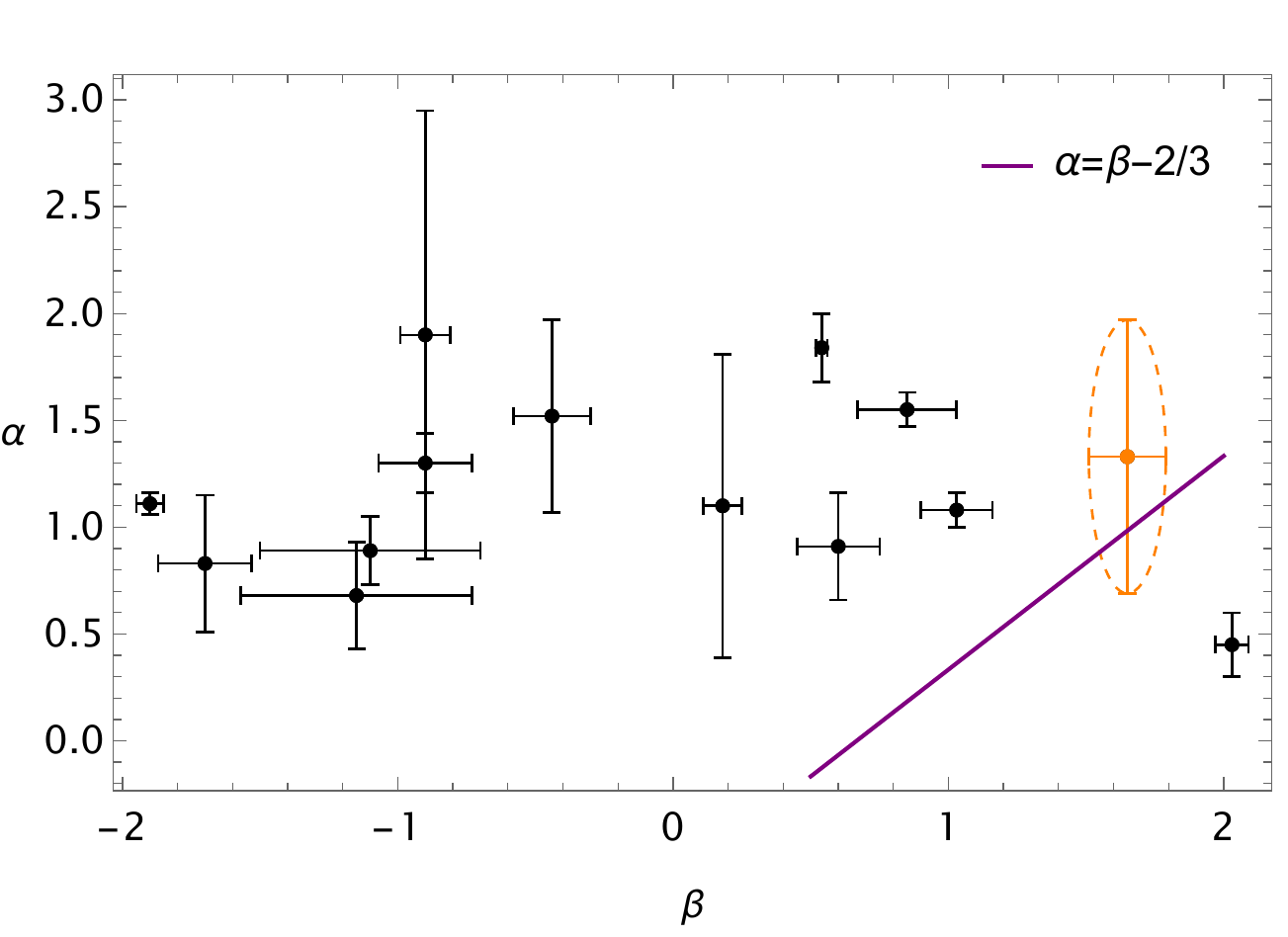} & 
    \includegraphics[width = 0.28\textwidth]{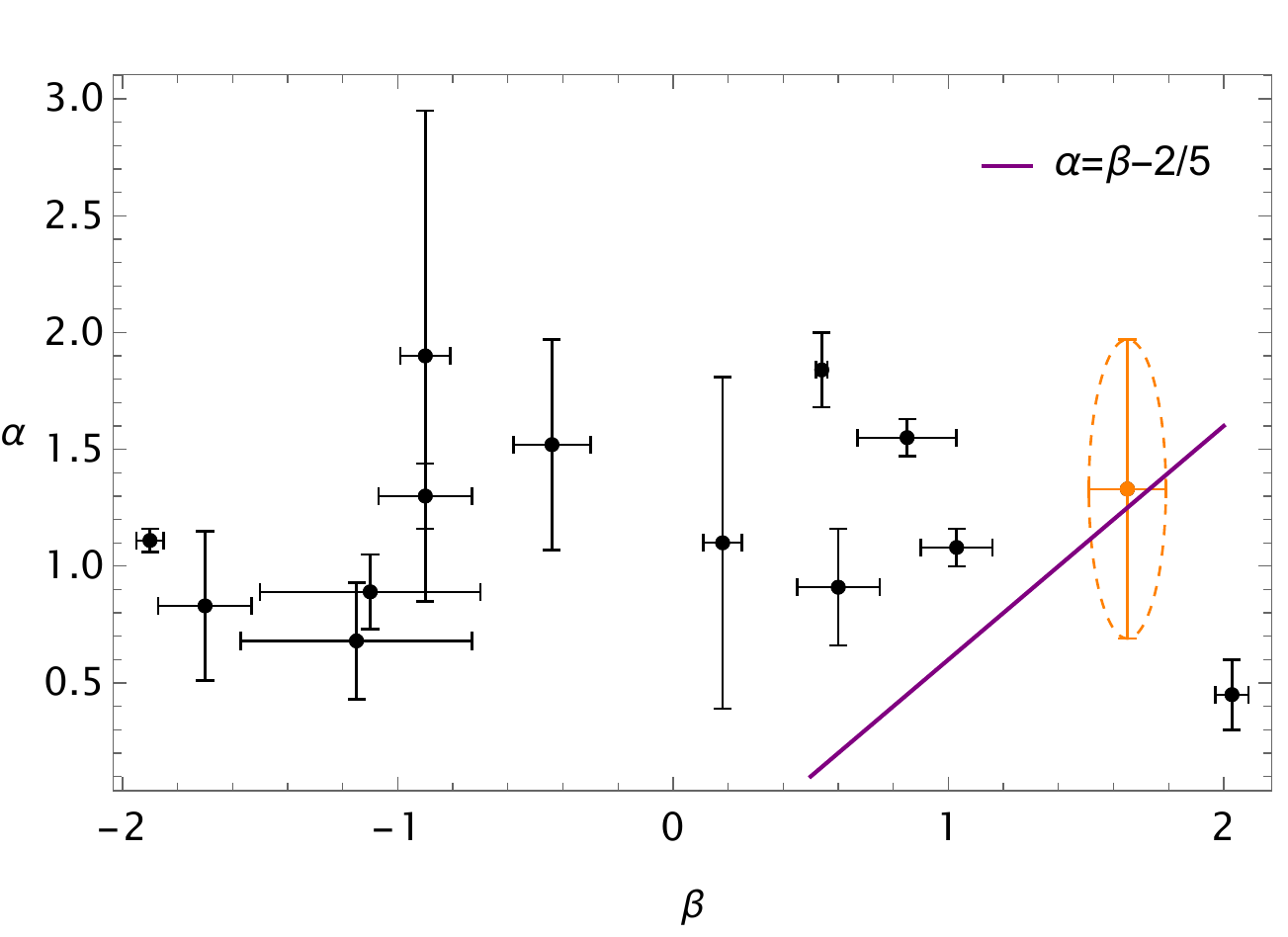} &
    \includegraphics[width = 0.28\textwidth]{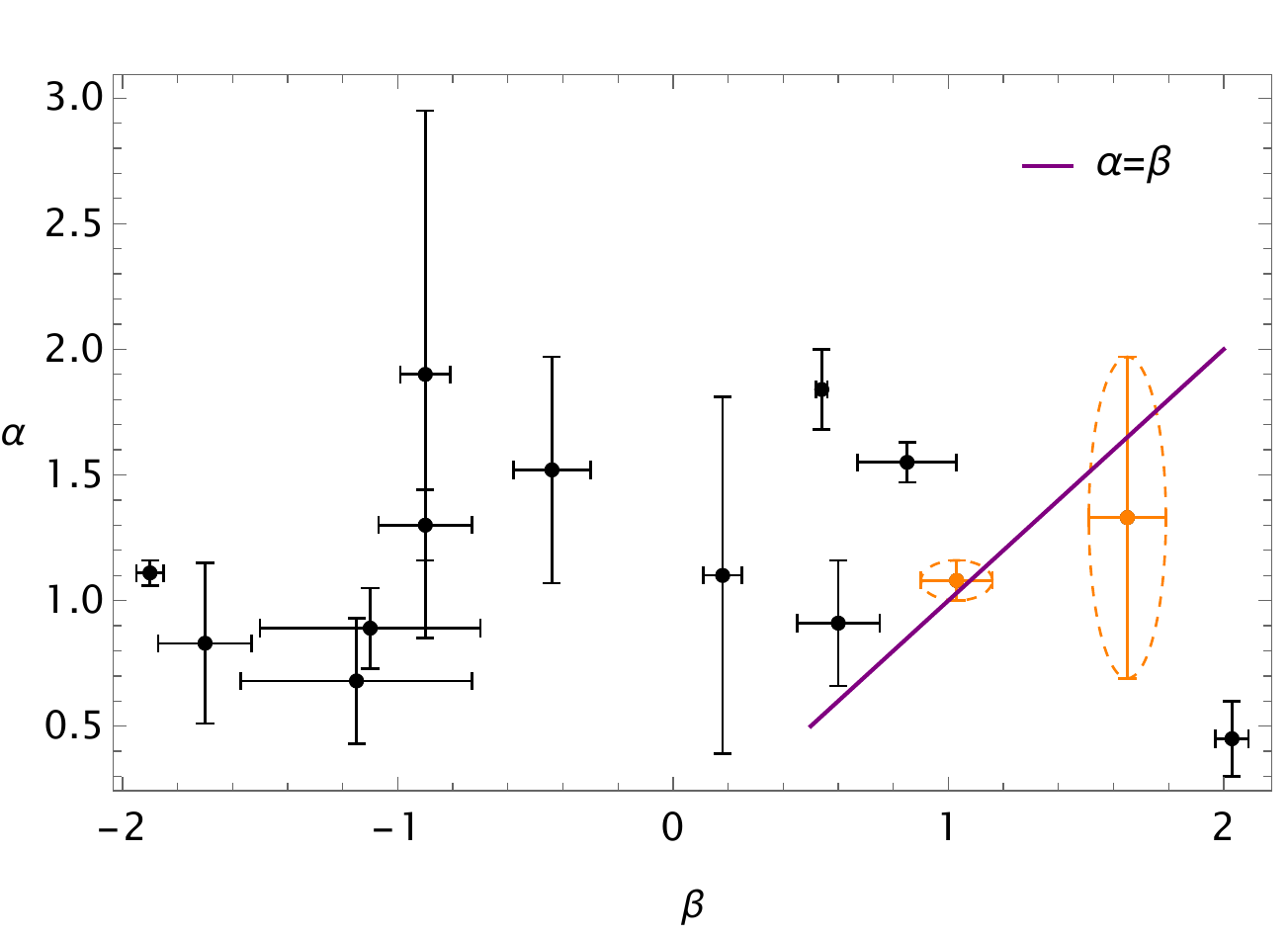} \\
    
    \includegraphics[width = 0.28\textwidth]{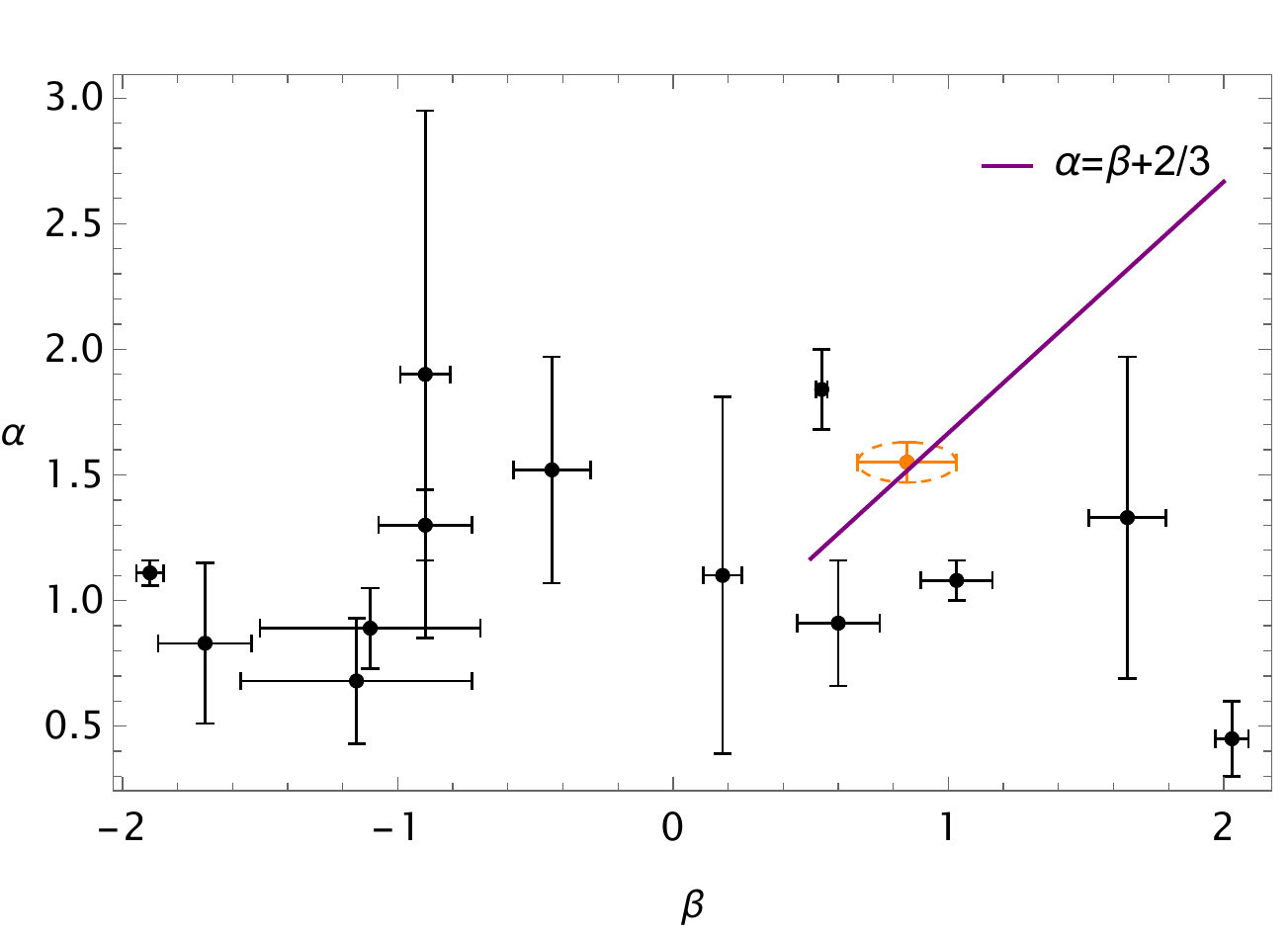} &
    \includegraphics[width = 0.28\textwidth]{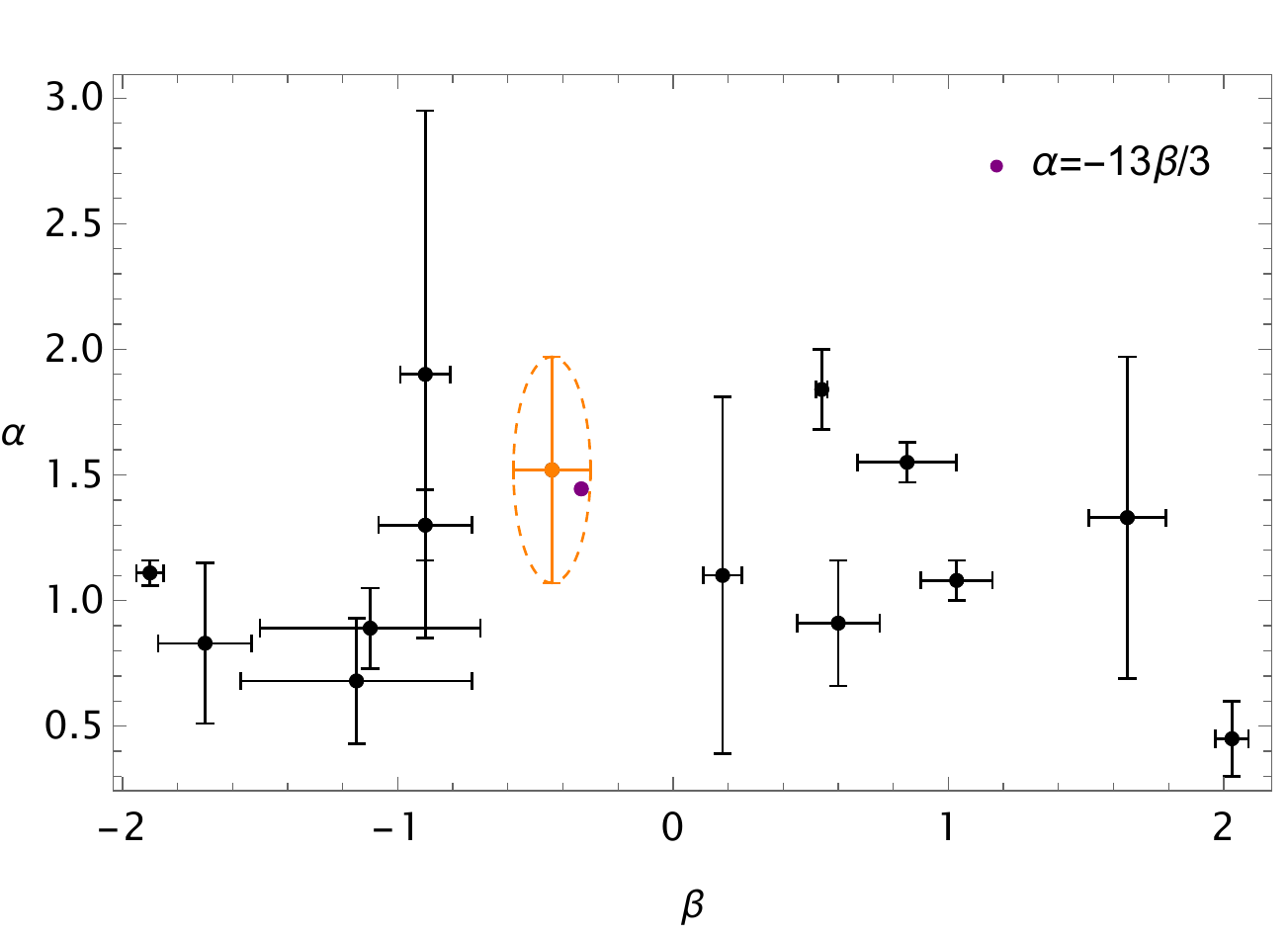} \\
\end{tabular}
\caption{Plots of fulfilled CRs, without energy injection, tested against plateau sample of 14 GRBs. Blue and purple colored lines represent the $1<p<2$ and $p>2$ spectral regimes, respectively. GRBs that fulfill the CR are shown in orange, GRBs that do not fulfill the CR are shown in black. Plots are as follows: (1) ISM, SC, $\nu_{\rm m} < \nu < \nu_{\rm c}$ CRs in Tables \ref{tab:2} and \ref{tab:3}; (2) Wind, SC, $\nu_{\rm m} < \nu < \nu_{\rm c}$ CRs in Tables \ref{tab:2} and \ref{tab:3}; (3) ISM, SC, $\nu_{\rm m} < \nu < \nu_{\rm c}$ CRs in Tables \ref{tab:4} and \ref{tab:5}; (4) Wind, SC, $\nu_{\rm m} < \nu < \nu_{\rm c}$ CRs in Tables \ref{tab:4} and \ref{tab:5}; (5) ISM, SC, $\nu_{\rm m} < \nu < \nu_{\rm c}$ CRs in Tables \ref{tab:6} and \ref{tab:7}; (6) Wind, SC, $\nu_{\rm m} < \nu < \nu_{\rm c}$ CRs in Tables \ref{tab:6} and \ref{tab:7}; (7) $k=1$, SC, $\nu_{\rm m} < \nu < \nu_{\rm c}$; (8) $k=1.5$, SC, $\nu_{\rm m} < \nu < \nu_{\rm c}$; (9) $k=2$, SC, $\nu_{\rm m} < \nu < \nu_{\rm c}$; (10) $k=2.5$, SC, $\nu_{\rm m} < \nu < \nu_{\rm c}$; (11) $k=2.5$, FC, $\nu < \nu_{\rm c}$.}
\label{fig:plat_noinj} 
\end{figure*}

\begin{figure*} 
\centering
\begin{tabular}{cc}
    \includegraphics[width = 0.45\textwidth]{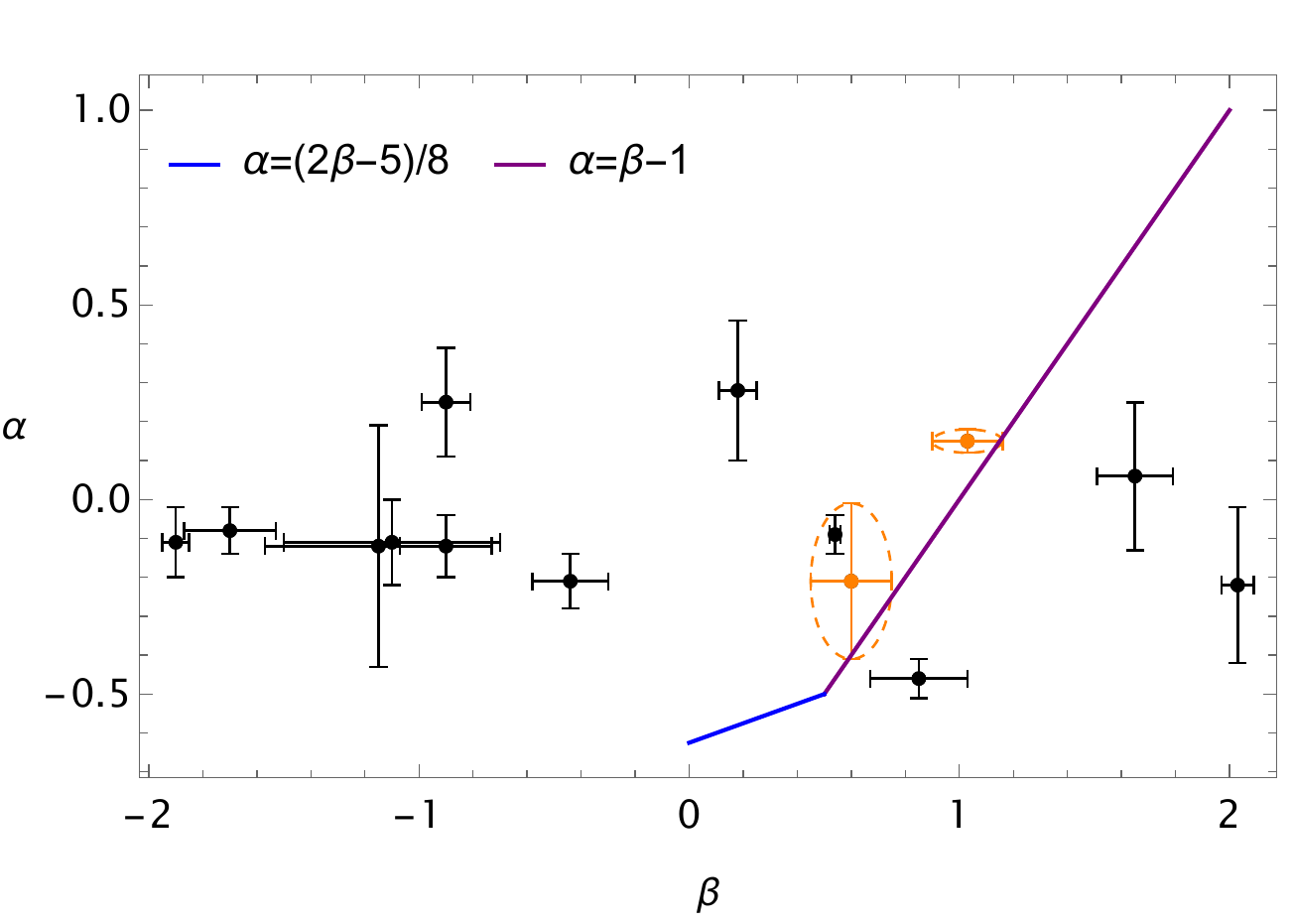} & 
    \includegraphics[width = 0.45\textwidth]{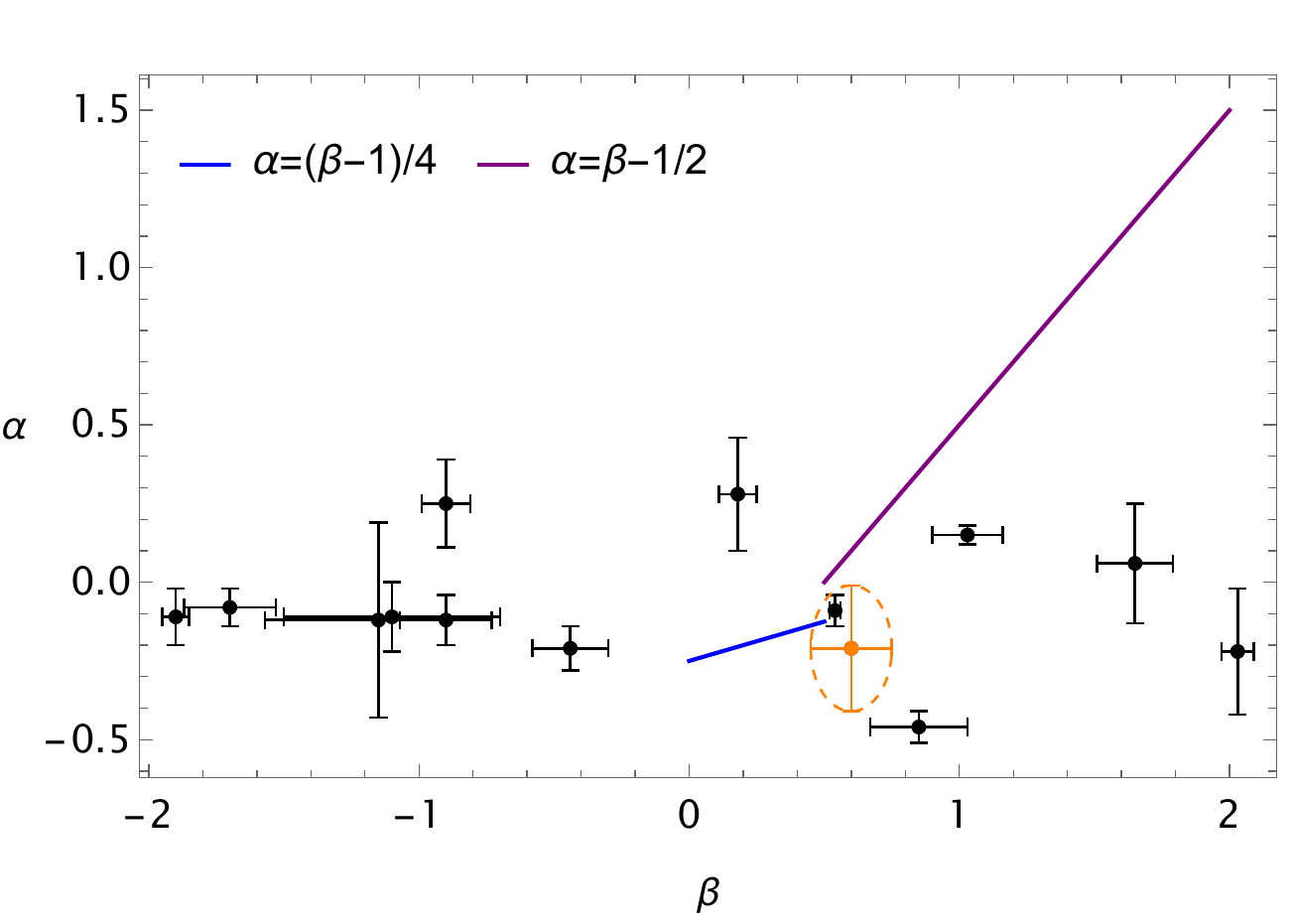}
\end{tabular}
\caption{Plots of fulfilled CRs, with energy injection, tested against plateau sample of 14 GRBs. Blue and purple colored lines represent the $1<p<2$ and $p>2$ spectral regimes, respectively. GRBs that fulfill the CR are shown in orange, GRBs that do not fulfill the CR are shown in black. The left plot shows the ISM, SC, $\nu_{\rm m} < \nu < \nu_{\rm c}$ CRs in Table \ref{tab:4} and the right plot shows the ISM, SC, $\nu_{\rm m} < \nu < \nu_{\rm c}$ in  Table \ref{tab:6}.
\label{fig:plat_inj}}
\end{figure*}

The fulfilled CRs for the plateau sample are shown in Figure \ref{fig:plat_noinj} for relations without injection and Figure \ref{fig:plat_inj} for relations with injection. The GRBs that do or do not fulfill the relations are shown in orange and black, respectively. Relations for $1 < p < 2$ and $p > 2$ are shown in blue, and purple, respectively.

\subsection{Full sample}\label{sec:full}

We test our full sample of 26 GRBs from Table \ref{tab:Sample} against the CRs presented in Tables \ref{tab:2},
\ref{tab:3}, \ref{tab:4}, \ref{tab:5},
\ref{tab:6},  \ref{tab:7}, and  \ref{tab:8} that contain scenarios with and without energy injection as well as the conditions of thin and thick shells within various synchrotron spectrum regimes. We find that 12/26 GRBs (46\%) in our sample satisfy at least one CR. 

When considering the 13 total relations that are fulfilled, we see that 9 of the relations are without energy injection and 2 of the relations are with energy injection. Of these, there is an average of 2.6 GRBs fulfilling the relations without injection, and an average of 2.5 GRBs fulfilling the CRs with injection, indicating a slight preference for cases without energy injection compared to the cases with injection.

\begin{figure*} 
\centering
\begin{tabular}{ccc}
    \includegraphics[width =0.28\textwidth]{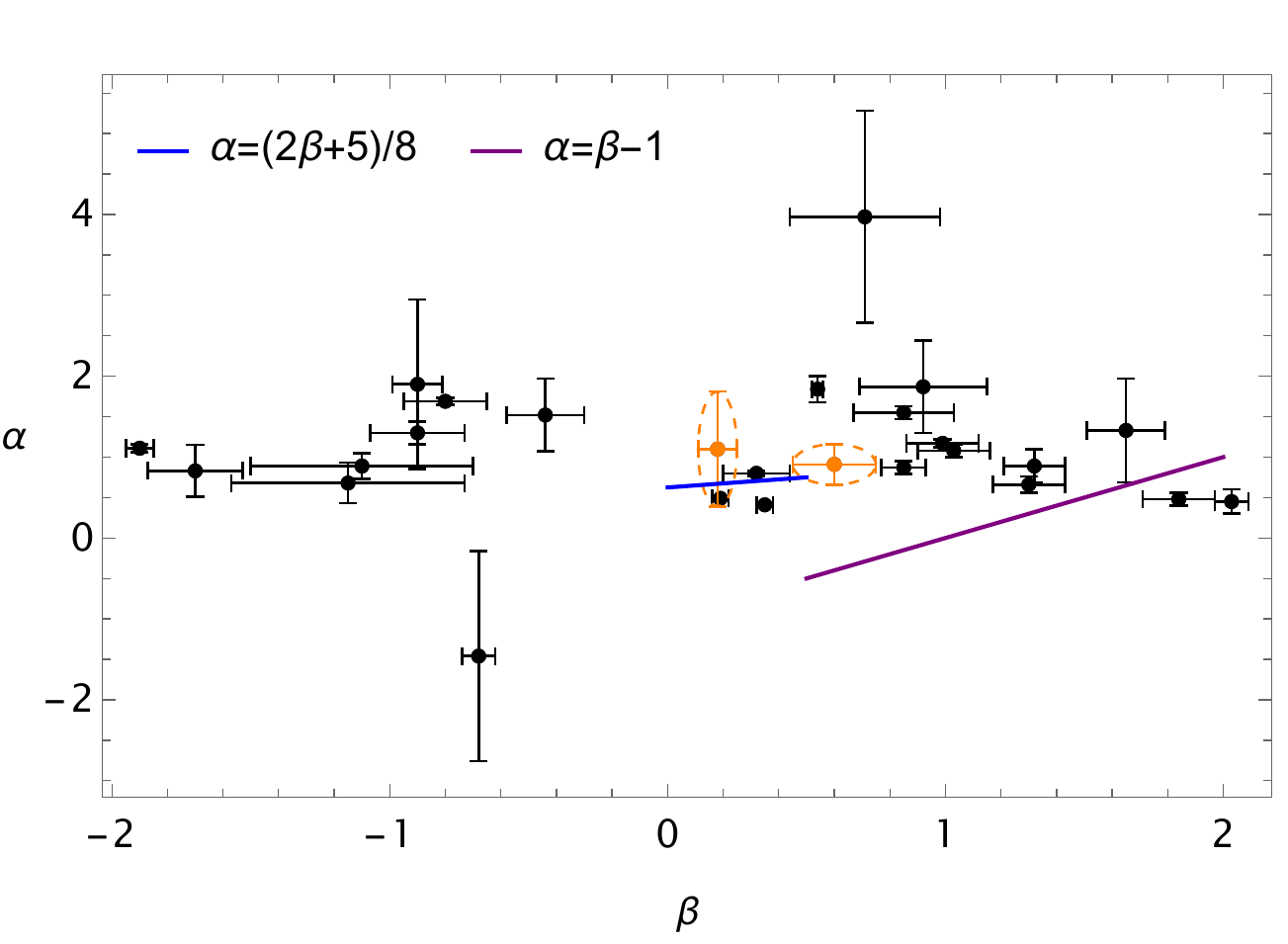} &
    \includegraphics[width=0.28\textwidth]{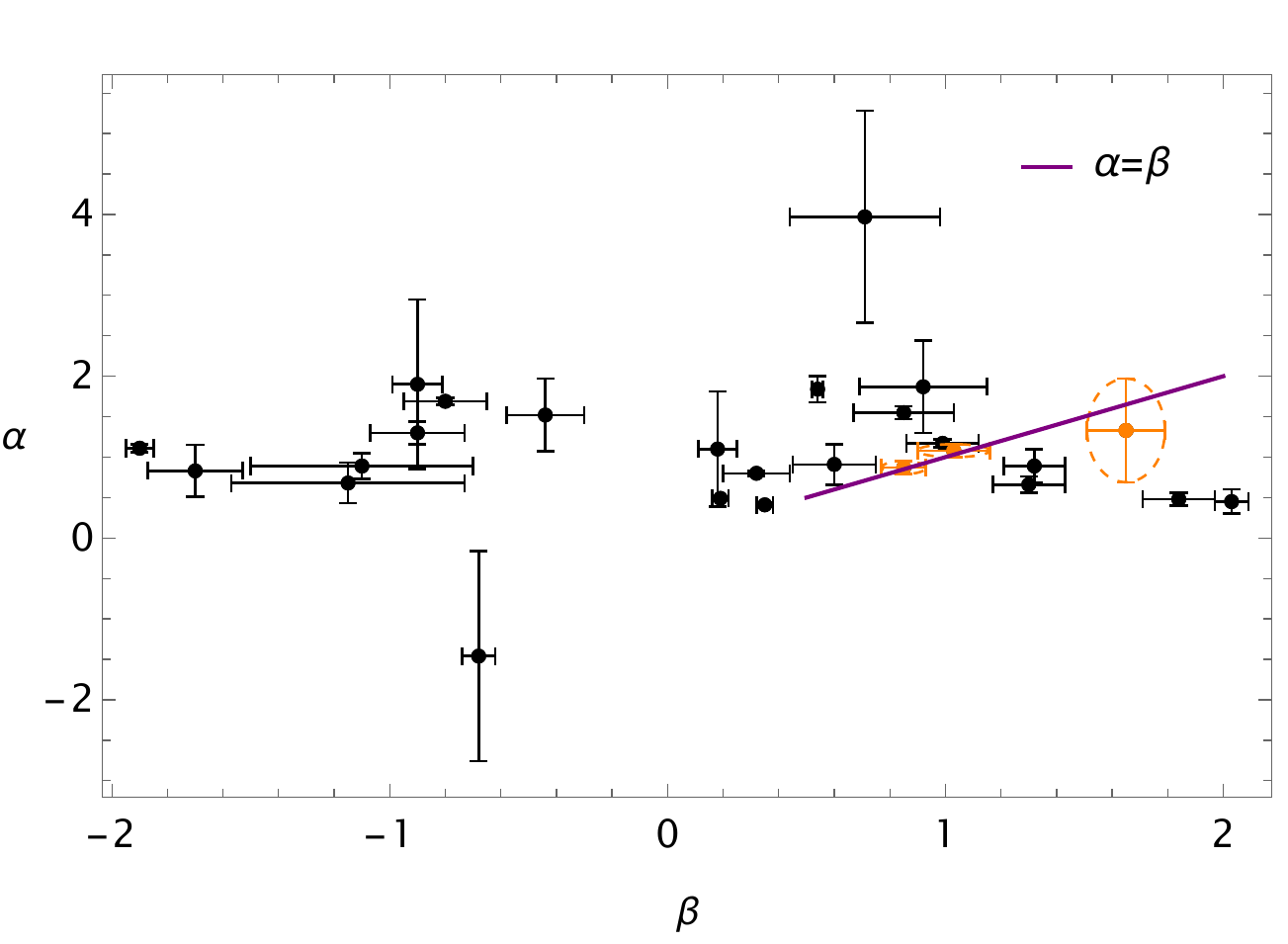} &
    \includegraphics[width = 0.28\textwidth]{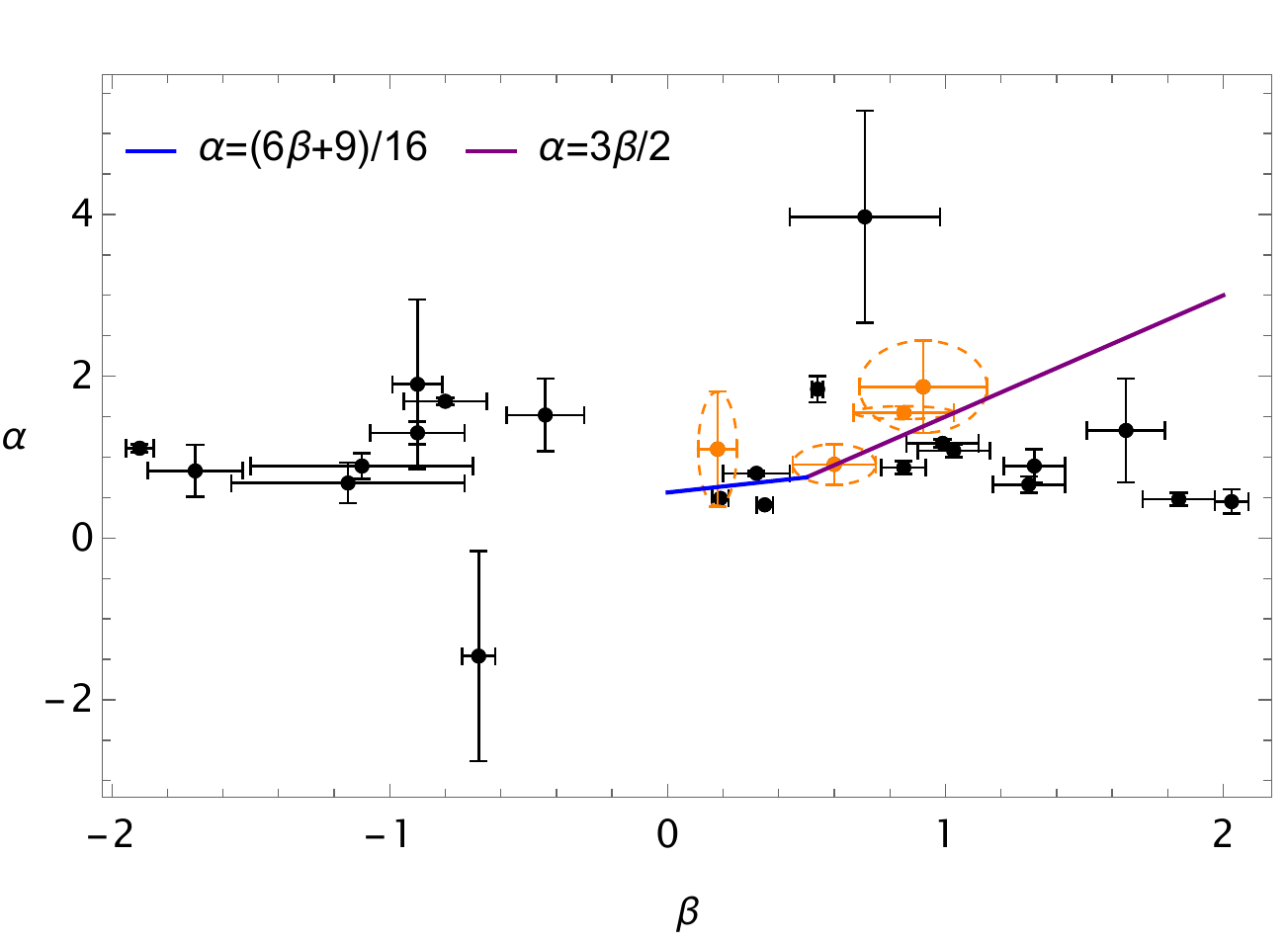} \\
    
    \includegraphics[width = 0.28\textwidth]{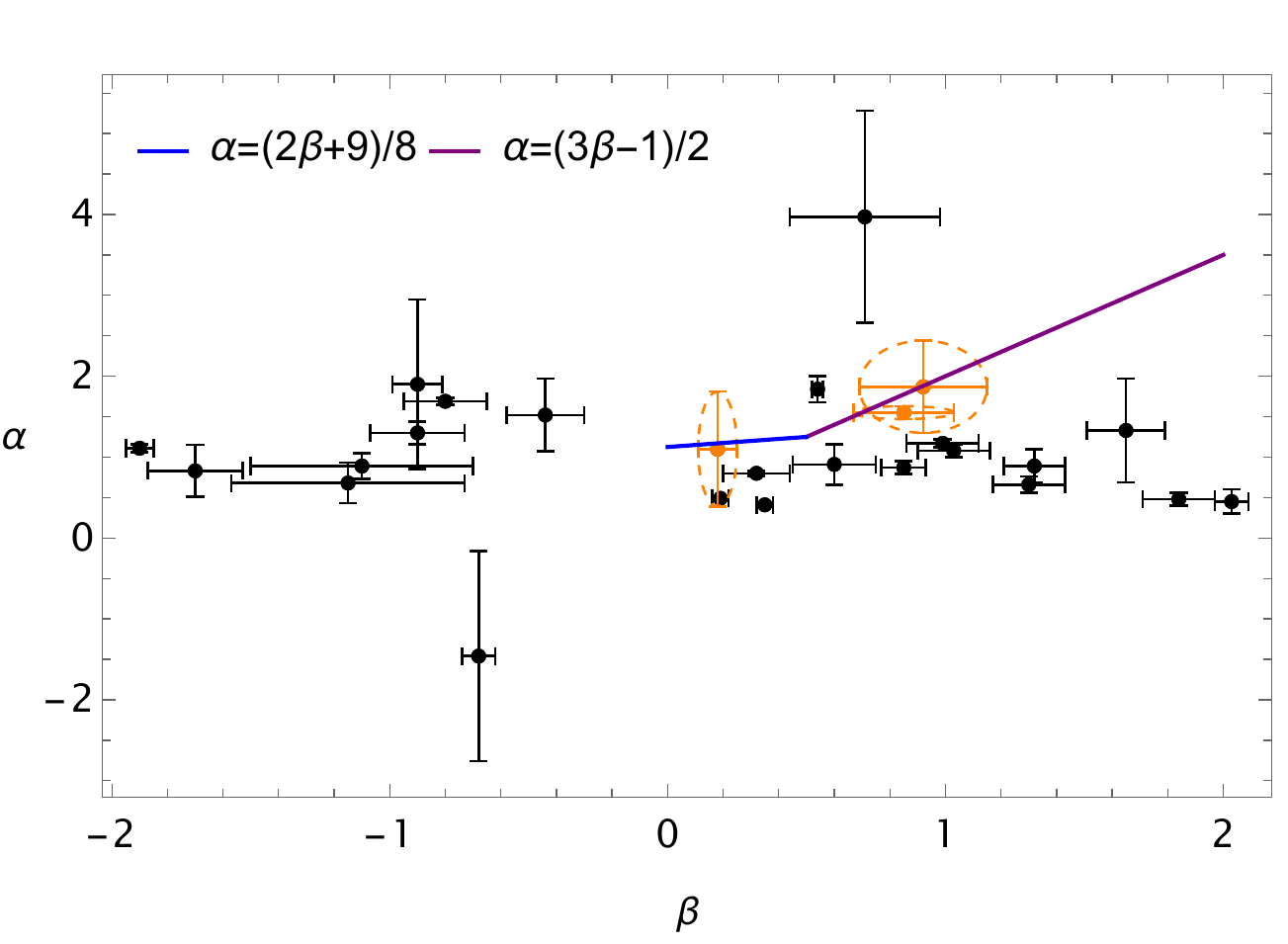} &
    \includegraphics[width = 0.28\textwidth]{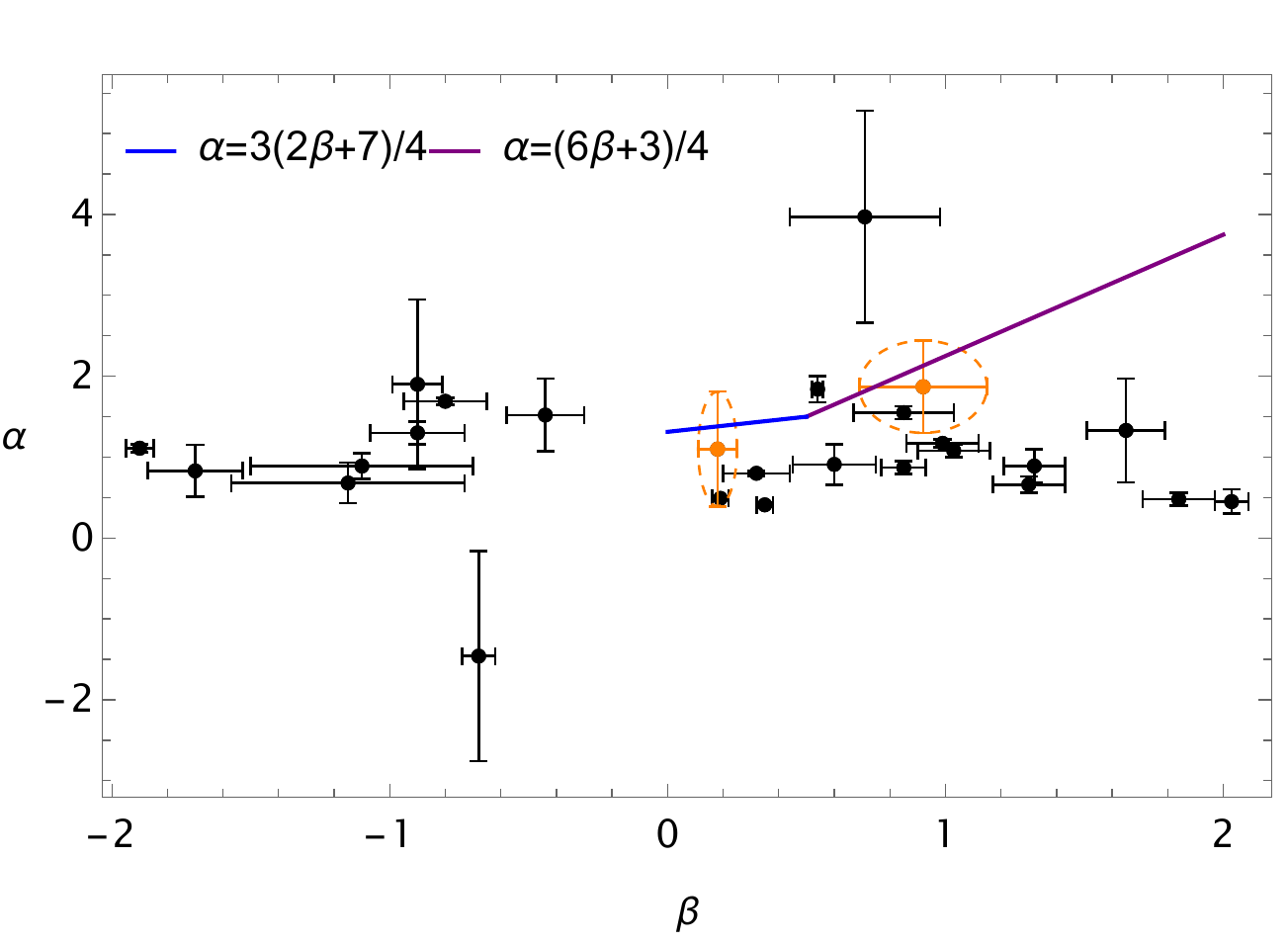} &
    \includegraphics[width = 0.28\textwidth]{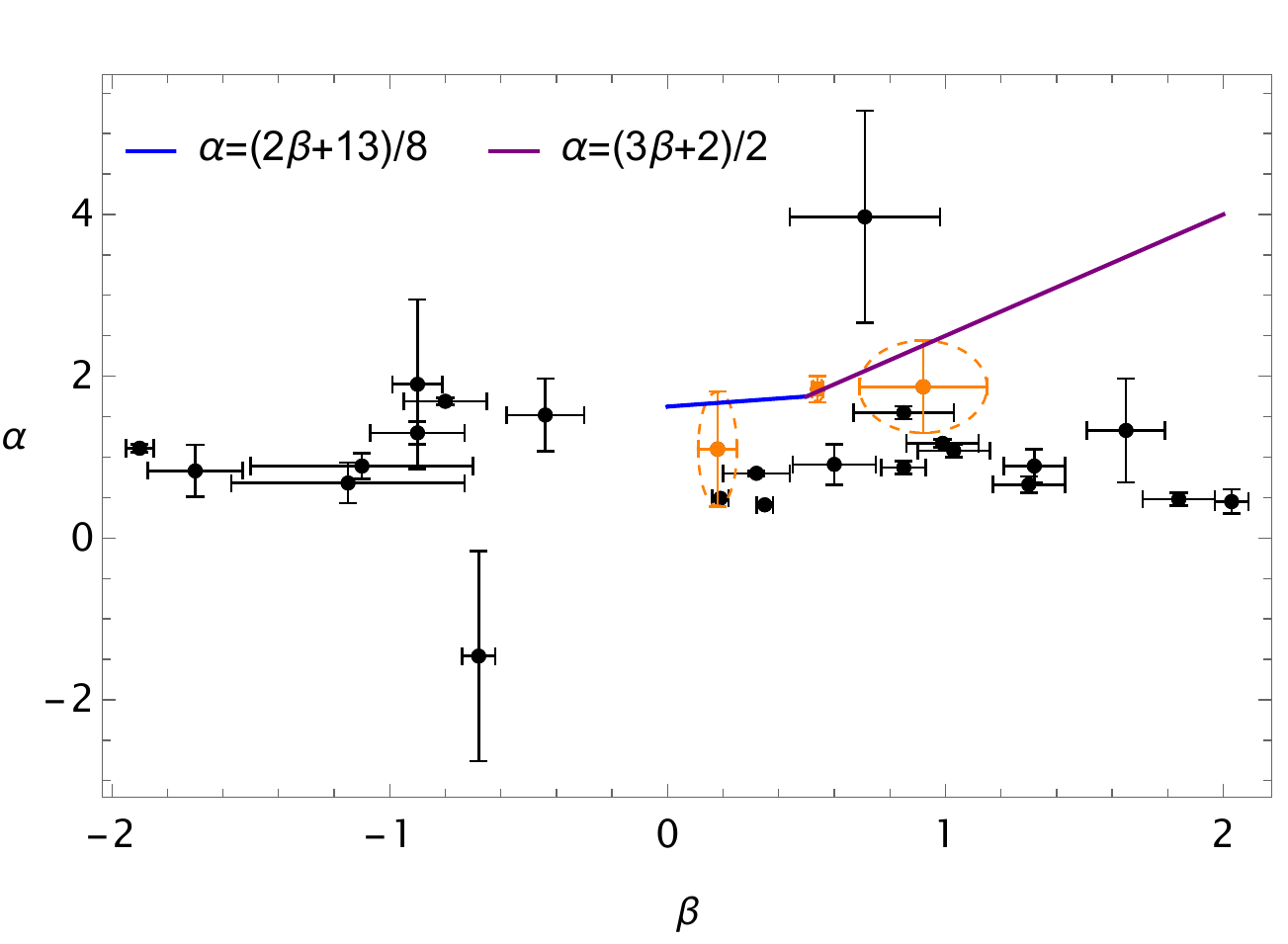} \\
    
    \includegraphics[width = 0.28\textwidth]{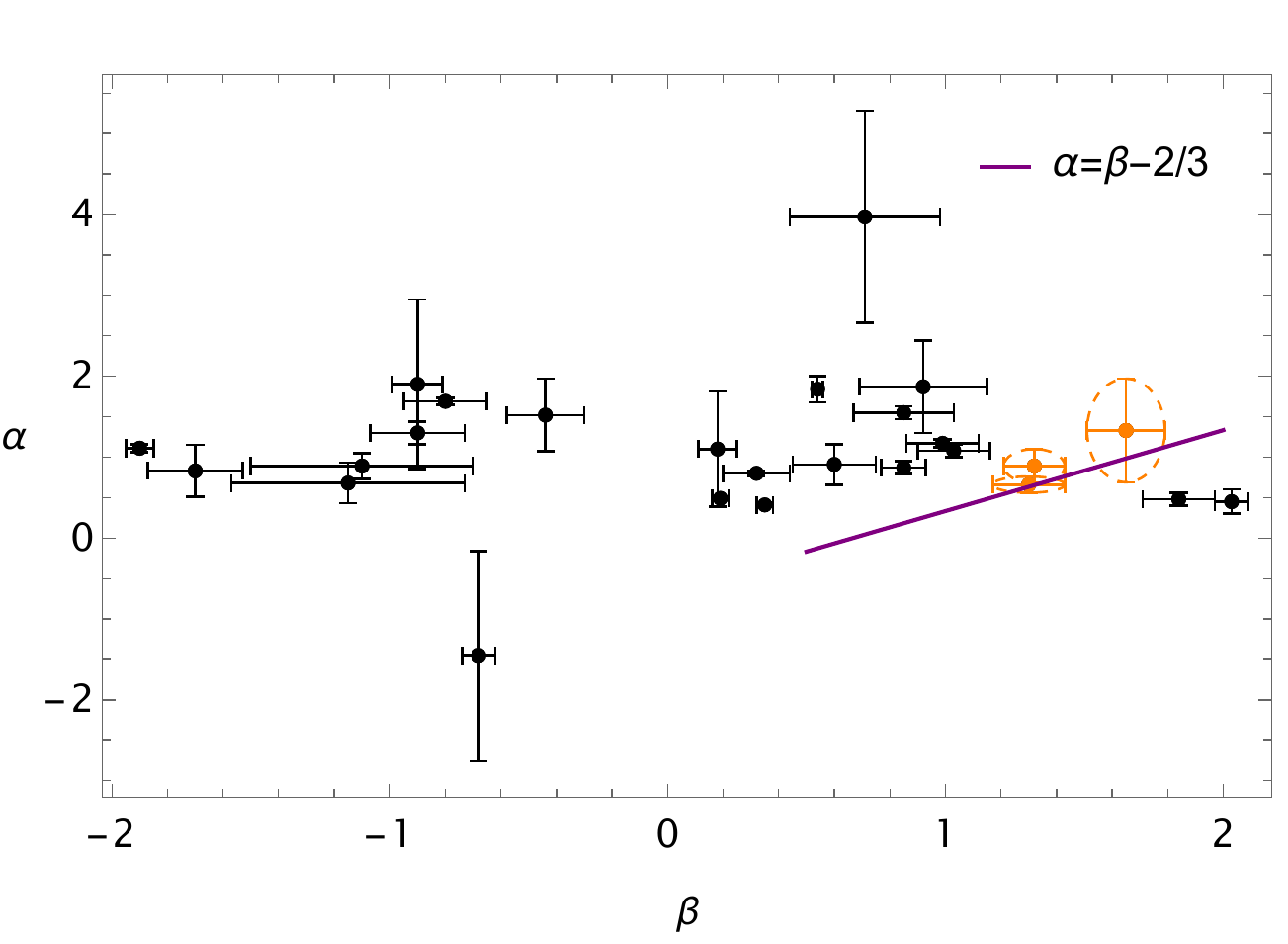} & 
    \includegraphics[width = 0.28\textwidth]{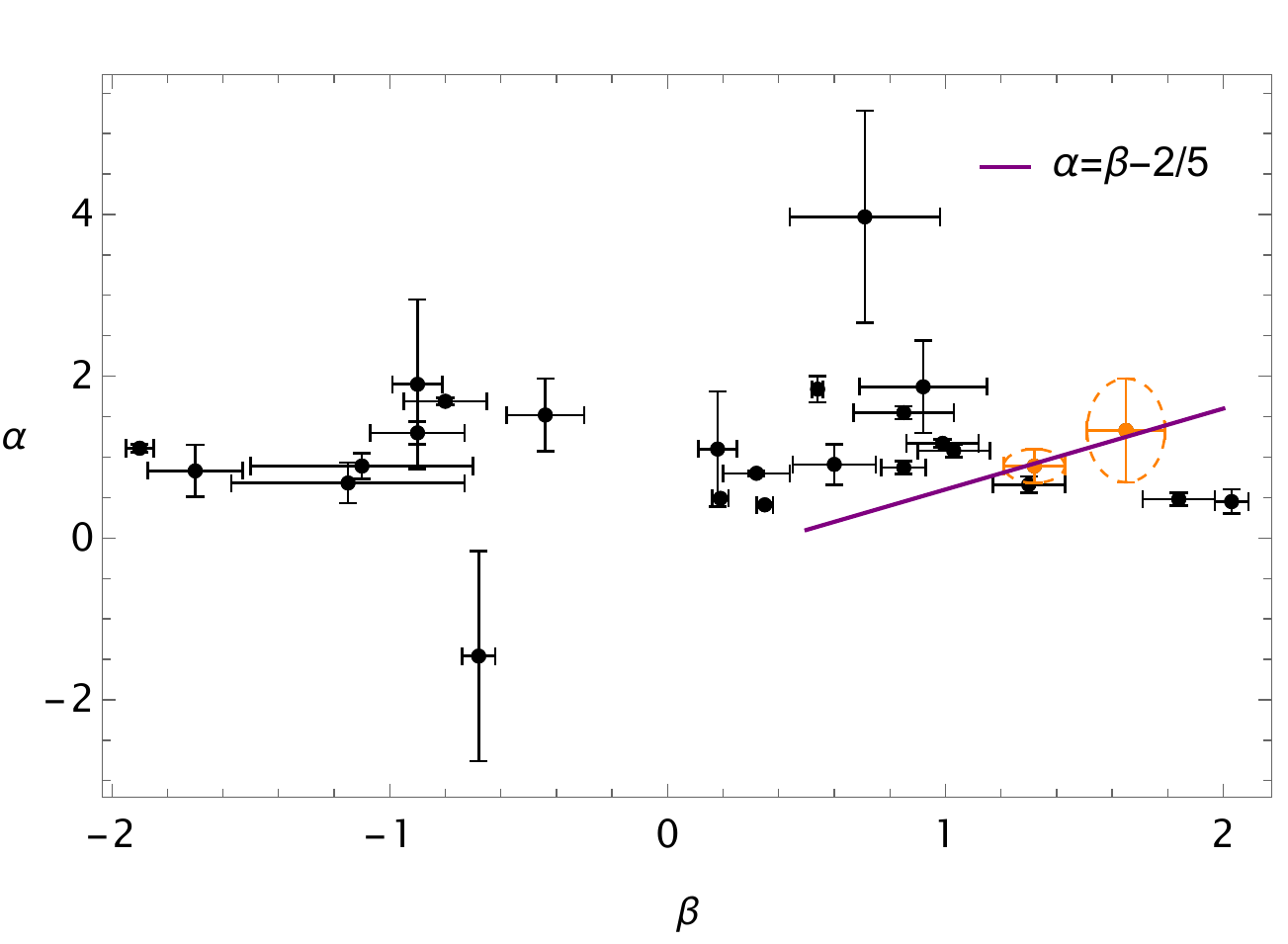} &
    \includegraphics[width = 0.28\textwidth]{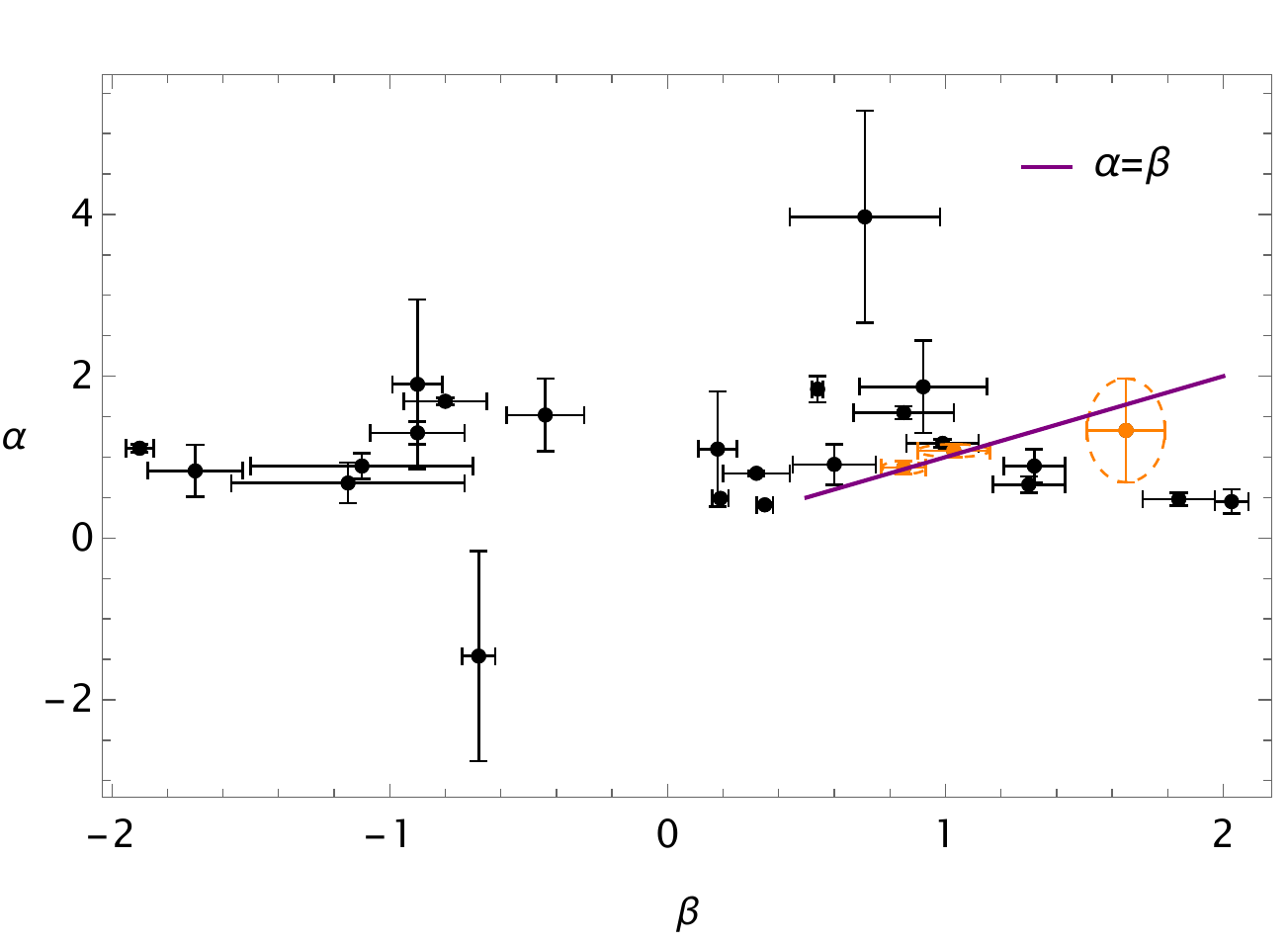} \\
    
    \includegraphics[width = 0.28\textwidth]{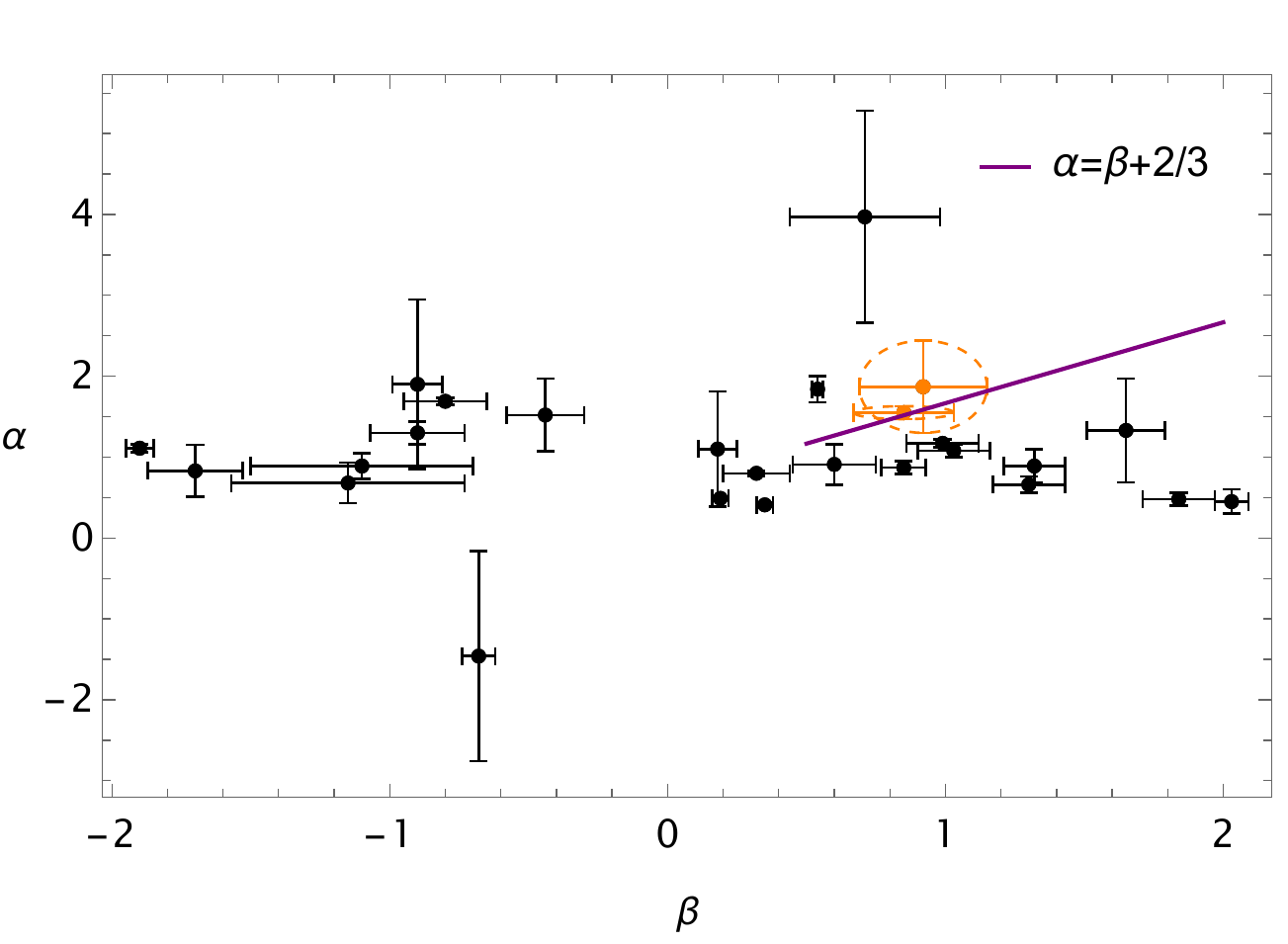} &
    \includegraphics[width = 0.28\textwidth]{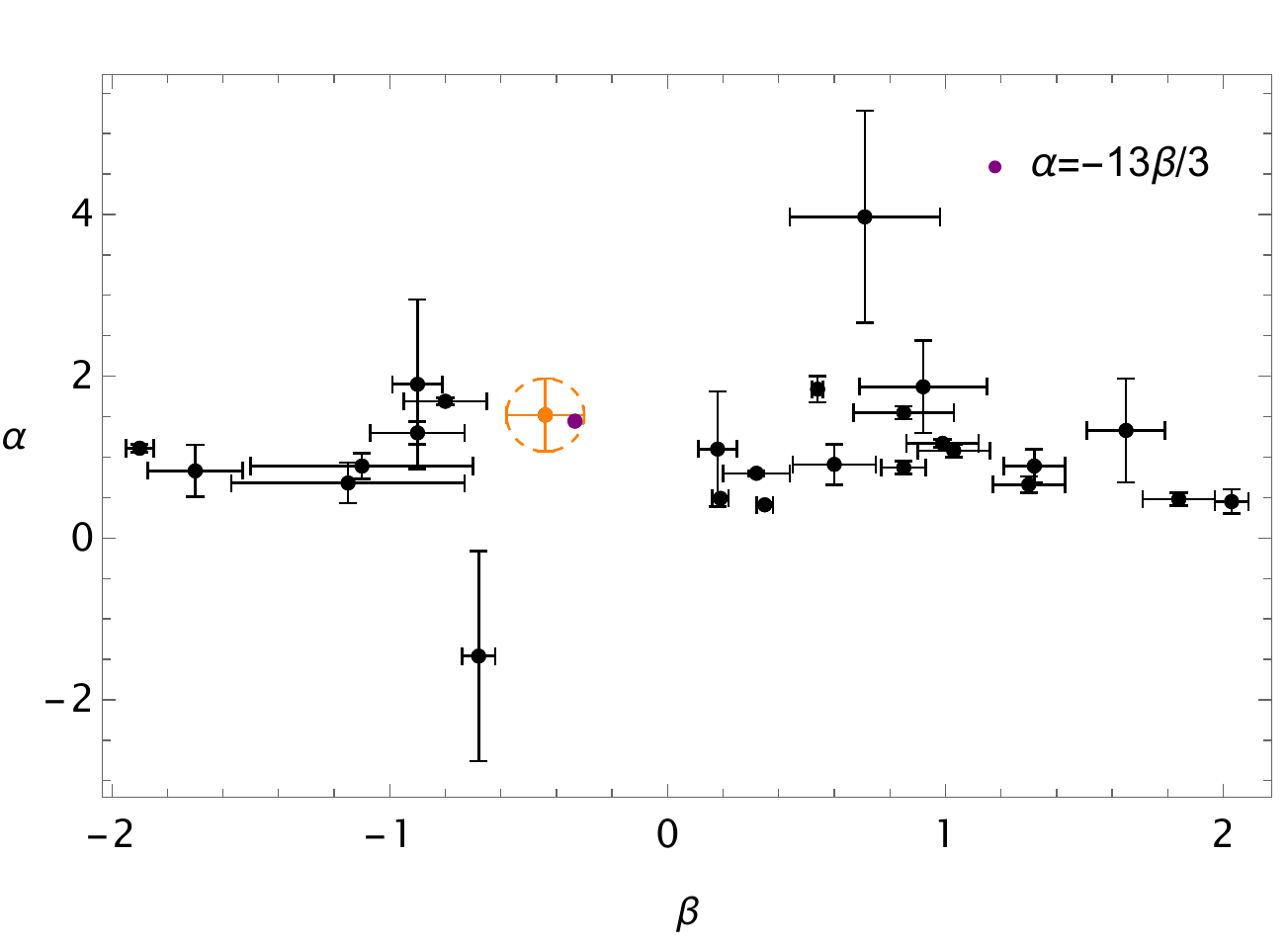} \\
\end{tabular}
\caption{Plots of fulfilled CRs, without energy injection, tested against full sample of 26 GRBs. Blue and purple colored lines represent the $1<p<2$ and $p>2$ spectral regimes, respectively. GRBs that fulfill the CR are shown in orange, GRBs that do not fulfill the CR are shown in black. Plots are as follows: (1) ISM, SC, $\nu_{\rm m} < \nu < \nu_{\rm c}$ CRs in Tables \ref{tab:2} and \ref{tab:3}; (2) Wind, SC, $\nu_{\rm m} < \nu < \nu_{\rm c}$ CRs in Tables \ref{tab:2} and \ref{tab:3}; (3) ISM, SC, $\nu_{\rm m} < \nu < \nu_{\rm c}$ CRs in Tables \ref{tab:4} and \ref{tab:5}; (4) Wind, SC, $\nu_{\rm m} < \nu < \nu_{\rm c}$ CRs in Tables \ref{tab:4} and \ref{tab:5}; (5) ISM, SC, $\nu_{\rm m} < \nu < \nu_{\rm c}$ CRs in Tables \ref{tab:6} and \ref{tab:7}; (6) Wind, SC, $\nu_{\rm m} < \nu < \nu_{\rm c}$ CRs in Tables \ref{tab:6} and \ref{tab:7}; (7) $k=1$, SC, $\nu_{\rm m} < \nu < \nu_{\rm c}$; (8) $k=1.5$, SC, $\nu_{\rm m} < \nu < \nu_{\rm c}$; (9) $k=2$, SC, $\nu_{\rm m} < \nu < \nu_{\rm c}$; (10) $k=2.5$, SC, $\nu_{\rm m} < \nu < \nu_{\rm c}$; (11) $k=2.5$, FC, $\nu < \nu_{\rm c}$.}
\label{fig:noinj} 
\end{figure*}

\begin{figure*} 
\centering
\begin{tabular}{cc}
    \includegraphics[width = 0.45\textwidth]{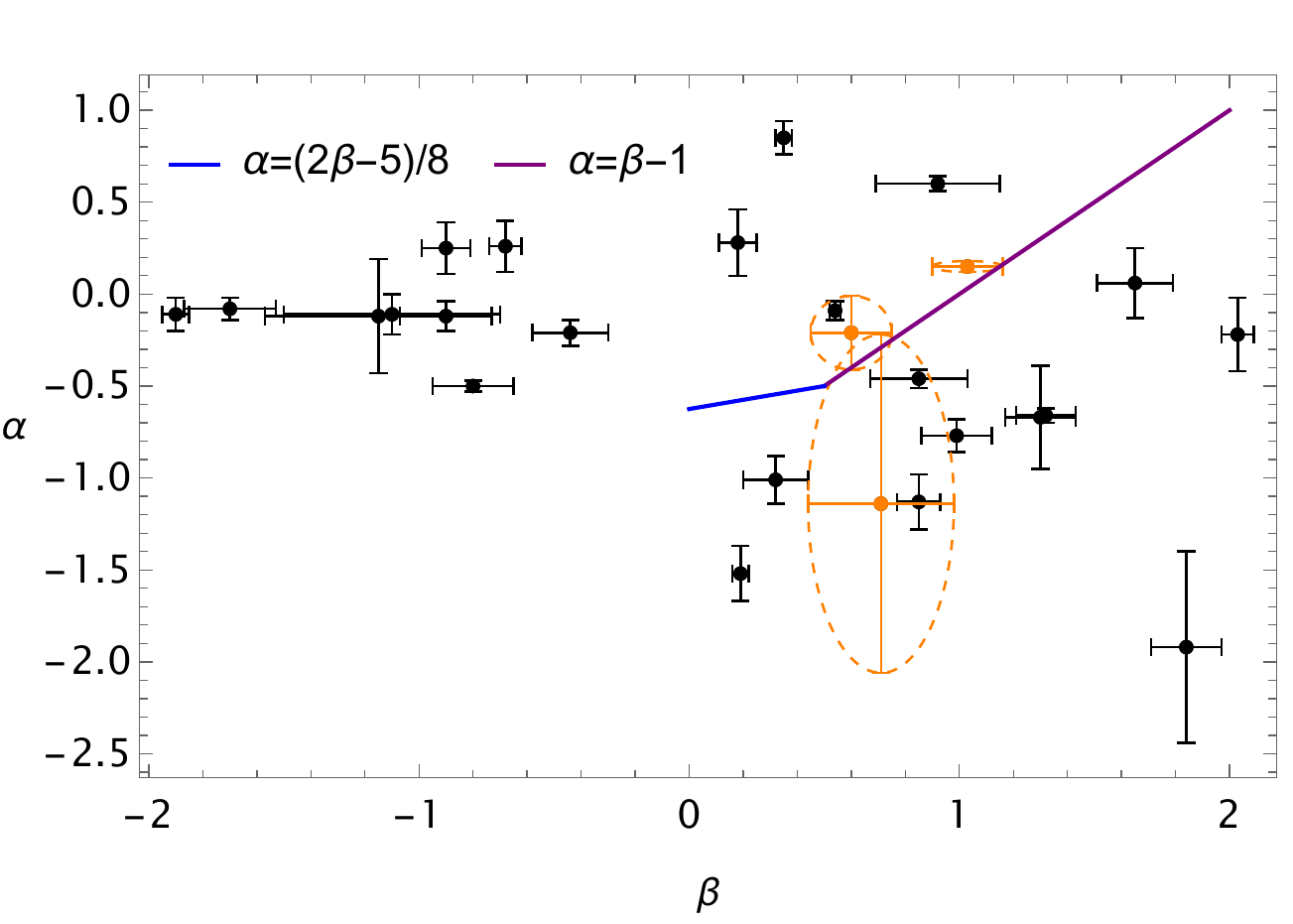} & 
    \includegraphics[width = 0.45\textwidth]{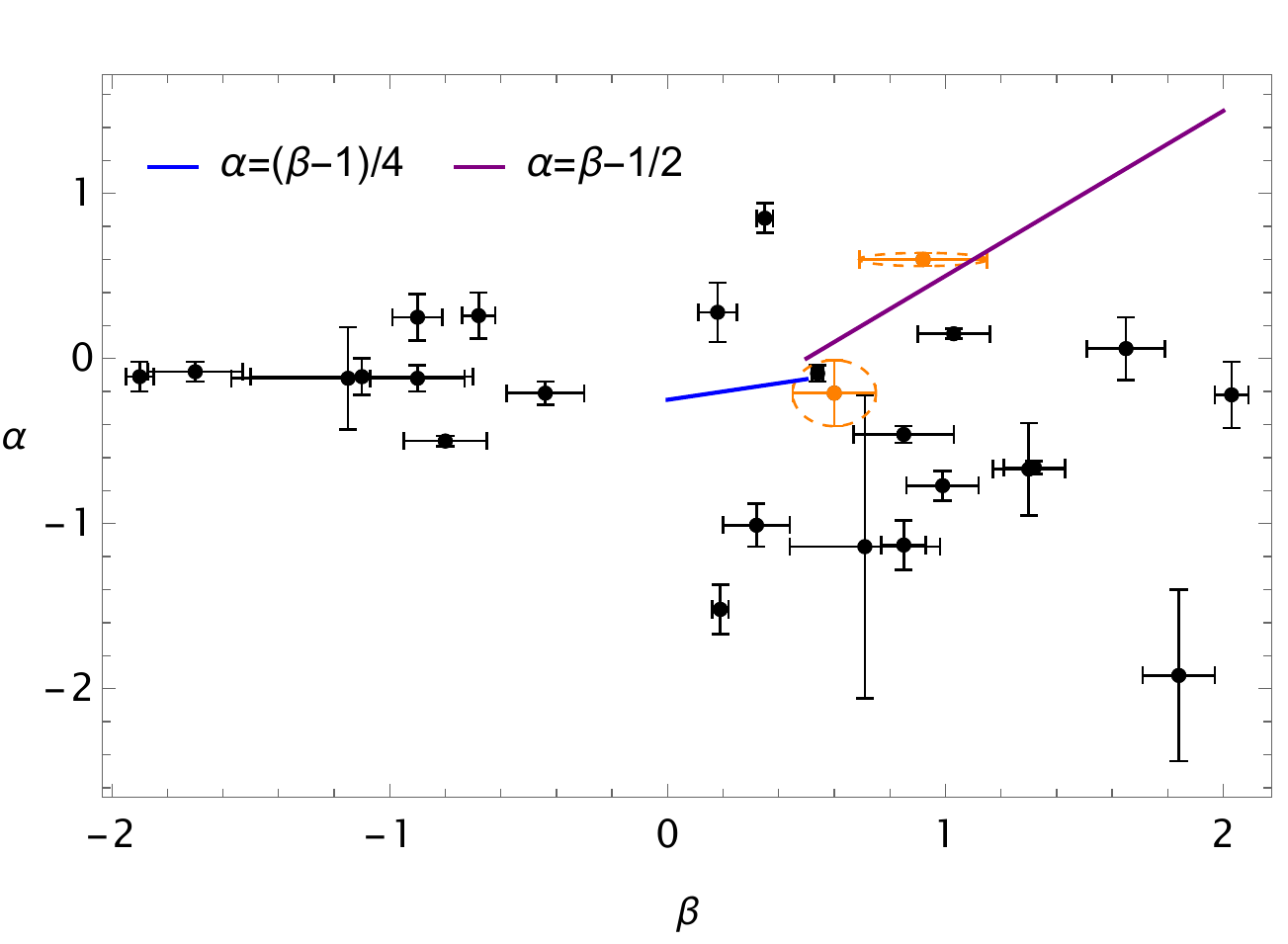}
\end{tabular}
\caption{Plots of fulfilled CRs, with energy injection, tested against full sample of 26 GRBs. Blue and purple colored lines represent the $1<p<2$ and $p>2$ spectral regimes, respectively. GRBs that fulfill the CR are shown in orange, GRBs that do not fulfill the CR are shown in black. The left plot shows the ISM, SC, $\nu_{\rm m} < \nu < \nu_{\rm c}$ CRs in Table \ref{tab:4} and the right plot shows the ISM, SC, $\nu_{\rm m} < \nu < \nu_{\rm c}$ in  Table \ref{tab:6}.
\label{fig:inj}}
\end{figure*}

Considering individual GRBs that satisfy at least one CR, but do not display a radio plateau, GRB 011121 fulfills the SC, $\nu_{\rm m} < \nu < \nu_{\rm c}$ regime without injection in Table \ref{tab:8} for $k=1$. 
GRB 060218 fulfills the SC, $\nu_{\rm m} < \nu < \nu_{\rm c}$ regime with energy injection in Tables \ref{tab:4} and \ref{tab:5} for an ISM environment. 
GRB 100814A fulfills the SC, $\nu_{\rm m} < \nu < \nu_{\rm c}$ regime without injection in Tables \ref{tab:2} and \ref{tab:3} for a Wind environment and the same regime in Table \ref{tab:8}, also for a Wind environment. 
GRB 110715A fulfills the SC, $\nu_{\rm m} < \nu < \nu_{\rm c}$ regime without injection in Table \ref{tab:8} for $k=1, 1.5$.
GRB 120326A fulfills the SC, $\nu_{\rm m} < \nu < \nu_{\rm c}$ regime without injection in Tables \ref{tab:4}, \ref{tab:5}, \ref{tab:6}, and \ref{tab:7} for both an ISM and Wind environment, as well as Table \ref{tab:8} for $k=2.5$. It also satisfies the same regime with energy injection in Tables \ref{tab:6} and \ref{tab:7} for an ISM environment.
The fulfilled CRs for the full sample of 26 GRBs are shown in Figure \ref{fig:noinj} for CRs without injection and Figure \ref{fig:inj} for CRs with injection.

\section{Discussion and Conclusions}
\label{sec:discussion}

We have conducted an analysis of CRs in radio wavelengths, with an investigation into the behavior of GRBs that present a radio plateau, as well as an investigation of a broader sample of GRBs that present a break in their LCs. We consider scenarios both with and without energy injection. We find that for the full sample of 26 GRBs that present a break in their LC, the majority of the LCs do not fulfill any CR within our set, indicating that they are incompatible with the expectations of the standard fireball model\footnote{We are cognizant of the small-scale statistics inherent to our sample size and limited availability of published spectral data within the radio regime.}. For the 12 GRBs that do satisfy at least one CR in our set, relations without energy injection are preferred over those with injection, with the most favored region being the SC, $\nu_{\rm m} < \nu < \nu_{\rm c}$. There is no clear preference for either an ISM or Wind environment.

For the subsample of 14 GRBs that display a radio plateau, we see that the results follow similar trends as the full sample, with roughly 50\% of GRBs fulfilling at least one CR in our set and with the SC, $\nu_{\rm m} < \nu < \nu_{\rm c}$ without injection being the most preferred environment. If we consider the ratio of GRBs with and without plateaus fulfilling the CR we can clearly see that the GRBs with plateau emission present a better fulfillment rate (7/12) compared to the GRBs which do not have plateaus. With 12 GRBs we indicate the total number of GRBs satisfying at least one CR and with 7 GRBs we indicate the number of GRBs satisfying at least one CR with radio plateaus. This indicates that though relations without energy injection appear to be preferred over relations with injection, implying that the consideration of energy injection does not necessarily improve the agreement of radio data with the standard fireball model, GRBs with a radio plateau do appear to be more likely to agree with the standard fireball model.

Regarding the timescales of the events, here they range from $10^{5.52}$ to $10^{6.96}$ seconds as detailed in Table \ref{tab:Sample}.
The decelerating material enters a non-relativistic phase once it has sufficiently swept the surrounding medium. It usually occurs at timescales from days to months, depending on the parameter values. The material's evolution is impacted by this transition, which in turn affects the synchrotron LC. In this case the CRs considered are deviated. It is worth noting that during this phase the dynamics of the decelerated material is described by the Sedov-Taylor solution instead of the Blandford-McKee considered here. 

Investigations done in the literature have also found compatible results with our analysis. \citet{2020ApJ...903...18S} compiled a set of 455 X-ray LCs from GRBs observed by \textit{Swift} and showed that the SC regime within both the stellar wind and ISM environment is favored when studying phase III for LCs with plateau emission. \citet{2015ApJS..219....9W} reached a similar conclusion using X-ray and optical observations and implementing the analysis used in \citet{2007ApJ...662.1093W}. \citet{2021PASJ..tmp...63D} used the same sample of 455 LCs and found that the majority of their sample fulfilled the stellar wind slow-cooling CRs when considering energy injection during the plateau emission phase using the W07 model. Furthermore, \citet{2021ApJS..255...13D} considered $\gamma$-ray emission for the three GRBs which have indication of the plateau emission (GRB 090510, GRB 090902B and GRB 160509) soon after the plateau observed by Fermi-LAT and found that the SC regime within both ISM and stellar wind environments is the most preferred regime.

In radio, KF21 used a sample of 21 GRBs observed by \textit{Swift} containing evidence of a jet break and found that the fireball model did not provide a good fit for radio LCs when compared to their optical and X-ray counterparts, underlining the importance of continued radio follow-up observations. Regarding the GRBs tested by KF21 present in our sample, the authors find that GRB141121A, observed at 15 GHz, can be modeled with a simple power law in both the Wind and ISM environments. However, the authors report that GRB141121A's LCs in lower frequencies are too complex to be modeled well and thus, they are incompatible with the standard fireball model; this is incompatible with our results since we use the radio LC observed at 13 GHz for the case of GRB 141121A. \citet{2021MNRAS.504.5685M} also analyzed the CRs with the radio afterglow of GRB190114C (not in our sample), detected by MAGIC. The authors analyzed radio and X-ray LCs with a simple power law model, and found the LC incompatible with the standard model.

\citet{2020ApJ...903...18S}, \citet{2021ApJS..255...13D}, and \citet{2021PASJ..tmp...63D}, investigated $\gamma$-ray and X-ray LCs, showing that the most favored model supports the SC regime, which is also the most preferred environment for the GRBs in our study. We also note that we have investigated the cases that show plateau emission in radio afterglows, which have not been an object of extensive study before.
In conclusion, it is challenging to draw a definite picture of the fulfillment or not of the CRs, although the main trend for relation both with and without energy injection and is the lack of agreement with the standard fireball model, as seen with other results in radio afterglows in the literature.

\section*{Acknowledgements}
The authors would like to acknowledge Debarpita Jyoti for her assistance in this analysis. This work was made possible in part by the United States Department of Energy, Office of Science, Office of Workforce Development for Teachers and Scientists (WDTS) under the Science Undergraduate Laboratory Internships (SULI) program. We thank Dr. Cuellar for managing the SULI program at Stanford National Accelerator Laboratory. We acknowledge the National Astronomical Observatory of Japan for their support in making this research possible through the Division of Science support. PC acknowledges support of the
Department of Atomic Energy, Government of India, under project no. 12-R\&D-TFR-1155 5.02-0700
M.G. Dainotti acknowledges the support of the Division of Science and NAOJ.
D. Levine acknowledge the support of the Division of Science and NAOJ and the United States Department of Energy in funding the Science Undergraduate Laboratory Internship (SULI) program. 

\section*{Data Availability}

The data used in this investigation is presented in the tables and the lightcurve data is taken from:  \citet{2012ApJ...746..156C, 2018A&A...616A.169M, 2013ApJ...767..161Z, 2015ApJ...814....1L, 2016ApJ...833...88L,  2018ApJ...858...65L, 2019ApJ...884..121L, 2015ApJ...812..122C, 2019MNRAS.486.2721B, 2017MNRAS.464.4624S, 2014ApJ...781...37P, 2014MNRAS.440.2059A, 2015ApJ...806...52S, 2019MNRAS.484.5245H, 2020ApJ...894...43K, 2020ApJ...891L..15C, 2017ApJ...845..152B, 2017Sci...358.1579H, 2018Natur.554..207M, 2018ApJ...867...57R, 2018ApJ...856L..18M, 2021ApJ...907...60M, 2020MNRAS.496.3326R}




\bibliographystyle{mnras}
\bibliography{fireball.bib}




\bsp	
\label{lastpage}
\end{document}